\documentclass[final,3p,times]{elsarticle}

\usepackage{epsfig}

\usepackage{amssymb}
\usepackage[utf8]{inputenc}

\usepackage{amsmath}
\usepackage{graphicx}
\usepackage{subcaption}
\usepackage{tikz}
\usetikzlibrary{shapes}
\usepackage{comment}

\newcounter{bla}

\journal{Computer Physics Communications}

\begin{document}

\begin{frontmatter}



\title{BROADCAST: A high-order compressible CFD toolbox for stability and sensitivity using Algorithmic Differentiation}


\author[a]{Arthur Poulain\corref{author}}
\author[a]{Cédric Content}
\author[a]{Denis Sipp}
\author[b]{Georgios Rigas}
\author[c]{Eric Garnier}

\cortext[author] {Corresponding author.\\\textit{E-mail address:} arthur.poulain@onera.fr}
\address[a]{DAAA, ONERA, Université Paris Saclay, F-92190 Meudon, France}
\address[b]{Imperial College London, South Kensington, London, United Kingdom}
\address[c]{Univ. Lille, CNRS, ONERA, Arts et Metiers Institute of Technology, Centrale Lille, UMR 9014 - LMFL - Laboratoire de Mécanique des Fluides de Lille - Kampé de Fériet, F-59000 Lille, France}

\begin{abstract}
The evolution of any complex dynamical system is described by its state derivative operators. However, the extraction of the exact N-order state derivative operators is often inaccurate and requires approximations. The open-source CFD code called BROADCAST discretises the compressible Navier-Stokes equations and then extracts the linearised N-derivative operators through Algorithmic Differentiation (AD) providing a toolbox for laminar flow dynamic analyses. Furthermore, the gradients through adjoint derivation are extracted either by transposition of the linearised operator or through the backward mode of the AD tool. The software includes base-flow computation and linear global stability analysis via eigen-decomposition of the linearised operator or via singular value decomposition of the resolvent operator. Sensitivity tools as well as weakly nonlinear analysis complete the package. The numerical method for the spatial discretisation of the equations consists of a finite-difference high-order shock-capturing scheme applied within a finite volume framework on 2D curvilinear structured grids. The stability and sensitivity tools are demonstrated on two cases: a cylinder flow at low Mach number and a hypersonic boundary layer.

\end{abstract}


\end{frontmatter}



{\bf PROGRAM SUMMARY}

\begin{small}
\noindent
{\em Program Title:} BROADCAST                                     \\
{\em CPC Library link to program files:} (to be added by Technical Editor) \\
{\em Code Ocean capsule:} (to be added by Technical Editor) \\
{\em Licensing provisions:} Mozilla Public License 2.0  \\
{\em Programming language:}  Python, Fortran                                 \\
{\em Nature of problem:} The extraction of the exact N-order state derivative operators useful to study complex dynamical systems is often inaccurate and requires approximations from the user. \\
{\em Solution method:} This program discretises the compressible Navier-Stokes equations and then extracts the linearised state derivative operators through Algorithmic Differentiation providing a toolbox for laminar flow dynamic analyses. 
\end{small}



\section{Introduction}

\begin{figure}[!ht]
\centering
\begin{tikzpicture}[scale=1.]
\tikzstyle{base} =[rectangle,draw,rounded corners=4pt,fill=blue!30]
\tikzstyle{basebis} =[rectangle,draw,dashed,rounded corners=4pt,fill=blue!15]
\tikzstyle{cyl} =[rectangle,draw,rounded corners=4pt,fill=red!30]
\tikzstyle{cylbis} =[rectangle,draw,dashed,rounded corners=4pt,fill=red!15]
\tikzstyle{suite}=[->,>=latex,thick,rounded corners=4pt]
\tikzstyle{suitebis}=[->,>=latex,dashed,thick,rounded corners=4pt]
\draw (-4,2.5) node[] {DYNAMICAL SYSTEM TOOLS};
\node[base] (1) at (-4,1.5) {Fixed-point};
\node[base] (2) at (-4,0) {Linear stability};
\node[base] (3) at (-4,-1.5) {Adjoint methods};
\node[base] (4) at (-4,-3) {Weakly nonlinear analysis};
\node[basebis] (5) at (-4,-4.5) {Periodic orbits and Floquet stab.};
\draw (4,2.5) node[] {APPLICATION TO AN OSCILLATOR FLOW};
\node[cyl] (11) at (2.9,1.5) {Base-flow};
\node[cyl] (22) at (3.5,0) {Global eigenmode};
\node[cyl] (33) at (3.3,-1.5) {Sensitivity map};
\node[cyl] (44) at (4.9,-3) {Sub- or super-criticality of bifurcation};
\node[cylbis] (55) at (3.6,-4.5) {Bifurcation diagram};
\draw[suite] (1) -- (2);
\draw[suite] (2) -- (3);
\draw[suite] (3) -- (4);
\draw[suitebis] (4) -- (5);
\draw[suitebis] (5) -- (55) node[midway,above] {HBM / TSM};
\draw[suite] (1) -- (11) node[midway,above] {$R(\overline{q}) = 0$} node[midway,below] {$N=0$ derivative};
\draw[suite] (2) -- (22) node[midway,above] {$A \hat{q} = \lambda \hat{q}$} node[midway,below] {$N=1$ derivative};
\draw[suite] (3) -- (33) node[midway,above] {$\nabla_{\overline{q}} \lambda = H'^* q^\dag$} node[midway,below] {$N=2$ derivative};
\draw[suite] (4) -- (44) node[midway,above] {$\frac{d \mathcal{A}}{dt_1} = \kappa \mathcal{A} - \mu \mathcal{A}|\mathcal{A}|^2$} node[midway,below] {$N=3$ derivative};
\end{tikzpicture}
\caption{Dynamical system analysis and its application to an oscillator flow.}
\label{graphsys3}
\end{figure}
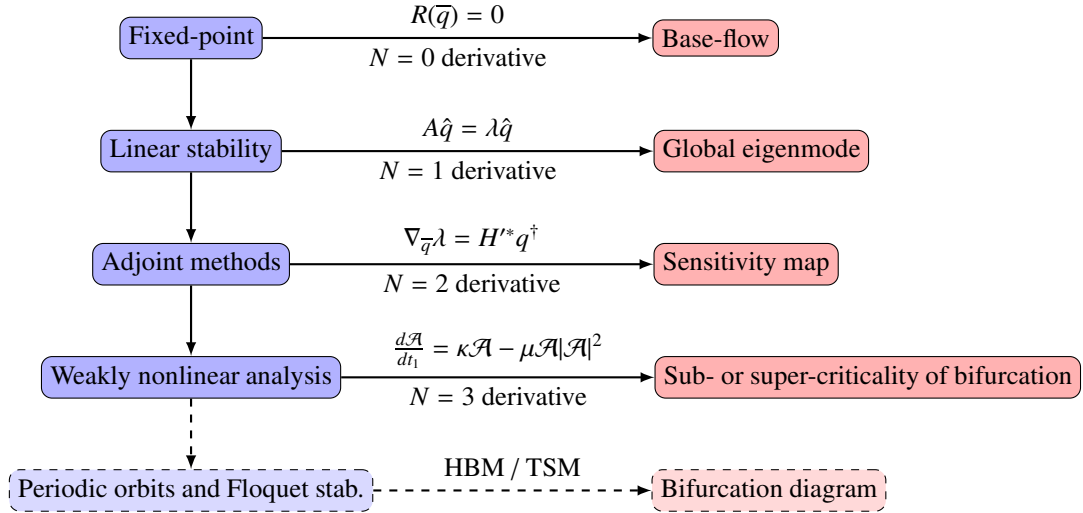

In high-speed aerodynamics, the governing flow equations involve complex physics, for example discontinuities with shock waves, propagation of small-amplitude acoustic waves, perturbations that grow due to various instability mechanisms. The accurate simulation of such physics requires high-order non-linear discretisation schemes. Hence, once the governing equations are spatially discretised, the flow-state $q$ is high-dimensional and the residual $ R(q) $, such that
\begin{equation} \label{eq:ns}
    \partial q / \partial t = R(q,\delta),
\end{equation}
is strongly nonlinear (both due to the physics and the spatial discretisation schemes). We have here made explicit the dependence of the dynamics with respect to an external control parameter $ \delta $, which can be for example the Reynolds or Mach numbers. Within the framework of dynamical systems, analysing the system consists in determining, as a function of the parameter $ \delta$, all fixed-points, periodic orbits, and their respective stability properties to establish bifurcation diagrams. The sensitivity and control of such systems, for example understanding how a weak forcing (uncertainty) or open/closed-loop control affects these bifurcation diagram, are crucial for the optimisation of transport vehicles.
The accurate evaluation of the $ N^{th}-$order derivative of the residual $ R(q) $ steps in as a key-enabler to deal with the points mentioned above (see Figure \ref{graphsys3} for overview).

The base-flow $ \overline{q} $ corresponds to a fixed-point of the zeroth order $N=0$ derivative of the residual:
\begin{equation} \label{Newton}    
R(\overline{q})=0.    
\end{equation}
Several methods are available to determine such points: global (explicit or implicit) time-stepping methods, local implicit time-stepping methods \citep{jameson1981implicit} or Newton-Raphson algorithms \citep{crivellini2011implicit} are most common. Many of these are based on the solution of a linear system involving the first-order $N=1$ derivative of the residual, the Jacobian matrix
$ A=\partial R/\partial q $.

Once a fixed-point is found, the Jacobian matrix is also used to assess its stability \citep{huerre1990local}. The asymptotic stability is evaluated by looking at the most unstable eigenvalues of $A$ \citep{jackson1987finite}:
\begin{equation} \label{eq:gm}
A\hat{q} = \lambda \hat{q}.
\end{equation}
If the real part of $ \lambda $ is positive, then the fixed point is unstable at large times. Pseudo-resonance properties \citep{trefethen1993hydrodynamic,sipp2013characterization} associated to $A$ are assessed by looking, for a given frequency $\omega$, at the largest eigenvalues $ \mu^2 $ of the Hermitian matrix:
\begin{equation} \label{eq:sv}
    {\mathcal R}{\mathcal R}^*  \check{q}=\mu^2\check{q},
\end{equation} 
where ${\mathcal R}=(i\omega I-A)^{-1} $ is the resolvent operator and the matrix ${\mathcal R}^*=(-i\omega I-A^*)^{-1}$ may be evaluated either from the adjoint discrete operator or with the conjugate transpose of $\mathcal R$ or $A$.
Globally-stable base-flows ($ \Re(\lambda) <0 $ for all global modes $ \hat{q}$ ) exhibiting strong pseudo-resonances ($ \mu^2 $ large for at least one resolvent mode $ \check{q} $) are noise-amplifiers according to \citet{huerre1990local}, while globally unstable flows (at least one global mode exhibits $ \Re(\lambda) >0 $) correspond to oscillator flows.  
The adjoint matrix $A^*$ is also a key-ingredient for gradient-based optimisation \citep{gunzburger1991analysis} and sensitivity studies \citep{giannetti2007structural,meliga2014sensitivity}.

Higher-order derivatives such as the second order derivative operator $N=2$ (called Hessian and written $H=\partial^2 R/\partial_{qq}$) allow the performance evaluation of open-loop control strategies \citep{marquet2008sensitivity,mettot2014computation}. It relies on the linear sensitivity of an unstable eigenvalue $ \lambda $ (oscillator flow \citep{marquet2008sensitivity}) or of an optimal gain $\mu^2$ (noise amplifier flow \citep{brandt2011effect}) to perturbations of the base-flow or to the introduction of small-amplitude steady forcings, which may provide qualitative control maps \citep{mettot2013linear}. 
According to \citet{sipp2007global}, the second $N=2$ derivative allows to evaluate the coefficient of the cubic nonlinear term in the Stuart-Landau amplitude equation that governs the behaviour of the instability in the vicinity of the critical parameter $ \delta \approx \delta_c$. Hence, it determines whether a bifurcation is sub- or super-critical \citep{chomaz2005global}, which is of major importance for the determination of the flight-envelope of a vehicle  \citep{crouch2007predicting,moulin2020flutter}. Finally, the computation of periodic orbits \citep{gopinath2005time,rigas2021nonlinear} and the assessment of their stability can be performed by nonlinear Harmonic-Balance (HBM) or Time-Spectral (TSM) methods \citep{canuto2012spectral} and Floquet or Hill-Floquet stability analyses \citep{thomas2010harmonic}.

First and second-order derivative operators are tractable in a continuous framework in the case of simple  governing equations and simple spatial discretisation scheme such as the incompressible Navier-Stokes equations at moderate Reynolds number with Taylor-Hood finite elements and no stabilization \citep{marquet2008sensitivity}. There, all the equations are first derived in a continuous form (direct first-order, adjoint, second-order derivatives) and then discretised. Specific numerical methods and analysis are required at each step to discretise the resulting equations. For instance, FreeFem++ \citep{hecht2012new} and FEniCS \citep{logg2010dolfin} can be used to obtain spatial discretisations of any partial-differential equation from its weak form. 
However, the continuous framework becomes difficult to implement for systems driven by more complex PDE or more complex spatial discretisations, e.g. high-order discretised finite-volume compressible Navier-Stokes equations for advection dominated flows. For example, the continuous linearised and adjoint operators need special treatment in presence of shock waves \cite{giles2001analytic}. Conversely, in a discrete framework, the Jacobian and adjoint matrices are computed from the already discretised (and well-behaved) residual and exhibit the same level of accuracy \cite{de2012efficient}: the discrete method consists in discretising the governing equations and boundary conditions with a given spatial scheme and then computing the linearised and adjoint operators on these discretised set of equations. For instance, the SU2 code is built within this framework \cite{palacios_stanford_2013}.

Within a discrete framework, the linearisation of the residual can be performed by finite difference methods \cite{mettot2014computation} but the derived operators are approximated up to an $\epsilon$ error ($\epsilon^2$ for a second-order finite difference). Moreover, $\epsilon$ is a user-input parameter depending on the normalisation \cite{knoll2004jacobian} and $N$ further $\epsilon$ parameters would be required to compute the $N$-derivative operators. An alternative method to compute the derivative operators is Algorithmic Differentiation (AD) \citep{griewank2008evaluating}. A Source-Transformation AD code reads the script lines of the governing equations and generates a new script with the algorithmically computed analytic differentiation of the discretised equations. The Algorithmic Differentiation retains the same approximation error as the residual. Once compiled, the computation time of AD is of the same order of magnitude as a finite difference method while it reaches machine-precision accuracy. Furthermore, it does not depend on any user-input parameter $\epsilon$. The useful feature of the AD method is that successive derivatives can be computed quickly without further parameters so that $N$-order derivative operators are extracted easily. For instance, the software TAPENADE \citep{hascoet2013tapenade} is a Source-Transformation AD code capable of linearising subroutines written in Fortran and C languages.

The objective of this paper is to present a general framework to solve and analyse large-scale discretised dynamical systems, representing conservative boundary value problems, through direct and adjoint derivatives computed with Algorithmic Differentiation. This method is developed here for the compressible Navier-Stokes equations discretised with high-order finite-volume methods to show how to efficiently and accurately find fixed-points, perform linear global stability and resolvent analysis, compute eigenvalue sensitivity for open-loop control and analyse the criticality of a given bifurcation.

The open-source CFD code BROADCAST, built within the discrete framework, is based on the compressible Navier-Stokes equations. It extracts the exact $N$-order derivatives by Algorithmic Differentiation (AD) through the TAPENADE \cite{hascoet2013tapenade} software. The spatial discretisation scheme relies on a high order shock-capturing numerical scheme for compressible flow simulations \cite{sciacovelli2021assessment}. As a toolbox oriented towards simple test case applications, BROADCAST is restricted to 2D curvilinear multi-block structured meshes and runs as a sequential code. Nonetheless, as BROADCAST is interfaced with PETSc \cite{balay2019petsc}, the Newton solver and stability tools work in parallel through MPI. Even if the base-flow is assumed two-dimensional, linear 3D perturbations are addressed on the same 2D domain by considering spanwise periodic perturbations of a given period, as done for example in Bugeat {\it et al.} \cite{bugeat20193d}. This extends the stability analyses to harmonic 3D linear disturbances at a slightly larger but affordable computational cost. The extension to linear 3D perturbations implemented in BROADCAST follows Bugeat {\it et al.} however, the linearisations are here performed by AD and not by a finite difference method.

The outline of this paper is the following. First, in \S \ref{sec:theory} the theoretical methods and tools included in BROADCAST are described along with their choice of implementation. Starting from the governing equations and the spatial discretisation method, the fixed-point solution computation is then presented. Details of the extraction of the Jacobian are also given. Then, global stability methods (\S \ref{sec:oscillator}) are explained, with a special focus on extraction of resolvent modes (\S \ref{sec:amplifier}) and extension towards 3D instabilities. The theoretical developments underlying linear sensitivity and building of weakly-nonlinear amplitude equations are also briefly recalled. Finally, the different features of the code are illustrated and validated on two test cases: a circular cylinder flow at low Mach number \citep{mettot2014computation, fabre2018practical} (\S \ref{sec:cyl}) and a hypersonic laminar boundary layer \citep{bugeat20193d} (\S \ref{sec:bl}).

\section{Theory and implementation} \label{sec:theory}

In the following, we will present the governing equations, discretisation schemes and various numerical algorithms implemented in BROADCAST.

\subsection{Program implementation}

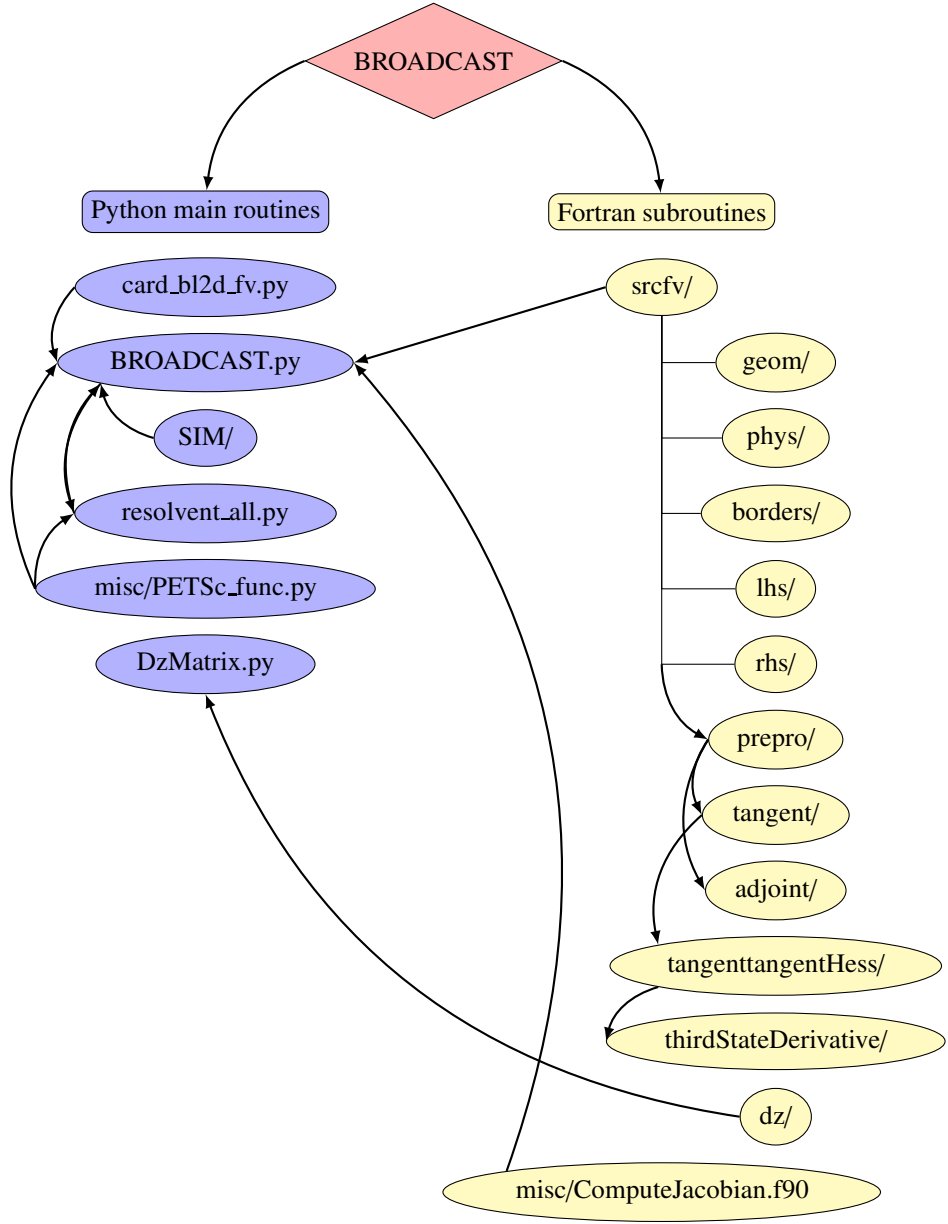
\begin{figure}[!ht]
\centering
\begin{tikzpicture}[scale=1.]
\tikzstyle{broadcast} =[diamond,draw,aspect=2.2,fill=red!30]
\tikzstyle{base} =[rectangle,draw,rounded corners=4pt,fill=yellow!30]
\tikzstyle{base2} =[rectangle,draw,rounded corners=4pt,fill=blue!30]
\tikzstyle{ell} =[ellipse,draw,fill=yellow!30]
\tikzstyle{ell2} =[ellipse,draw,fill=blue!30]
\tikzstyle{suite}=[->,>=latex,thick,rounded corners=4pt]
\node[broadcast] (1) at (0,5) {BROADCAST};
\node[base2] (2) at (-3,3) {Python main routines};
\node[base] (3) at (3,3) {Fortran subroutines};
\draw[suite] (1.west)to[bend right] (2);
\draw[suite] (1.east)to[bend left] (3);
\node[ell] (4) at (3,2) {srcfv/};
\node[ell2] (5) at (-3,2) {card\_bl2d\_fv.py};
\node[ell2] (6) at (-3,1) {BROADCAST.py};
\node[ell2] (7) at (-3,-3) {DzMatrix.py};
\node[ell2] (8) at (-3,-1) {resolvent\_all.py};
\node[ell2] (9) at (-3,0) {SIM/};
\node[ell2] (10) at (-3,-2) {misc/PETSc\_func.py};
\node[ell] (11) at (4.5,1) {geom/};
\node[ell] (12) at (4.5,0) {phys/};
\node[ell] (13) at (4.5,-1) {borders/};
\node[ell] (14) at (4.5,-2) {lhs/};
\node[ell] (15) at (4.5,-3) {rhs/};
\node[ell] (16) at (4.5,-4) {prepro/};
\node[ell] (17) at (4.5,-5) {tangent/};
\node[ell] (18) at (4.5,-6) {adjoint/};
\node[ell] (19) at (4.5,-9) {dz/};
\node[ell] (20) at (4.5,-7) {tangenttangentHess/};
\node[ell] (21) at (4.5,-8) {thirdStateDerivative/};
\node[ell] (22) at (3,-10) {misc/ComputeJacobian.f90};
\draw[] (4) |- (11);
\draw[] (4) |- (12);
\draw[] (4) |- (13);
\draw[] (4) |- (14);
\draw[] (4) |- (15);
\draw[suite] (3,-3)to[bend right] (16.west);
\draw[suite] (16.west)to[bend right] (17.west);
\draw[suite] (16.west)to[bend right] (18.west);
\draw[suite] (17.west)to[bend right] (20.north west);
\draw[suite] (20.south west)to[bend right] (21.west);
\draw[suite] (5.west)to[bend right] (6.west);
\draw[suite] (10.west)to[bend left] (8.west);
\draw[suite] (10.west)to[bend left] (6.west);
\draw[suite] (6.south west)to[bend right] (8.west);
\draw[suite] (8.west)to[bend left] (6.south west);
\draw[suite] (9.west)to[bend left] (6.south west);
\draw[suite] (4.west)to[] (6.east);
\draw[suite] (22.north west)to[bend right] (6.east);
\draw[suite] (19.west)to[bend left] (7.south);
\end{tikzpicture}
\caption{Organisation of BROADCAST code. Description of the programs in Table \ref{table:tabprog}.}
\label{graphorg}
\end{figure}

As a development-oriented tool, one of the objective of BROADCAST is fast and easy modifications to plug different numerical models or methods. This leads the programming language choice to sub-routines written in Fortran for computational efficiency called by main programs written in Python for components modularity and to ease new developments. This functional approach, which is at the opposite of a black-box code, has however the drawback not to offer a friendly user-interface. The organisation of the code and a short description of the program files are given in Figure \ref{graphorg} and Table \ref{table:tabprog}.

\begin{table}[!ht]
\begin{center}
\begin{tabular}{ c|l } 
\hline
\multicolumn{2}{c}{\textbf{PYTHON PROGRAMS}} \\
\hline
card\_bl2d\_fv.py & Card to run BROADCAST (Num. parameters, options...) \\
\hline
BROADCAST.py & Main program (mesh, init., dim. choice, BC location...) \\
\hline
SIM/ & Folder including self-similar profile generator \\
\hline
resolvent\_all.py & Main functions to perform 2D resolvent analysis \\
\hline
resolvent\_all3D\_1block.py & Likewise for 3D resolvent \\
\hline
misc/PETSc\_func.py & All the useful PETSc functions \\
\hline
DzMatrix.py & Compute the transverse contributions of the Jacobian \\
\hline
biglobal\_cyl.py & 2D/3D global stability analysis for cylinder \\
\hline
Hessian\_cyl.py & Compute the Hessian operator applied to a given mode \\
\hline
\multicolumn{2}{c}{\textbf{FORTRAN PROGRAMS}} \\
\hline
geom/ & Metrics computation \\
\hline
phys/ & Primitives and viscosity computations \\
\hline
borders/ & Boundary conditions \\
\hline 
lhs/ & Basic matrix-free inversion for time-stepping method \\
\hline
rhs/ & Spatial numerical scheme for residual computation \\
\hline
prepro/ & Preprocessed files for AD \\
\hline
tangent/ & Linearised files computed by AD \\
\hline
adjoint/ & Linearised files computed by AD in backward mode \\
\hline
tangenttangentHess/ & Twice linearised files computed by AD \\
\hline
thirdStateDerivative/ & Third time linearised files computed by AD \\
\hline
dz/ & Transverse contributions of the Jacobian operator \\
\hline
misc/ComputeJacobian.f90 & Functions to construct the Jacobian \\
\hline
\end{tabular}
\caption{Main program files description (not exhaustive).}
\label{table:tabprog}
\end{center}
\end{table}

\subsection{Governing equations}

Any conservative system for a state $q$ can be written in the form:
\begin{equation}
 \frac{\partial q}{\partial t} + \nabla \cdot F(q) = 0 
 \label{NS1}
\end{equation}
We consider here the conservative variables of a fluid  $q = (\rho, \rho \mathbf{v}, \rho E)$ and the associated compressible Navier-Stokes equations for the definition of the fluxes $ F(q) $. The variables $\rho$, $\mathbf{v}=(u,v,w)$ and $E$ are respectively the density, velocity vector and total energy. However most of the following developments remain valid for any conservative system.

The Navier-Stokes equations read:

\begin{equation}
\frac{\partial \rho}{\partial t} + \nabla \cdot \left( \rho \mathbf{v} \right) = 0 
\label{NS0}
\end{equation}
\begin{equation}
\frac{\partial (\rho \mathbf{v})}{\partial t} + \nabla \cdot \left( \rho \mathbf{v} \mathbf{v} + p \mathbf{I} - \mathbf{\tau} \right) = 0 
\label{NS00}
\end{equation}
\begin{equation}
\frac{\partial (\rho E)}{\partial t} + \nabla \cdot \left( (\rho E +p) \mathbf{v} - \mathbf{\tau} \cdot \mathbf{v} - \lambda \nabla T \right) = 0, 
\label{NS000}
\end{equation}
with $E = p / (\rho (\gamma -1)) + \frac{1}{2}\mathbf{v} \cdot \mathbf{v} $,  $\mathbf{\tau} = \mu ( \nabla \mathbf{v} + (\nabla \mathbf{v})^T ) - \frac{2}{3} \mu (\nabla \cdot \mathbf{v}) \mathbf{I} $, $\mathbf{I}$ the identity matrix, $\lambda = \mu c_p / Pr$ and $P_r$ the Prandtl number ($P_r = 0.72$).


To close the system, two more equations are required. First, one assumes a homogeneous, thermally and calorically perfect gas. The perfect gas law is:

\begin{equation}
p = \rho r T
\label{NS2}
\end{equation}
with $r = 287.1$ the specific gas constant.


Then, the Sutherland's law is selected to link viscosity to temperature \cite{sutherland1893lii}.

\begin{equation}
\mu (T) = \mu_{ref} \left( \frac{T}{T_{ref}} \right)^{3/2} \frac{T_{ref} + S}{T + S}
\label{NS3}
\end{equation}
with $S = 110.4$ K the Sutherland's temperature, $\mu_{ref} = 1.716 \time 10^{-5}$ $\textrm{kg}.\textrm{m}^{-1}.\textrm{s}^{-1}$ and $T_{ref} = 273.15$ K.


\subsection{Spatial Discretisation}

The finite volume framework, the associated convective and viscous flux and the implementation of the boundary conditions are presented below.

\subsubsection{Finite volumes}

In order to work within the finite volume framework, the conservative equation \eqref{NS1} is written in its weak formulation:

\begin{equation}
\int_\Omega \frac{\partial q}{\partial t} \, d\Omega+  \int_\Omega \nabla \cdot F(q) \, d\Omega = 0, 
 \label{FV1}
\end{equation}
which becomes using the Green-Ostrogradski theorem:

\begin{equation}
 \int_\Omega \frac{\partial q}{\partial t} \, d\Omega +  \int_\Gamma  F(q) \cdot n \, d\Gamma = 0,
 \label{FV2}
\end{equation}
with $\Omega$ the domain volume, $\Gamma$ the boundaries and $n$ the outer normal.

The equation \eqref{FV2} is discretised and applied to each cell volume $\Omega_{i,j}$ and its edges $\partial \Omega_{i,j}$ (considering 2D meshes):

\begin{equation}
\lvert \Omega_{i,j} \rvert \frac{\text{d} q_{i,j} }{\text{d} t} + \sum_{\Gamma \in \partial \Omega_{i,j}} \lvert \Gamma \rvert h_\Gamma (q, n) = 0 
 \label{FV3}
\end{equation}
with $h_\Gamma$ the flux through the edge $\Gamma$, $\lvert \Omega_{i,j} \rvert$ the cell volume and $\lvert \Gamma \rvert$ the edge area. Here the term $q_{i,j}$ can represent either the average value in the cell or the cell-centered value. Both values are identical up to an order 2 accuracy \cite{maugars2016methodes}: 
\begin{equation}
q_{i,j} = \frac{1}{\lvert \Omega_{i,j} \rvert} \int_{\Omega_{i,j}} q \,d\Omega_{i,j} = q (x_{C_{i,j}}) + O(h^2),
 \label{FV4}
\end{equation}
with $x_{C_{i,j}}$ the cell center. 
Equation \eqref{FV3} is therefore 2nd order accurate. Then the approximation method used to compute equation \eqref{FV3} in BROADCAST relies on an extension of the finite difference framework. The time-derivative component is approximated by its cell-centered value and the flux is derived from a high order finite difference scheme. As shown in equation \eqref{FV4}, this method is strictly at most order 2 in the finite volume sense, however, Rezgui {\it et al.} \cite{rezgui2001third} showed that a high order accuracy in the finite difference sense is reached on Cartesian and smoothly deformed (curvilinear) meshes.

Equation \eqref{FV3} is finally recast into:
\begin{equation}
\frac{\text{d} q}{\text{d} t} = R(q),    \label{Newton0}    
\end{equation}
 with $R$ as the nonlinear residual defined as the sum of the fluxes through the edges of a cell divided by its volume. The fluxes $F$ are split into two components, inviscid and viscous.

\subsubsection{Convective flux}

The space discretisation for the inviscid flux follows the FE-MUSCL (Flux-Extrapolated-MUSCL) scheme \cite{cinnella2016high}. This is a high order accurate upwind scheme resulting from an upwind recursive correction to the leading truncation error term of a centered second order scheme \cite{lerat2003approximations,lerat2006}. It is equivalent to a MUSCL scheme constructed with a N-order extrapolation to the fluxes. This high-order convective scheme has been assessed by Sciacovelli {\it et al.} \cite{sciacovelli2021assessment} showing excellent results in accuracy and shock capturing features in hypersonic flow simulations. Thanks to its definition, the FE-MUSCL scheme can be readily extended to various orders. In BROADCAST, numerical scheme orders from 3rd to 9th are implemented. Description of the FE-MUSCL scheme and its implementation are detailed in \ref{sec:convflux}.

\subsubsection{Viscous flux}

The viscous fluxes are computed on a five-point compact stencil inside the domain (fourth order accurate) and on a three-point compact scheme near the solid boundaries (second order accurate) in order to get only one layer of ghost cells at a wall interface. The five-point compact scheme was first suggested by Zingg {\it et al.} \cite{zingg2000comparison} but this conservative scheme was not fourth order accurate for all the second order derivative viscous terms. The viscous flux discretisation implemented in BROADCAST is the one offered by Shen {\it et al.} \cite{shen2009high} which computed new coefficients to reach fourth order accuracy for all derivative terms. These coefficients are modified to move from a cell-centered definition to a face location one.

\subsubsection{Boundary conditions and ghost cells}

Boundary conditions are enforced with layers of ghost cells around the domain. The number of layers depends on the order of the numerical scheme and is equal to $(\text{order+1}) / 2 $. The residual $R$ is therefore computed from the ``augmented" state vector $q_{large} = q + q_{\text{ghost}}$ and returns a vector $R(q_{large})$ of the size of the initial state vector $q$.

Solid boundary conditions such as walls fill ghost cells from the information inside the domain to force zero velocity or no flux exchange at the boundary for instance. Permeable boundary conditions fill ghost cells with information either from inside the domain or the external user-fixed state depending on the direction given by the characteristic relations \cite{poinsot1992boundary}. For example, Dirichlet boundary condition is prescribed when all characteristics are ingoing (supersonic inlet). Conversely, an extrapolation is made when all characteristics are outgoing (supersonic outlet).

\subsection{Time-stepping}

BROADCAST includes time-stepping methods to compute base-flow. An explicit low-storage Runge-Kutta \cite{bogey2004family} and an implicit LU-SGS in which we use an approximated Jacobian \cite{yoon1991three} are available. However, in this paper, any base-flow is computed by a Newton algorithm described in the following section. The time-steppers can then be used to get an initial guess.

\subsection{Base-flow}

To find a fixed-point $ \overline{q} $ satisfying equation \eqref{Newton}, 
different algorithms are possible. First, the Newton method consists of an iterative method, where from a state $ q^n $, we build $ q^{n+1} = q^n + \delta q^n $, with 
\begin{equation}
 A(q^n)\, \delta q^n = - R(q^n),
\label{Newton2}    
\end{equation}
where $A(q^n) =\left.  \frac{\partial R}{\partial q} \right|_{q^n} $ is the Jacobian operator evaluated at $ q=q^n$. The Newton algorithm must be initialised by a guess.
A constant reference state may perform well for simple flows. If it fails, it is better to initialise the algorithm with approximate solutions, that have been generated by other numerical simulations or with a time-stepping method.   
A way to speed up the algorithm consists in keeping the same Jacobian for different Newton iterations (pseudo-Newton method). A criterion is thus required to decide when the Jacobian must be computed again. However, as it requires a further user-parameter to be tuned, we decided not to follow this path and recompute the Jacobian at each iteration.

To ensure convergence regardless of the initial state, a pseudo-transient continuation method (or relaxation method) has been implemented following Crivellini {\it et al.} \cite{crivellini2011implicit}:
\begin{equation}
\left( \frac{I}{\Delta t} + A(q^n) \right) \, \delta q^n = - R(q^n),
\label{Newton4}    
\end{equation}
with $\Delta t$ a pseudo time-step derived from an adaptive CFL number, $CFL=(|| \mathbf{v}||+\sqrt{\gamma p/\rho})\Delta t/\Delta x$, $\Delta x$ being the local mesh size:
\begin{eqnarray}
CFL_n &=& \frac{CFL_{\,0}}{r_n}  \\
r_n &=& \max \left( \frac{\lVert R(q^n) \rVert_{L^2}}{\lVert R(q^0) \rVert_{L^2}}, \frac{\lVert R(q^n) \rVert_{L^\infty}}{\lVert R(q^0) \rVert_{L^\infty}} \right), 
\end{eqnarray}
with $\lVert \cdot \rVert_{L^2} $ and $\lVert \cdot \rVert_{L^\infty} $ being respectively the $L^2$ and $L^\infty$ norms. During the first iterations, term $I/\Delta t$ in equation \eqref{Newton4} penalizes the norm of $ \delta q$ in the transient phase. Then, from an initial $CFL_{\,0}$, the CFL number increases with the decrease of the residual norm and the additional term disappears so as to recover the Newton algorithm and quadratic convergence properties.

\begin{table}[!ht]
\begin{center}
\begin{tabular}{ll}
Steps & Base-flow algorithm \\
\hline      
1 & \textbf{Initialisation:} \\
1.1 & \quad\quad Geometry \& Boundaries \\
1.2 & \quad\quad Mesh \\
1.3 & \quad\quad Reference parameters \\
1.4 & \quad\quad State normalisation \\
2 & \textbf{Solution initialisation (self-similar or uniform flow)} \\
3 & \textbf{Base-flow computation:} \\
 & For iteration $ = 1 \rightarrow M $: \\
3.1 & \quad \quad Update ghost cells with boundary conditions (BC) \\
3.2 & \quad \quad Compute residual $R(q)$ \\
3.3 & \quad \quad Compute residual norm \& relaxation coefficient \\
3.4 & \quad \quad Construct Jacobian $A$ (see Table \ref{tab:codejac}) \\
3.5 & \quad \quad Solve linear system $A\,\delta q = - R$ with LU algorithm \\
3.6 & \quad \quad Update state vector $q$ \& check residual convergence \\
4 & \textbf{Write state vector \& Jacobian.}
\end{tabular}
\end{center}
\caption{Base-flow pseudo-algorithm}
\label{tab:codebase}
\end{table}


The pseudo-algorithm to compute the base-flow is described in Table \ref{tab:codebase}. Step 3 requires $M$ Newton iterations to converge. Typically, starting from a self-similar solution for a boundary layer, $M \approx 10$.


The Jacobian operator construction method is similar to the one offered by Mettot \cite{mettot2013linear}, except that the evaluation of the derivative is performed through AD instead of finite-difference. The Jacobian matrix corresponds to the linearised discrete residual and is extracted by successive matrix-vector products and stored in a sparse format. 


For the explicit construction of the matrix of dimension $n$, by choosing the vector $v$ as a canonical vector, $n$ different matrix-vector products $Av$ would be required to fill in all the entries $A_{ij}$ of the Jacobian $A$. Nevertheless, with a finite-volume discretisation, the Jacobian is sparse and exhibits a block-diagonal shape: the vector $v$ can be therefore chosen such that different entries of the Jacobian are computed in a single matrix-vector product. This strategy significantly reduces the number of operations. The choice of the vector $v$ depends only on the stencil of the discretisation scheme with a high-order scheme requiring more matrix-vector products to be performed as its stencil is wider.

Using a finite-difference technique, the accuracy of the Jacobian depends on a real parameter, $\epsilon$, whose optimal value cannot be known a priori \cite{knoll2004jacobian}.
In BROADCAST, instead of computing $Av$ by a first order finite difference, the linearised residual is computed by Algorithmic Differentiation (AD) \cite{griewank2008evaluating} through the Source-Transformation AD code \textit{TAPENADE} \cite{hascoet2013tapenade}. This results in faster computation than a high-order finite difference method (which requires the computation of several residuals) while keeping the analytical accuracy. Also, successive derivatives can be computed robustly and accurately (finite differences method would require to adapt several case-dependent $\epsilon$ parameters). Another advantage of Algorithmic Differentiation is that it can also run in backward mode, enabling the calculation of the adjoint derivative $ A^* v$ for any $ v $, while finite difference methods are limited to the calculation of $ Av$. 

\begin{table}[!ht]
\begin{center}
\begin{tabular}{ll}
Steps & Construction of Jacobian $A$\\
\hline      
1 & \textbf{Compute non-zero entries of the Jacobian} \\
 & For number $ = 1 \rightarrow p $:\\
1.1 & \quad \quad Compute a test-vector $v$ \\
1.2 & \quad \quad Update ghost cells of $v$ with the linearised BC by AD\\
1.3 & \quad \quad Compute $Av$ with the linearised residual by AD\\
1.4 & \quad \quad Convert $Av$ to Jacobian entry $A_{ij}$\\
2 &  \textbf{Remove the zero entries computed}\\
3 & \textbf{Construct Jacobian $A$ in CSR format} \\
\end{tabular}
\end{center}
\caption{Jacobian extraction pseudo-algorithm}
\label{tab:codejac}
\end{table}

The pseudo-algorithm to extract the Jacobian operator is described in Table \ref{tab:codejac}. Step 1 requires $p$ Matrix-vector products. The scalar $p$ depends on the number of equations ($5$ for the compressible Navier-Stokes equations) and the stencil of the discretisation scheme. Here, $p = 5 \times (\text{stencil width})^2$. For the 7th order scheme, $p = 5 \times (9)^2 = 405$. However as the inviscid scheme is not compact, some matrix-vector products computed at step 1 contain zero entries. Step 2 is then necessary to remove the Jacobian zero entries computed so that the Jacobian in CSR format stores only non-zero entries.

All linear systems involving sparse matrices are solved using the PETSC software \cite{balay2019petsc} based on the direct sparse LU solver MUMPS \cite{amestoy2001fully}. BROADCAST code being written in Python language, the petsc4py version is used \cite{dalcin2011parallel}.

\subsection{Oscillator flows} \label{sec:oscillator}

In the following, the global stability, linear sensitivity of eigenvalues and weakly nonlinear analysis are described for oscillator flows.

\subsubsection{Global modes}

The linear global stability of the 2D base-flow $\overline{q}(x,y)$, which is a solution of the steady compressible Navier-Stokes equations, is assessed by the calculation of the global modes.


Small amplitude perturbations represented by the vector $q'(t)=e^{\lambda t} \hat{q}$ are added to the base-flow $\overline{q}$.
Linearising eq. \eqref{eq:ns}, we obtain the eigen-problem in eq. \eqref{eq:gm}.
If there exists one eigenvalue $ \lambda$ whose real part is positive, then the base-flow is linearly unstable and either the system should be time-stepped or a weakly nonlinear analysis as described below should be performed to determine the final saturated nonlinear state.

Here, the largest magnitude eigenvalues and corresponding eigenvectors are calculated with a matrix-free Arnoldi algorithm. For this, we rewrite eigen-problem \eqref{eq:gm} as:
\begin{equation} \label{eq:shiftedgm}
    (A-sI)^{-1}\hat{q}=\frac{1}{\lambda-s} \hat{q},
\end{equation}
where $ s $ is a shift-parameter to indicate where the eigenvalues should be looked for.
Hence, the eigenvalues of $A$ close to $s$ are also the largest magnitude eigenvalues of $ C=(A-sI)^{-1}$.
Arnoldi methods rely on the construction of Krylov subspaces, which are obtained by computing recursively matrix-vector products ($Cv, C^2v, \cdots, C^pv,\cdots$) with $v$ a Krylov vector.  

We use the SLEPc library \cite{roman2015slepc} which implements various Krylov-Schur methods \cite{hernandez2007krylov} whose Arnoldi algorithm. Computation of the global modes is performed with the steps described in Table \ref{tab:codeglobal}.

\begin{table}[!ht]
\begin{center}
\begin{tabular}{ll}
Steps & Global mode algorithm \\
\hline      
1 & \textbf{Initialisation:} \\
1.1 & \quad\quad Read geometry and state vector \\
1.2 & \quad\quad Read shift $s$ \\
1.3 & \quad\quad Construct Jacobian $A$ (see Table \ref{tab:codejac}) \\
2 & \textbf{Definition of the shift-inverted matrix $C$:} \\
2.1 & \quad\quad Construct matrix $ s I - A$ \\
2.2 & \quad\quad Define matrix-vector product $Cv$ ($C=(sI-A)^{-1}$ is never computed)\\
3 & \textbf{Solve largest magnitude eigenvalue of eq. \eqref{eq:shiftedgm}} \\
3.1 & \quad\quad Arnoldi method with matrix-free algorithm \\
3.2 & \quad\quad Store the largest magnitude eigenvalue $\lambda'$ and its eigenvector $\hat{q}$ \\
4 & \textbf{Write eigenvalue $\lambda=\lambda'^{-1}+s$ and global mode $\hat{q}$.}
\end{tabular}
\end{center}
\caption{Global mode pseudo-algorithm}
\label{tab:codeglobal}
\end{table}


\subsubsection{Adjoint global modes}

The adjoint operator $A^*$ gives access to the adjoint modes through the equation \eqref{adj1}.

\begin{equation}
A^* \tilde{q} = \lambda^*  \tilde{q}
\label{adj1}
\end{equation}
As the Jacobian $A$ is real, the discrete adjoint operator is directly computed by transposition of $A$. The adjoint mode corresponding to its direct global mode is targeted with the same shift-and-invert method written in Eq. \eqref{eq:shiftedgm} replacing $A$ by $A^*$ and $\lambda$ by ${\lambda^*}$. 

Furthermore, to associate an adjoint mode $\tilde{q}$ with its direct mode $\hat{q}$, a scalar product must be chosen for normalisation. A first choice leads to the $L^2$ inner product $\langle u, v \rangle = u^*v = \sum_i u_i v_i$ and then $\langle \tilde{q}, \hat{q} \rangle = 1$. However as the direct and adjoint modes generally have a different spatial support (direct mode is located downstream while adjoint mode is upstream), the local mesh size should be taken into account in the inner product. This gives the discrete inner product defined by the matrix $Q$ in equation \eqref{innerprod}.

\begin{equation}
\left. \langle u, v \rangle \right|_Q = u^* Q v = \int_{\Omega} u^* v \,d\Omega
\label{innerprod}
\end{equation}
The adjoint mode $q^\dag$ normalised with the discrete inner product given by Eq. \eqref{innerprod} ($\left.\langle q^\dag, \hat{q} \rangle \right|_Q = 1$) is then $q^\dag = Q^{-1} \tilde{q}$.

\subsubsection{Sensitivity of eigenvalues} \label{sec:sensi}

Linear sensitivity of an unstable eigenvalue $\lambda $ (globally unstable flow) provides knowledge on the evolution of an eigenvalue with a small variation of the base-flow. It is studied following the method described by Mettot {\it et al.} \cite{mettot2014computation}. The eigenvalue $\lambda$ is assumed to depend on the base-flow $\overline{q}$ \cite{bottaro2003effect} so that a base-flow modification $\delta \overline{q}$ produces a variation of the eigenvalue $\delta \lambda$, that is

\begin{equation}
\delta \lambda = \langle \nabla_{\overline{q}} \lambda, \delta \overline{q} \rangle
\label{grad}
\end{equation}

The sensitivity of the eigenvalue with respect to base-flow variation written $\nabla_{\overline{q}} \lambda$ above is a complex vector whose real part represents the sensitivity of the amplification rate (real part of the eigenvalue $\lambda$) while the imaginary part represents the angular frequency (imaginary part of the eigenvalue $\lambda$). The inner product $\langle \cdot,\cdot \rangle$ is the one associated with the $L^2$ norm. From Sipp {\it et al.} \cite{sipp2010dynamics}, any variation of the Jacobian $\delta A$ generates the variation $\delta \lambda$

\begin{equation}
\delta \lambda = \langle \tilde{q}, \delta A \hat{q} \rangle
\label{gradjac}
\end{equation}
where $\tilde{q}$ is the adjoint mode.

For a given global mode $\hat{q}$, the variation of the Jacobian $\delta A$ is assumed to be induced by the variation of the base-flow $\delta \overline{q}$:

\begin{equation}
\delta A \hat{q} = \left.\frac{\partial (A \hat{q})}{\partial q}\right|_{\overline{q}} \delta \overline{q}
\label{hes1}
\end{equation}
Then the sparse Hessian operator $H$ is:

\begin{equation}
H_{ijk} = \left. \frac{\partial^2 R_i}{\partial q_j \partial q_k} \right|_{\overline{q}}
\label{hes3}
\end{equation}

Noticing that the Hessian is the second derivative of the residual around the base-flow, it can be computed easily applying a second time the Algorithmic Differentiation tool to the linearised residual. Introducing the Hessian $H$ into the equation \eqref{hes1}, it can be recast as

\begin{equation}
\delta A \hat{q} = H(\hat{q}, \delta \overline{q})
\label{hes2}
\end{equation}
with $H(\hat{q}, \delta \overline{q}) = H' \delta \overline{q}$ and $H'$ defined as:

\begin{equation}
H'_{ik} = \sum_j H_{ijk} \hat{q}_j
\label{hes4}
\end{equation}
This leads to $\delta A \hat{q} = H' \delta \overline{q}$ which can be substituted into the equation \eqref{gradjac} and results in

\begin{equation}
\delta \lambda = \langle \tilde{q}, H' \delta \overline{q} \rangle = \langle H'^* \tilde{q}, \delta \overline{q} \rangle
\label{hes5}
\end{equation}
By comparing equations \eqref{grad} and \eqref{hes5}, the gradient of the eigenvalue is:

\begin{equation}
\nabla_{\overline{q}} \lambda = H'^* \tilde{q}
\label{hes6}
\end{equation}

Linear sensitivity to other parameters with additional constraints (divergence-free modification, steady source term...) can also be investigated. The sensitivity to a steady source term in the Navier-Stokes equation is here studied by considering a small amplitude force term $\delta f$. Writing $R(\overline{q} + \delta \overline{q}) + \delta f = 0$, one gets at the first order:

\begin{equation}
\delta \overline{q} = - A^{-1} \delta f
\label{hes7}
\end{equation}
Then introducing equation \eqref{hes7} into equation \eqref{hes5} yields

\begin{equation}
\delta \lambda = \langle H'^* \tilde{q}, - A^{-1} \delta f \rangle = \langle -A^{*-1} H' \tilde{q}, \delta f \rangle
\label{hes8}
\end{equation}
By comparing with $\delta \lambda = \langle \nabla_f \lambda, \delta f \rangle$, the sensitivity with respect to a steady source term is:

\begin{equation}
\nabla_f \lambda = -A^{*-1} H'^* \tilde{q}
\label{hes9}
\end{equation}

Then, continuous and discrete approaches can be linked. The inner product associated with the $L^2$ norm is replaced by the discrete inner product defined with the Hermitian matrix Q (Equation \eqref{innerprod}). The sensitivity becomes $\delta \lambda = \left. \langle \nabla_f \left. \lambda\right|_Q, \delta f \rangle \right|_Q$ and then:

\begin{equation}
\nabla_f \left. \lambda\right|_Q = - Q^{-1} A^{*-1} H'^*  q^\dag
\label{hes10}
\end{equation}
with the associated adjoint mode $ q^\dag  = Q^{-1} \tilde{q}$.

To compute the sensitivity of an eigenvalue with respect to base-flow variation or to a steady source term in Navier-Stokes equations, one requires to have the direct and corresponding adjoint mode, the Jacobian and the Hessian operators. The eigenmodes are computed with PETSc while the operators are derived through the AD method.

\subsubsection{Amplitude equations}

The weakly nonlinear analysis for cylinder bifurcation was described by Sipp and Lebedev \cite{sipp2007global} for incompressible flow. The same method is here extended to compressible flows, see details in \ref{sec:wnl}. In the vicinity of the critical Reynolds number $Re_c$, introducing a small parameter $\epsilon$ defined as $\epsilon^2 = Re_c^{-1} - Re^{-1}$ with $0 < \epsilon \ll 1$, one may predict whether a bifurcation is sub-critical or super-critical by computing the sign of the sum of the coefficients $\mu + \nu + \xi$ of the Stuart-Landau amplitude $\mathcal{A}$ equation \eqref{landau}.

\begin{equation}
\frac{d \mathcal{A}}{dt} = \epsilon^2 \kappa \mathcal{A} - \epsilon^2 (\mu + \nu + \xi) \mathcal{A} |\mathcal{A}|^2
\label{landau}
\end{equation}

These coefficients rely on five different modes computed at the critical frequency $\omega$: the first direct mode $\hat{q}_{1}$, the corresponding adjoint mode $\tilde{q}_1$, the zeroth (mean flow) harmonic $\hat{q}_{20}$, the second harmonic $\hat{q}_{22}$ and the base-flow modification due to an $\epsilon$ Reynolds increase $\hat{q}_{21}$ (Equations \eqref{eq:mode0}-\eqref{eq:mode3}).

\begin{eqnarray}
i\omega \hat{q}_{1} - A \hat{q}_{1} &=& 0 \label{eq:mode0} \\
-i\omega \tilde{q}_{1} - A^* \tilde{q}_{1} &=& 0 \label{eq:mode0b} \\
2i\omega \hat{q}_{22} - A \hat{q}_{22} &=& \frac{1}{2} H(\hat{q}_{1}, \hat{q}_{1}) \label{eq:mode1} \\
-A \hat{q}_{20} &=& H (\hat{q}_{1},\overline{\hat{q}_{1}}) \label{eq:mode2} \\
- A \hat{q}_{21} &=& \partial_{\epsilon} R(q_0,0) \label{eq:mode3}
\end{eqnarray}

Following the same method described in Sipp and Lebedev \cite{sipp2007global} in the incompressible framework, by developing the Taylor expansion of the residual $R = R(q,\epsilon)$ up to the third order, analytical expressions of the coefficients $\mu$, $\nu$ and $\xi$ are found for the compressible framework (Equations \eqref{mu}-\eqref{xi}). They depend on the discrete Hessian $H$ defined previously (the component $i$ of the vector $H(q,q)$ is equal to $H_i(q,q) = \sum_k \sum_j \partial_{q_jq_k} (R_i(\overline{q},0)) q_j q_k$) and the discrete third-order derivative $T$ ($T_i(q,q,q) = \sum_l \sum_k \sum_j \partial_{q_jq_kq_l} (R_i(\overline{q},0)) q_j q_k q_l$) operators where $\partial_q(\cdot) = \partial (\cdot) / \partial q$. These derivative operators are computed by AD returning directly the vectors $H(q,q)$ or $T(q,q,q)$, preventing then any explicit storage of these operators.

\begin{eqnarray}
\kappa&=&\frac{\langle \tilde{q}_{1}, H(\hat{q}_{1}, \hat{q}_{21}) \rangle}{\langle \tilde{q}_{1}, \hat{q}_{1} \rangle} + \frac{\langle \tilde{q}_{1}, \partial_{q\epsilon} R(q_0,0) \hat{q}_{1} \rangle}{\langle \tilde{q}_{1}, \hat{q}_{1} \rangle} \label{kappa} \\
\mu & =& - \frac{\langle \tilde{q}_{1}, H(\hat{q}_{1},\hat{q}_{20})\rangle}{\langle \tilde{q}_{1},\hat{q}_{1} \rangle}
\label{mu} \\
\nu &=& - \frac{\langle \tilde{q}_{1}, H(\overline{\hat{q}_{1}},\hat{q}_{22})\rangle}{\langle \tilde{q}_{1},\hat{q}_{1} \rangle}
\label{nu} \\
\xi &=& - \frac{1}{2} \frac{\langle \tilde{q}_{1},T(\hat{q}_{1}, \hat{q}_{1}, \overline{\hat{q}_{1}}) \rangle}{\langle \tilde{q}_{1}, \hat{q}_{1} \rangle}.
\label{xi}
\end{eqnarray}

The coefficient $\mu$ consists of the response arising from the interaction of the first harmonic with the zeroth harmonic. The coefficient $\nu$ consists of the response arising from the interaction of the complex conjugate part of the first harmonic with the second harmonic. The coefficient $\xi$ consists of the response arising from the interaction of the complex conjugate part of the first harmonic twice with the first harmonic (itself). This last coefficient exists only at compressible regime due to the triadic form of the equations.

\subsection{Amplifier flows} \label{sec:amplifier}

The equivalent of global stability analysis is presented below for amplifier flows (globally stable). Optimal response (resp. forcing) is linked with global direct (resp. adjoint) mode. Then, linear sensitivity of the optimal gain is studied.

\subsubsection{Resolvent modes}

If all of the eigenvalues $ \lambda $ are stable, then it is relevant to perform a resolvent analysis.
For this we consider an additional small-amplitude forcing field $f'(t)=e^{i\omega t} \check{f}$ on the right-hand-side of equation \eqref{eq:ns}, a response $q'(t)=e^{i\omega t} \check{q}$, so that after linearisation: 
\begin{equation}
\check{q} = \mathcal{R} \check{f}
\label{gstab4}    
\end{equation}
with $\mathcal{R} = \left( i\omega I - A \right)^{-1}$ the linear global resolvent operator and $I$ the identity matrix. The resolvent corresponds to a transfer function between the input (forcing) and the response (perturbations). The resolvent analysis is often called input-output analysis for this reason. 

The optimal forcings / responses are then computed by optimising the gain between the energy of the response and the energy of the forcing:
\begin{equation}
\mu^2 = \sup_{\check{f}} \frac{\| \check{q} \|^2_E}{\| \check{f} \|^2_F},  
\label{gstab5}    
\end{equation}
with $\|\cdot\|_E$ and $\|\cdot\|_F$ the user-selected measures to evaluate the amplitude of the fluctuations and the forcing. These measures are defined with their associated discrete Hermitian matrices $Q_E$ (semi-norm) and $Q_F$ (norm):
\begin{equation}
\| \check{q} \|^2_E = \check{q}^* Q_E \check{q} \;\; , \;\;
\| \check{f} \|^2_F = \check{f}^* Q_F \check{f}
\label{gstab6}    
\end{equation}

Solving for $ \mu^2$ over a range of frequencies $\omega$ provides the most receptive frequency (where $ \mu(\omega)$ is largest) and the associated optimal forcing mode $ \check{f} $. The optimal response $\check{q}$ is then retrieved from the optimal forcing using equation \eqref{gstab4}.

The energy of the perturbations defined by $Q_E$ is generally the kinetic energy in incompressible flows. However, in compressible cases, in order to take into account the density and temperature disturbances, Chu's energy \cite{george2011chu} is often used \cite{bugeat20193d}:
\begin{equation}
E_\text{Chu} = \check{q}^* Q_E \check{q} = \frac{1}{2} \int_\Omega \left( \overline{\rho}|\mathbf{\check{v}}|^2 + \frac{\overline{T}}{\overline{\rho} \gamma M^2} \check{\rho}^2 + \frac{\overline{\rho}}{(\gamma -1)\gamma M^2\overline{T}} \check{T}^2 \right) \,d\Omega. 
\label{gstab7}    
\end{equation}
Chu's energy is the sum of the kinetic energy and the thermodynamic disturbances with the appropriate coefficients so that the conservative compression work is not included as a disturbance in the total energy \cite{hanifi1996transient}.
Matrix $Q_E$ for Chu's energy norm is block-diagonal and may be written with conservative variables, as detailed for instance in Bugeat {\it et al.} \cite{bugeat20193d}. It is a Hermitian matrix.

To study the receptivity of the flow, it is interesting to restrict the forcing field to specific regions of the flow or to specific components.
For this we introduce the prolongation/restriction matrix $P$
and replace $\check{f} $ by $ P \check{f} $, so that equation \eqref{gstab4} becomes:
\begin{equation}
\check{q} = \mathcal{R} P \check{f}.
\label{gstab8}    
\end{equation}
If $ \check{f} $ is of size $m \leq n$, then matrix $P$ is $n \times m$.
Similarly to Bugeat {\it et al.} \cite{bugeat20193d} and Sartor {\it et al.} \cite{sartor2015unsteadiness}, we choose here to apply the forcing only on the momentum equations and measure $Q_F$ as the discrete inner product ($Q_F=Q$ in equation \eqref{innerprod}).

\begin{equation}
\| \check{f} \|^2_F = \int_\Omega \check{f}^* \check{f} \,d\Omega.  
\label{Qfbugeat}
\end{equation}
Another scalar product definition could have been the one given by Eq. \eqref{Qf} which results in a dimension-less gain $ \mu(\omega) U_\infty / L_{\text{ref}}$. 

\begin{equation}
\| \check{f} \|^2_F = \int_\Omega \bar{\rho}^{-1} \check{f}^* \check{f} \,d\Omega.  
\label{Qf}
\end{equation}

However for sake of comparison with Bugeat {\it et al.} in \S \ref{sec:bl}, the discrete scalar product from Eq. \eqref{Qfbugeat} is chosen here.
Matrix $Q_F$ is $m \times m$ and positive-definite. Similarly, the perturbation energy can also be restricted to a region of the flow (or to specific components) which might be different from those of the forcing. A simple method to perform this consists in replacing some columns and rows of the matrix $Q_E$ by zero so that these perturbations outside the target region do not contribute to the optimisation. Matrix $Q_E$ remains $n \times n$ and Hermitian. 
   


The optimal gain in eq. \eqref{gstab5} becomes:
\begin{equation}
\mu^2 = \sup_{ \check{f}} \frac{\check{q}^* {Q_E} \check{q} }{\check{f}^* Q_F \check{f}} = \sup_{ \check{f}}  \frac{(\mathcal{R}P \check{f})^* {Q_E} (\mathcal{R}P \check{f})}{(P \check{f})^* Q_F (P \check{f})} =  \sup_{ \check{f}}  \frac{ \check{f}^* P^* \mathcal{R}^* {Q_E} \mathcal{R}P \check{f}}{ \check{f}^* {Q_F} \check{f}}
\label{gstab9}    
\end{equation}
The optimisation problem defined by equation \eqref{gstab9} is the Rayleigh quotient. It is equivalent to the generalised Hermitian eigenvalue problem:
\begin{equation}
\underbrace{P^* \mathcal{R}^* {Q_E} \mathcal{R}P}_{=C} \check{f} = \mu^2 {Q_F} \check{f},
\label{gstab11}    
\end{equation}
which is related to equation \eqref{eq:sv}.
Its largest eigenvalue is $\mu^2$ and the associated eigenvector is $\check{f}$.

From the definition of the Hermitian matrix $C = P^* \mathcal{R}^* {Q_E} \mathcal{R}P $, the different steps to compute $ C v $ are detailed in Table \ref{tab:codematrixfree}. Computation of the resolvent modes is performed following the steps described in Table \ref{tab:coderesolvent}. As one may notice, the frequency $\omega$ appears in matrix $ \mathcal{R}=(i\omega I-A)^{-1} $. So, to compute the optimal gain and its associated forcing and response over a range of $n$ frequencies, the steps from 2 to 5 have to be reproduced $n$ times in Table \ref{tab:coderesolvent}. Martini {\it et al.} \cite{martini2021efficient} offered an alternative resolvent approach based on a time-stepping method to compute simultaneously the optimal gains associated with several frequencies.


\begin{table}[!ht]
\begin{center}
\begin{tabular}{ll}
Steps & Matrix-free algorithm \\
\hline 
1 & \textbf{Compute $C v$:} \\
1.1 & \quad \quad Compute $v_1 = P \,v$ \\
1.2 & \quad \quad Solve linear system $  v_2 = \mathcal{R} v_1 $ with sparse LU algorithm \\
1.3 & \quad \quad Compute $ v_3 = {Q_E}\, v_2 $ \\
1.4 & \quad \quad Solve linear system $ v_4 = \mathcal{R^*} v_3 $ (LU decomposition of $\mathcal{R}$ done at step 1.2) \\
1.5 & \quad \quad Compute $ v_5 = P^*\, v_4 $ \\
1.6 & \quad \quad Update $v = v_5 $.
\end{tabular} \label{alg:matrixfree}
\end{center}
\caption{Matrix-free pseudo-algorithm}
\label{tab:codematrixfree}
\end{table}

\begin{table}[!ht]
\begin{center}
\begin{tabular}{ll}
Steps & Resolvent algorithm \\
\hline      
1 & \textbf{Initialisation:} \\
1.1 & \quad\quad Read geometry and state vector \\
1.2 & \quad\quad Compute norm matrices $Q_E$ \& $Q_F$ \\
1.3 & \quad\quad Compute prolongation/restriction matrix $P$ \\
1.3 & \quad\quad Construct Jacobian $A$ (see Table \ref{tab:codejac}) \\
2 & \textbf{Definition of Hermitian matrix $C$:} \\
2.1 & \quad\quad Construct matrix $ i \omega I - A$ \\
2.2 & \quad\quad Define matrix-vector product $Cv$ (see Table \ref{tab:codematrixfree}) \\
3 & \textbf{Solve generalised eigenvalue problem (eq. \eqref{gstab11}):} \\
3.1 & \quad\quad Arnoldi method with matrix-free algorithm \\
3.2 & \quad\quad Store the largest eigenvalue $\mu^2$ and its eigenvector $\check{f}$ \\
4 & \textbf{Solve linear system $\check{q} = \mathcal{R} P \check{f}$} \\
5 & \textbf{Write optimal gain $\mu$, forcing $\check{f}$ and response $\check{q}$.}
\end{tabular}
\end{center}
\caption{Resolvent mode pseudo-algorithm}
\label{tab:coderesolvent}
\end{table}


\subsubsection{Sensitivity of optimal gains}

Linear sensitivity of an optimal gain to base-flow modification (Eq. \eqref{sensigain}) or to steady forcing (Eq. \eqref{sensigain2}) may be computed following a similar method as the one described in \S \ref{sec:sensi}. Details of the calculation are given in Mettot \cite{mettot2013linear}.

\begin{equation}
\nabla_{\overline{q}} \left. \mu^2\right|_Q = 2 \mu^2 \,\text{Real} \,\left( Q^{-1}  \check{H}'^* Q_F \check{f} \right)
\label{sensigain}
\end{equation}

\begin{equation}
\nabla_f \left. \mu^2\right|_Q = - 2 \mu^2 \,\text{Real}\, \left( Q^{-1} A^{*-1} \check{H}'^* Q_F \check{f} \right)
\label{sensigain2}
\end{equation}
with $\check{H}'$ defined as $H(\check{q},\delta \overline{q}) = \check{H}' \delta \overline{q}$.

\subsection{Linear 3D perturbations}

The extension of global stability analysis to linear 3D perturbations follows Bugeat {\it et al.} \cite{bugeat20193d}, however, the linearisation is here performed by Algorithmic Differentiation and not by finite difference method.
The base-flow being assumed homogeneous in the $z$-direction, the perturbation field can be searched under the form:
\begin{equation}
q'(x,y,z,t) = \hat{q}(x,y) e^{\lambda t + i\beta z},
\label{gstab3}
\end{equation}
where $ \beta $ is the real wavenumber in the $ z$-direction.
A similar form can be assumed for the optimal response and forcing in the case of an input-output resolvent analysis.
These perturbations can therefore be studied on the same 2D mesh without discretisation of the $ z$-direction. The $z$-dependency of the forcing and response are taken into account analytically. One can split the 3D residual $R_\text{3D}$ as the sum of the 2D discretised residual $R$ and its z-derivative components $R_z$:
\begin{equation}
R_\text{3D} = R + R_z.
\label{3d}
\end{equation}
For the compressible Navier-Stokes equations, the $R_z$ residual can be written as the sum of three functions: 
\begin{equation}
R_z (q) = \sum_k \alpha_k(q) \frac{\partial a_k(q)}{\partial z} + \sum_l \lambda_l(q) \frac{\partial^2 b_l(q)}{\partial z^2} + \sum_m \gamma_m(q) \frac{\partial c_m(q)}{\partial z} \odot \frac{\partial d_m(q)}{\partial z},
\label{3d2}
\end{equation}
where the explicit definitions of the various introduced functions are given in \ref{sec:3D}. Notation $\odot$ refers to the element-wise product of two matrices or vectors (Hadamard product). The linearisation of equation \eqref{3d2} yields:

\begin{multline}
R_z (\overline{q} + q') = R_z (\overline{q}) + \left.\frac{\partial R_z}{\partial q}\right|_{\overline{q}} q' = \sum_k \left( \alpha_k(\overline{q}) + \left.\frac{\partial \alpha_k}{\partial q}\right|_{\overline{q}} q' \right) \frac{\partial}{\partial z} \left( a_k(\overline{q}) + \left.\frac{\partial a_k}{\partial q}\right|_{\overline{q}} q' \right) + \\ \sum_l \left( \lambda_l(\overline{q}) + \left.\frac{\partial \lambda_l}{\partial q}\right|_{\overline{q}} q' \right) \frac{\partial^2}{\partial z^2} \left( b_l(\overline{q}) + \left.\frac{\partial b_l}{\partial q}\right|_{\overline{q}} q' \right) + \\ \sum_m \left( \gamma_m(\overline{q}) + \left.\frac{\partial \gamma_m}{\partial q}\right|_{\overline{q}} q' \right) \frac{\partial }{\partial z} \left( c_m(\overline{q}) + \left.\frac{\partial c_m}{\partial q}\right|_{\overline{q}} q' \right) \odot \frac{\partial}{\partial z} \left( d_m(\overline{q}) + \left.\frac{\partial d_m}{\partial q}\right|_{\overline{q}} q' \right)
\label{3d3}
\end{multline}
As the base-flow is homogeneous in the $z$-direction, all $z$-derivatives of $\overline{q}$ in equation \eqref{3d3} vanish. Furthermore, only first-order terms for small fluctuations $q'$ are kept so it reduces to:
\begin{equation}
R_z (\overline{q} + q') = A_z q' = \sum_k \alpha_k(\overline{q}) \left.\frac{\partial a_k}{\partial q}\right|_{\overline{q}} \frac{\partial q'}{\partial z} + \sum_l \lambda_l(\overline{q}) \left.\frac{\partial b_l}{\partial q}\right|_{\overline{q}}\frac{\partial^2 q'}{\partial z^2}
\label{3d4}
\end{equation}
Taking into account equation \eqref{gstab3}, the full 3D Jacobian $A_\text{3D}$ therefore reads:
\begin{equation}
A_\text{3D}\, q' = (A + A_z )\, q' = \left(A + i\beta \sum_k \alpha_k(\overline{q}) \left.\frac{\partial a_k}{\partial q}\right|_{\overline{q}} - \beta^2 \sum_l \lambda_l(\overline{q}) \left.\frac{\partial b_l}{\partial q}\right|_{\overline{q}} \right) q'.
\label{3d5}
\end{equation}
Both matrices $\sum_k \alpha_k(\overline{q}) \left.\frac{\partial a_k}{\partial q}\right|_{\overline{q}}$ and $\sum_l \lambda_l(\overline{q}) \left.\frac{\partial b_l}{\partial q}\right|_{\overline{q}}$ are computed using Algorithmic Differentiation.

Numerically, the 3D linear extension only requires to compute two additional matrices of much smaller size than the 2D Jacobian and to fill the Jacobian with complex entries. This is an affordable computation cost which results in practice in approximately a 30\% increase in RAM (60\% increase in CPU time).

Computing global modes and resolvent modes consist in replacing in Tables \ref{tab:codeglobal} and \ref{tab:coderesolvent}, the 2D Jacobian $A$ by the 3D Jacobian $A_\text{3D}$ defined in equation \eqref{3d5} where the wavenumber $\beta$ appears. Then, if one studies the most unstable global mode over a range of $M$ wavenumbers $\beta$ then this adds a further loop of $M$ iterations on the steps from 2 to 4 in Table \ref{tab:codeglobal} (steps from 2 to 5 in Table \ref{tab:coderesolvent}).

Linear sensitivity described in \S \ref{sec:sensi} may also be extended to 3D perturbations. However, the following developments are correct only for 2D sensitivity (homogeneous gradient in $z$-direction) of eigenvalue/optimal gain of 3D modes.
Similarly to the equation \eqref{3d5}, the 3D Hessian operator can be written as $H_\text{3D}(\hat{q},\overline{q}) = H(\hat{q},\overline{q}) + H_z(\hat{q},\overline{q})$. One should notice that the base-flow $\overline{q}$ remains 2D and only the eigenmode $\hat{q}$ brings a new 3D contribution. From the equations \eqref{hes1} and \eqref{hes2}:

\begin{equation}
H_\text{3D}(\hat{q},\overline{q}) = \left.\frac{\partial \left( A_\text{3D} \hat{q} \right)}{\partial q}\right|_{\overline{q}}
\label{hess3D}
\end{equation}
From the equation \eqref{3d5}, the following expression may be derived:

\begin{equation}
H_\text{3D}(\hat{q},\overline{q}) = H(\hat{q},\overline{q}) + \left[ i\beta \sum_k \left( \left.\frac{\partial \alpha_k}{\partial q}\right|_{\overline{q}} \left.\frac{\partial a_k}{\partial q}\right|_{\overline{q}} + \alpha_k(\overline{q}) \left.\frac{\partial^2 a_k}{\partial q^2}\right|_{\overline{q}} \right) - \beta^2 \sum_l \left( \left.\frac{\partial \lambda_l}{\partial q}\right|_{\overline{q}} \left.\frac{\partial b_l}{\partial q}\right|_{\overline{q}} + \lambda_l(\overline{q}) \left.\frac{\partial^2 b_l}{\partial q^2}\right|_{\overline{q}} \right)  \right] \hat{q}
\label{hess3D2}
\end{equation}
The additional matrices are again computed with AD. Therefore, the 2D sensitivity of a 3D mode is given by the same equations as the one of a 2D mode but by replacing the 2D Hessian by the 3D Hessian written in equation \eqref{hess3D2}.

\section{Validation}

The BROADCAST code and its stability tools are validated on a circular cylinder at low Mach number \cite{sipp2007global, fabre2018practical} and on a hypersonic boundary layer developing on an adiabatic flat plate \cite{bugeat20193d}. These cases validate the solver on both Cartesian and curvilinear grids as well as for both hypersonic and incompressible regimes. Computations are performed on HPC with the following features: 128 GB RAM \& 2 sockets of 12 cores (Broadwell Intel Xeon CPU E5-2650 v4 at 2.2 GHz).

\subsection{Cylinder flow} \label{sec:cyl}

In the following, the base-flow around a circular cylinder at low Mach number, global stability, sensitivity of growth rate and weakly nonlinear analysis at the bifurcation critical point are performed.

\subsubsection{Geometry, mesh, boundary conditions and base-flow}

The circular cylinder is studied for low-Mach number flow to allow validation with incompressible results. The Navier-Stokes equations are non-dimensionalised by the triplet ($\rho_\infty,\: U_\infty,\: T_\infty $). The freestream Mach number is $M=0.1$ and the freestream temperature is $T_\infty = 288$ K and several Reynolds numbers are computed between $Re=40$ and $Re=100$ with the Reynolds number defined as $Re= \rho_\infty U_\infty D / \mu_\infty$ with $D$ the diameter of the cylinder.

The computational domain is circular with an outer radius of $100 D$. The O-mesh is curvilinear with a stretching in the wall-normal direction. The mesh is fine enough in the wall-tangent direction to get a smooth curvilinear grid where the high-order numerical scheme remains valid \cite{rezgui2001third, fosso2010curvilinear}. The reference mesh size is 630 points in the wall-tangent direction and 300 points in the wall-normal direction. An adiabatic no-slip wall condition is prescribed at the surface of the cylinder. This is a Dirichlet condition for velocity and a boundary layer assumption resulting in a zero wall-normal pressure gradient at the interface (gradient computed at second order under the assumption of a Cartesian regular grid at the interface). A non-reflecting condition is imposed at the outer radius with the free-stream conditions taken as reference. It is based on the characteristic relations and it determines pressure, density and velocity depending on the directions of the ingoing/outgoing characteristics \cite{poinsot1992boundary}. Connectivity conditions are added (through additional ghost cells) to close the mesh in the upstream part.

\begin{figure}[!ht]
\centering
\includegraphics[width=0.5\textwidth]{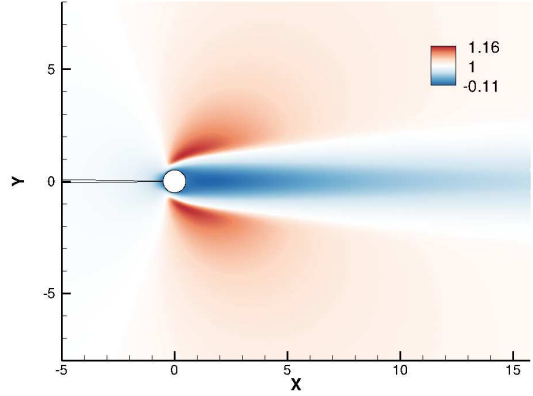}
\caption{Streamwise velocity of the base-flow at $Re = 46.8$ and $M=0.1$.}
\label{bsfcyl}
\end{figure}

We choose the fifth order numerical spatial scheme and the base-flow (Figure \ref{bsfcyl}) was converged up to a decrease of 12 orders of magnitude of the $L^2$ norm of the residual in 11 Newton iterations starting from a constant reference state for several Reynolds numbers around the first bifurcation.

\subsubsection{Global modes}

\begin{figure}[!ht]
\centering
\includegraphics[width=0.5\textwidth]{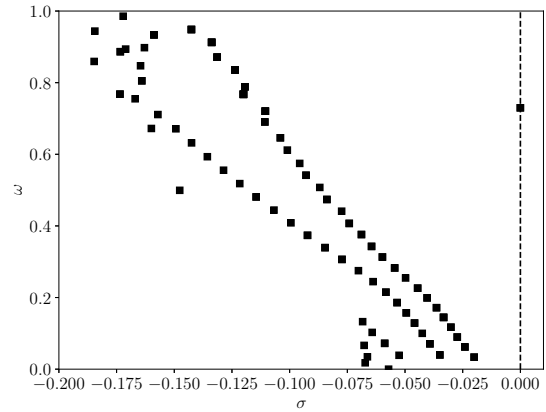}
\caption{Discretised spectrum (partial view) of the eigenvalues at $Re = 46.8$ and $M=0.1$.}
\label{spectrum}
\end{figure}

\begin{figure}[!ht]
\centering
\begin{subfigure}{0.49\textwidth}
\includegraphics[width=0.99\textwidth]{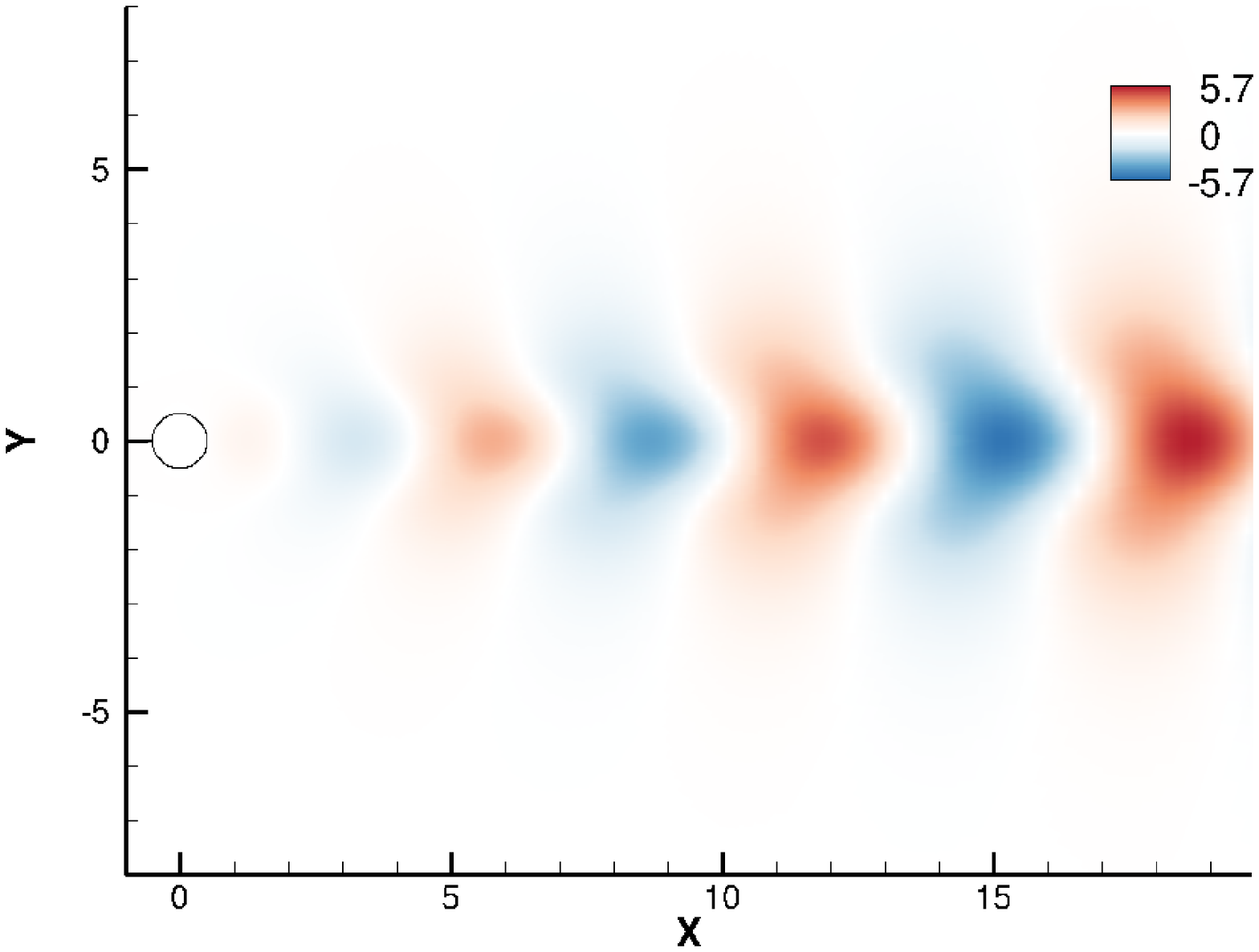}
\caption{Real part of the normal velocity of the direct mode $\hat{q}_{1}$ at $Re=46.8$ and $M=0.1$.}
\label{qdir}
\end{subfigure}
\begin{subfigure}{0.49\textwidth}
\includegraphics[width=0.99\textwidth]{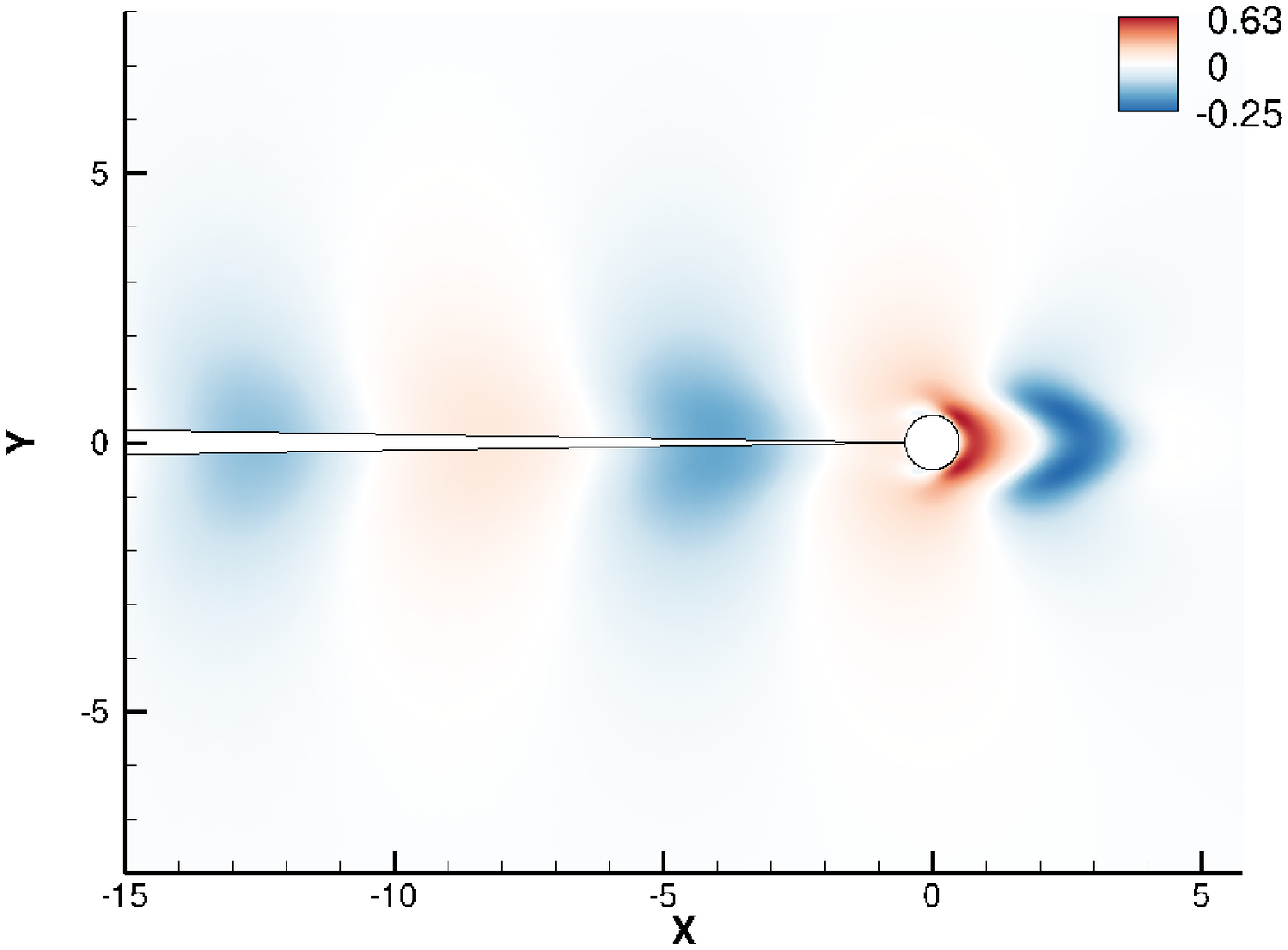}
\caption{Real part of the normal velocity of the adjoint mode ${q}_{1}^\dag = Q^{-1} \tilde{q}_1$ at $Re=46.8$ and $M=0.1$.}
\label{qadj}
\end{subfigure}
\caption{Direct and adjoint modes. Same normalisation as in Sipp and Lebedev \cite{sipp2007global}.}
\label{harmonics}
\end{figure}

The eigenvalue problem \eqref{eq:gm} for global stability is solved at different Reynolds numbers to find the most unstable global mode. The spectrum at $Re = 46.8$ is plotted in Figure \ref{spectrum}, which corresponds to criticality: $\lambda = \sigma + i \omega$, with $ \sigma=0$. The eigenvalues $\lambda$ are normalised by $U_\infty / D$. The most unstable mode and its associated adjoint mode at criticality are plotted in Figure \ref{harmonics}.

\begin{figure}[!ht]
\centering
\begin{subfigure}{0.49\textwidth}
\includegraphics[width=0.99\textwidth]{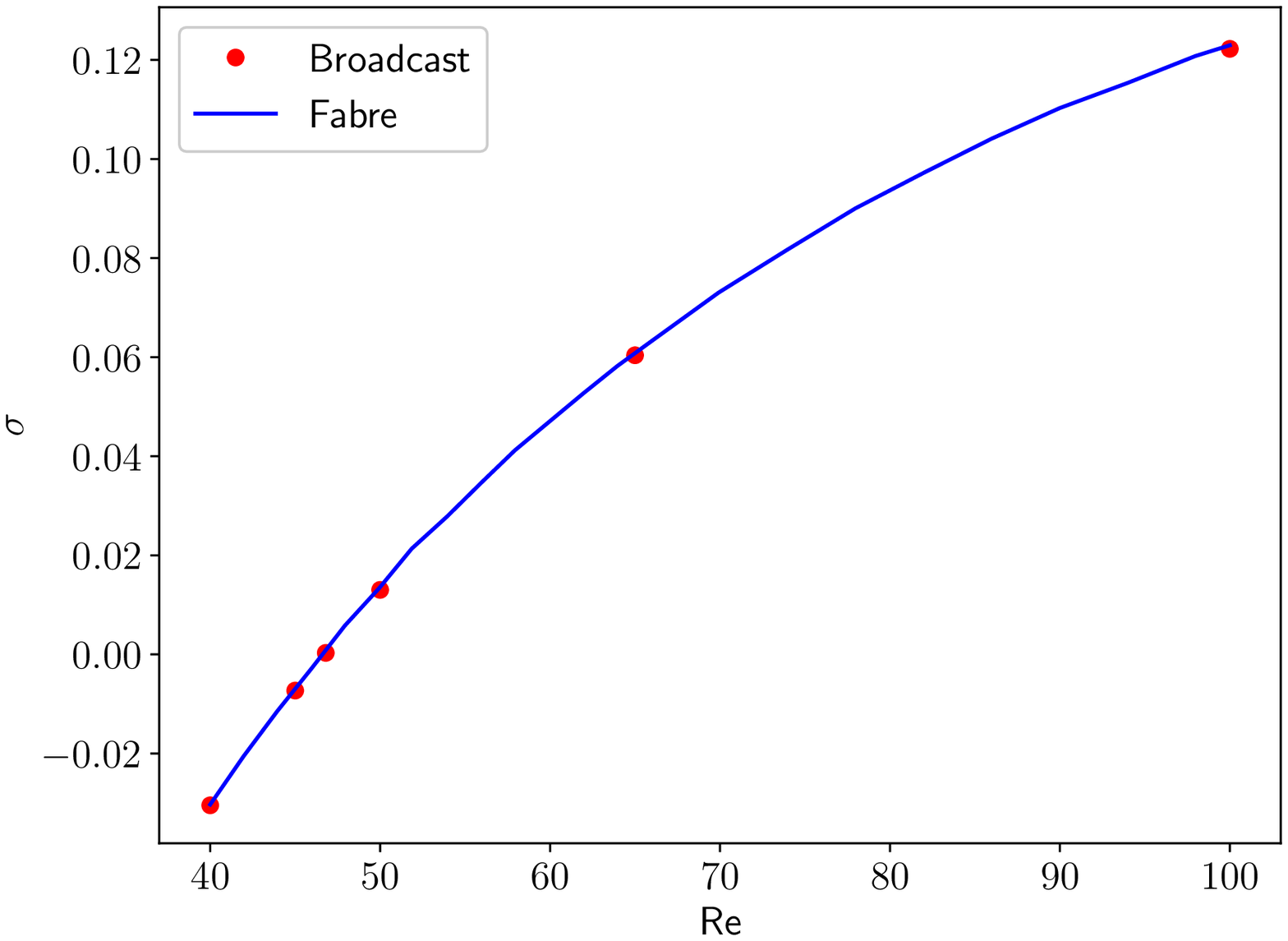}
\caption{Evolution of the growth rate  with Reynolds number for $ \beta=0$ and $ M=0.1$. Comparison with the incompressible results obtained by Fabre {\it et al.} \cite{fabre2018practical}.}
\label{cyl2D}
\end{subfigure}
\begin{subfigure}{0.49\textwidth}
\includegraphics[width=0.99\textwidth]{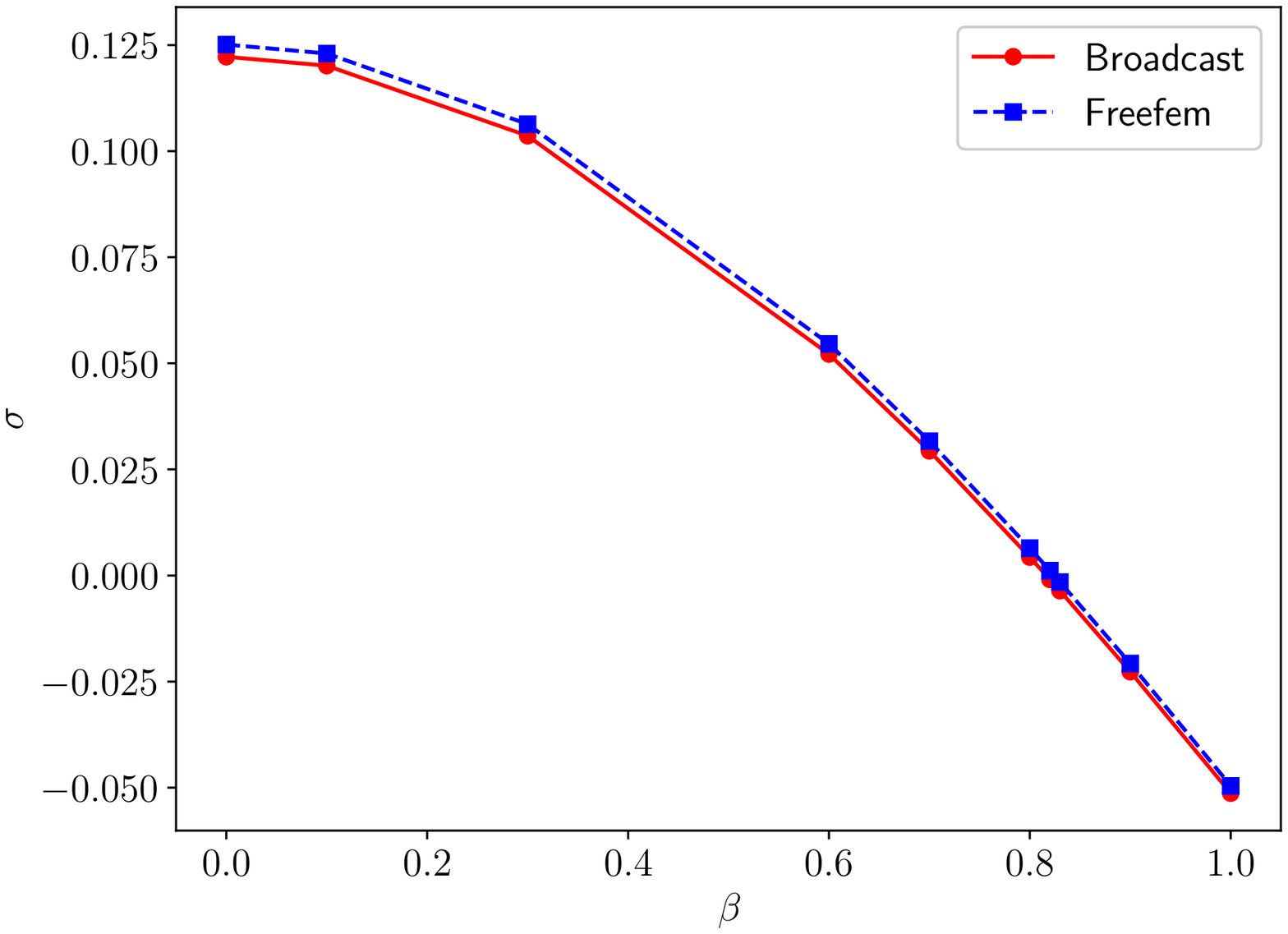}
\caption{Evolution of the growth rate with the wavenumber $\beta$ at $Re = 100$ and $ M=0.1$. Comparisons done with incompressible results obtained with FreeFEM software \cite{hecht2012new,marquet2008amplifier}.}
\label{cyl3D}
\end{subfigure}
\caption{Growth rate as a function of Reynolds number and spanwise wavenumber for cylinder flow at $ M=0.1$.}
\label{cyl}
\end{figure}

Evolution of the growth rate of the most unstable eigenvalue with Reynolds number is plotted in Figure \ref{cyl2D} and compared with the calculations of Fabre \cite{fabre2018practical}, which was obtained with an incompressible solver. Overall, good agreement is achieved. Furthermore, 3D instabilities computed with the 3D Jacobian $A_\text{3D}$ at $ Re=100 $ show the same evolution of the growth rate with the spanwise wavenumber $\beta$ (Figure \ref{cyl3D}) normalised by the diameter $D$ as the one provided by incompressible Navier-Stokes equations discretised with FreeFEM software \cite{hecht2012new,marquet2008amplifier} (mesh of 138708 triangles with P2 and P1 elements).

\subsubsection{Linear sensitivity of eigenvalue}

\begin{figure}[!ht]
\centering
\includegraphics[width=0.5\textwidth]{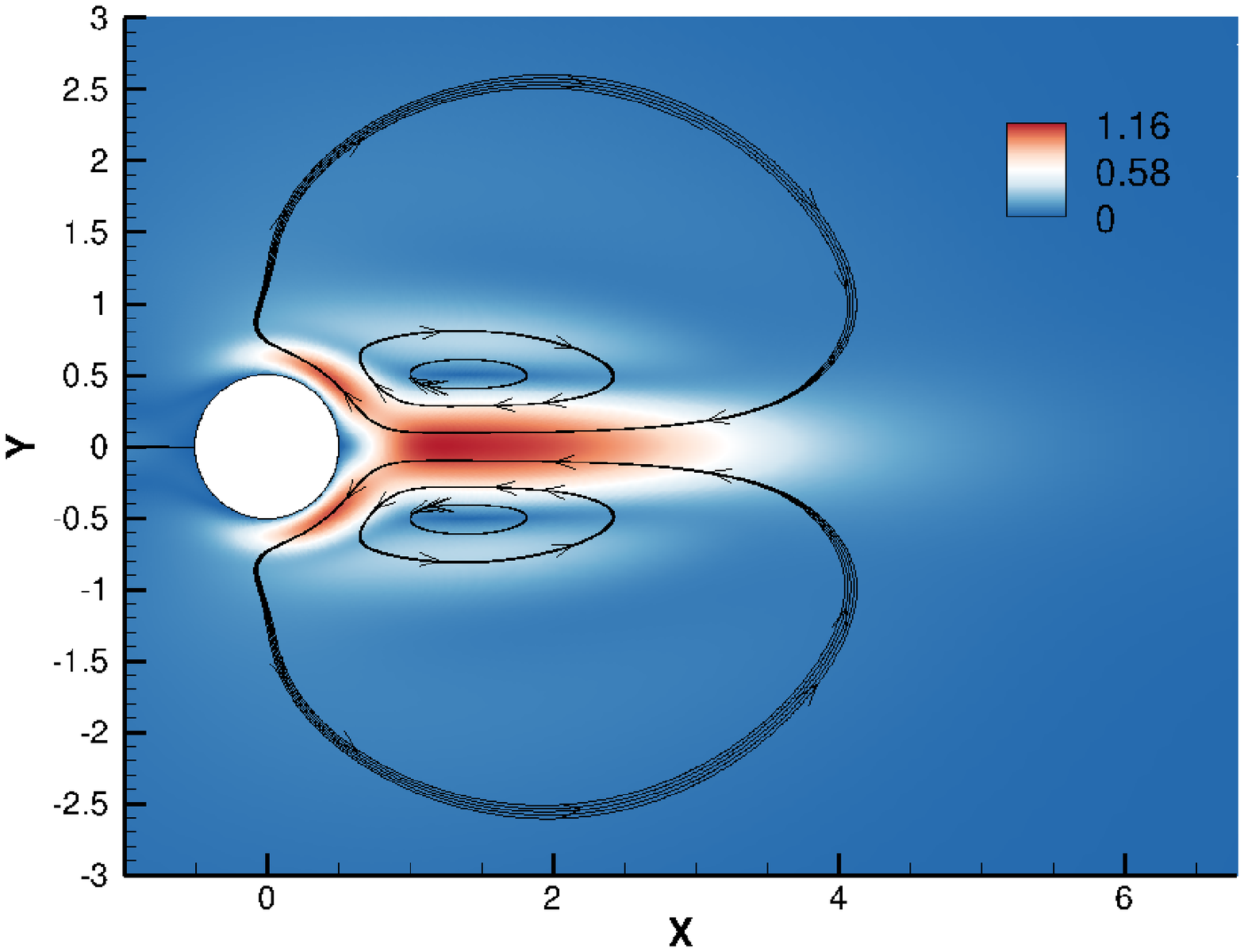}
\caption{Intensity and direction of the momentum component of the growth rate sensitivity to a steady source term $\left. \nabla_f \sigma \right|_Q$ computed with BROADCAST code. Similar to the Figure 15(a) in Mettot {\it et al.} \cite{mettot2014computation}.}
\label{sensitivity}
\end{figure}

In Figure \ref{sensitivity}, we show the sensitivity of the critical growth rate to a steady source term $\left. \nabla_f \sigma \right|_Q$ (see eq. \eqref{hes10}). We recover the same sensitivity regions and bounds as found in Figure 15(a) by Mettot {\it et al.} \cite{mettot2014computation}.

\subsubsection{Amplitude equations}

\begin{figure}[!ht]
\centering
\begin{subfigure}{0.49\textwidth}
\includegraphics[width=0.99\textwidth]{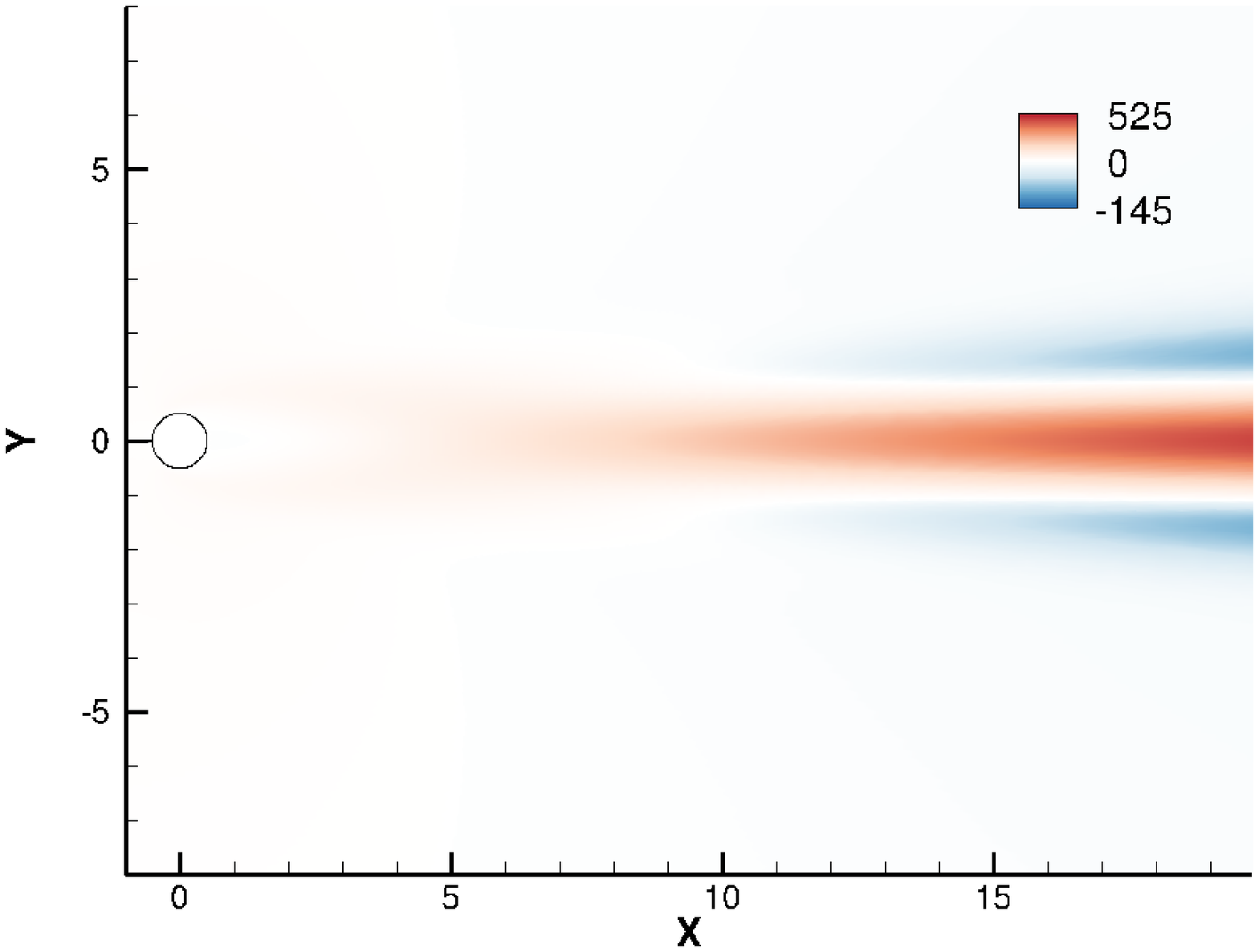}
\caption{Streamwise velocity of the zeroth harmonic $\hat{q}_{20}$ at $Re=46.8$ and $M=0.1$.}
\label{q20}
\end{subfigure}
\begin{subfigure}{0.49\textwidth}
\includegraphics[width=0.99\textwidth]{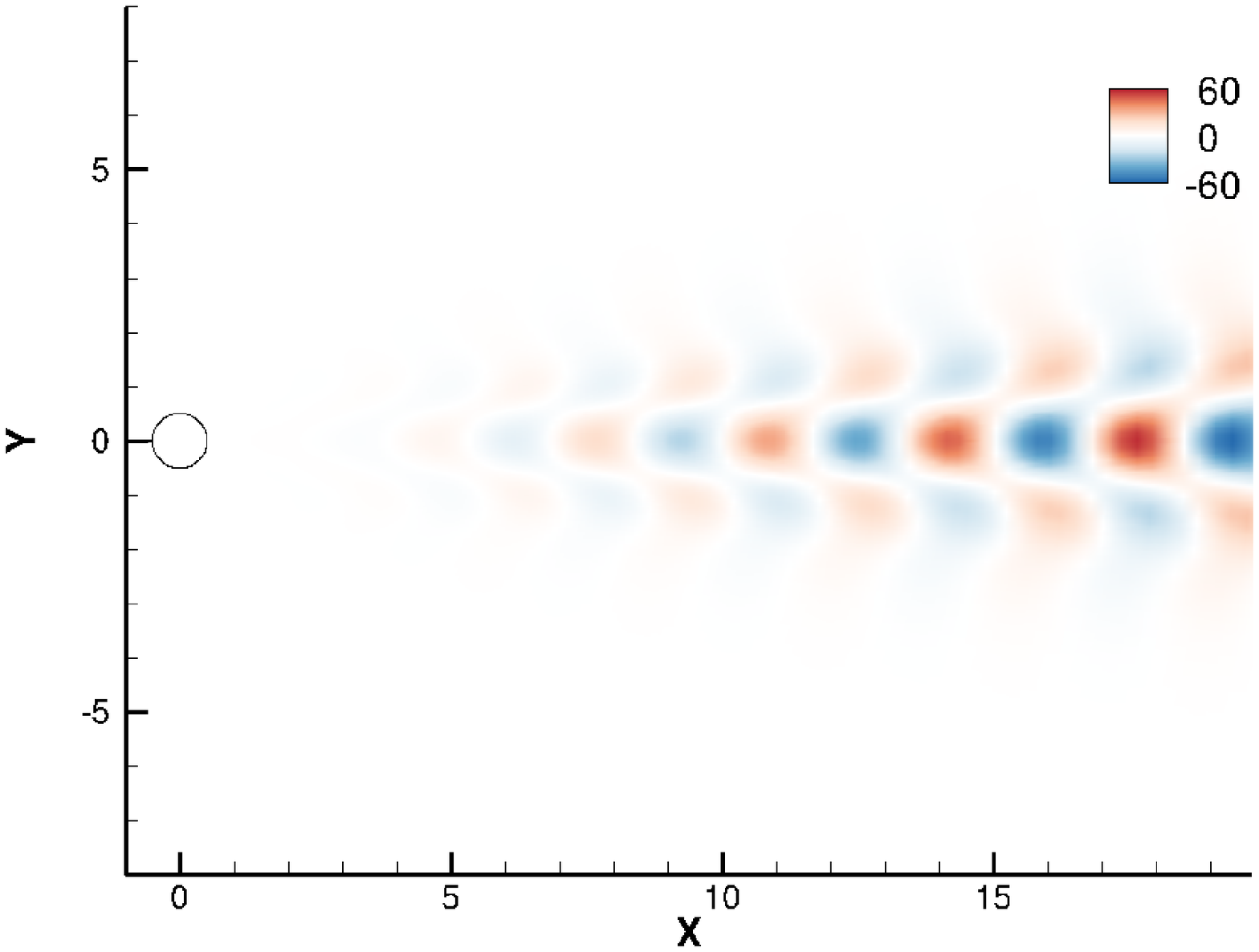}
\caption{Real part of the streamwise velocity of the second harmonic $\hat{q}_{22}$ at $Re=46.8$ and $M=0.1$.}
\label{q22}
\end{subfigure}
\caption{Higher harmonics used to compute the Stuart-Landau coefficients $\mu$ and $\nu$. Same normalisation as in Sipp and Lebedev \cite{sipp2007global}.}
\label{harmonics2}
\end{figure}

The weakly nonlinear analysis is compared to the results given in Sipp and Lebedev \cite{sipp2007global}. We focus on the coefficients of the cubic term, since its sign determines the super-critical (real part of $ \mu+\nu+\xi>0 $, see \eqref{mu}-\eqref{xi}) or sub-critical (real part of $ \mu+\nu+\xi<0 $) nature of the bifurcation. They are computed from the different modes $\hat{q}_1$, $\tilde{q}_1$, $\hat{q}_{20}$ and $\hat{q}_{22}$ (Eq. \eqref{eq:mode0}-\eqref{eq:mode2}). The modes are plotted in Figures \ref{harmonics} and \ref{harmonics2}. The coefficients, computed for $Re = 46.8$, $M=0.1$ and $ \beta=0$ with the same normalisation scheme as in Sipp {\it et al.}, are the following:

\begin{eqnarray}
\mu &= &9.24 - 33.2i \\
\nu &=& -0.31 - 0.87i \\
\xi &=& 0.00032 - 0.00095i
\end{eqnarray}

We note that the coefficient $\xi$ involving the third-order derivative term is negligible with respect to the two other coefficients $\mu$ and $\nu$ (ratio about $10^{-3}$ between $\nu$ and $\xi$). This stems from the fact that $ M=0.1$ is close to the incompressible regime, for which this constant is strictly zero. This is due to the quadratic nature of the incompressible Navier-Stokes equations, so that the third-order derivative of the residual is strictly zero. However this contribution would not disappear for higher Mach numbers and would therefore be mandatory to assess the nature of a bifurcation in transsonic and supersonic configurations.
The $ \mu $ and $ \nu $ coefficients can be compared to the incompressible values $\mu_\text{REF} = 9.4-30i$ and $\nu_\text{REF} = -0.30-0.87i$ \cite{sipp2007global}. The discrepancy remains small (maximum $10\%$ for $Im(\mu)$) but cannot be reduced by any grid refinement or domain enlargement. It rather appears linked to the behaviour of the compressible Navier-Stokes equations as the Mach number decreases. Indeed, compressibility has a strong impact on the global modes, which deviate from the incompressible solutions when the Mach number is above $ 0.1$. Conversely, at Mach numbers lower than $M = 0.02$, the system of equations is not well-conditioned making the calculation of the eigenmodes inaccurate. This comes from the division by a quasi-zero Mach number in the perfect gas law (eq. \eqref{NS2}).

\subsection{Adiabatic flat plate} \label{sec:bl}

In the following, the base-flow of a hypersonic boundary layer over an adiabatic flat plate, the resolvent analysis and the sensitivity of the optimal gain are performed.

\subsubsection{Geometry, mesh and boundary conditions}

The adiabatic flat plate is studied in the hypersonic regime and validated against Bugeat's study \cite{bugeat20193d}.
The Navier-Stokes equations are again non-dimensionalised by the triplet ($\rho_\infty,\: U_\infty,\: T_\infty $). The freestream Mach number is $M=4.5$, the freestream temperature is $T_\infty = 288$ K and the unit Reynolds number is $Re_{\text{unit}}=3.4 \times 10^6$ with the unit Reynolds number defined as $Re_{\text{unit}}= \rho_\infty U_\infty / \mu_\infty$.

The flat plate geometry is studied in a rectangular computational domain. The domain starts with a thin boundary layer profile at $Re_{x,\text{in}} = 8160$ and ends at $Re_{x,\text{out}} = 2 \times 10^6$. The height of the domain is high enough in order not to affect the development of the boundary-layer or the stability analysis. In practice, the domain height is about $9\,\delta^*_\text{out}$ with $\delta^*_\text{out}$ the compressible displacement thickness at the outlet. This gives $Re_{y,\text{top}} = 119000$. The Cartesian mesh is equispaced in the x-direction and stretched in the wall-normal direction (y-direction). The stretching has the following properties: maximum $y^+ \leq 1$, cell height geometric growth rate of $2 \%$ from $y=0$ till $y=3\delta^*_\text{out}$ where an equivalent $ \Delta y^+ \approx 10$ and then a growth rate increase from $2 \%$ till $10 \%$ from $y=3\delta^*_\text{out}$ till $y=L_y$ where $\Delta y^+ \approx 130$. The reference mesh has the size $(N_x, N_y) = (1000, 150)$ which gives $N = 150 000$ grid points.


Four different boundary conditions are applied around the rectangular domain. First, at the inlet, a Dirichlet boundary condition is applied since the flow is supersonic. The imposed flow profile corresponds to a compressible self-similar solution for $u$, $v$, $\rho$ and $T$. A self-similar solver for compressible boundary layer flow without pressure gradient is implemented in BROADCAST. It uses the Levy-Lee transformation, the Chapman-Rubesin \cite{chapman1949temperature} linear law for viscosity and a shooting method to solve the self-similar equations \cite{begou2018prevision,cho2004similarity}. One should note that the wall-normal velocity $v$ computed from self-similar equations is small but not zero in the free-stream. This velocity is imposed at the inlet without any further treatment. At the outlet, an extrapolation boundary condition is applied (the flow is overall assumed supersonic). It is known that the flow is actually not supersonic close to the wall at the inlet and outlet. A specific inlet boundary condition splitting the flow into a supersonic part and a subsonic part and applying respectively a complete Dirichlet and a non-reflecting treatment was implemented but did not alter the boundary layer development. The supersonic inlet and outlet boundary conditions are kept for robustness in supersonic computations. Then, an adiabatic no-slip wall is prescribed at the bottom while a non-reflecting condition is employed at the top boundary.

\subsubsection{Base-flow}

The compressible self-similar solution is taken as initial state for the Newton method. On the reference mesh, at $ M=4$, with the seventh order numerical scheme, the Newton algorithm converges in 7 iterations (high initial CFL as the self-similar solution is close to the base-flow solution) and decreases the residual $L^2$ norms by 12 orders of magnitude. 

\begin{figure}[!ht]
\centering
\begin{subfigure}{0.49\textwidth}
\includegraphics[width=0.99\textwidth]{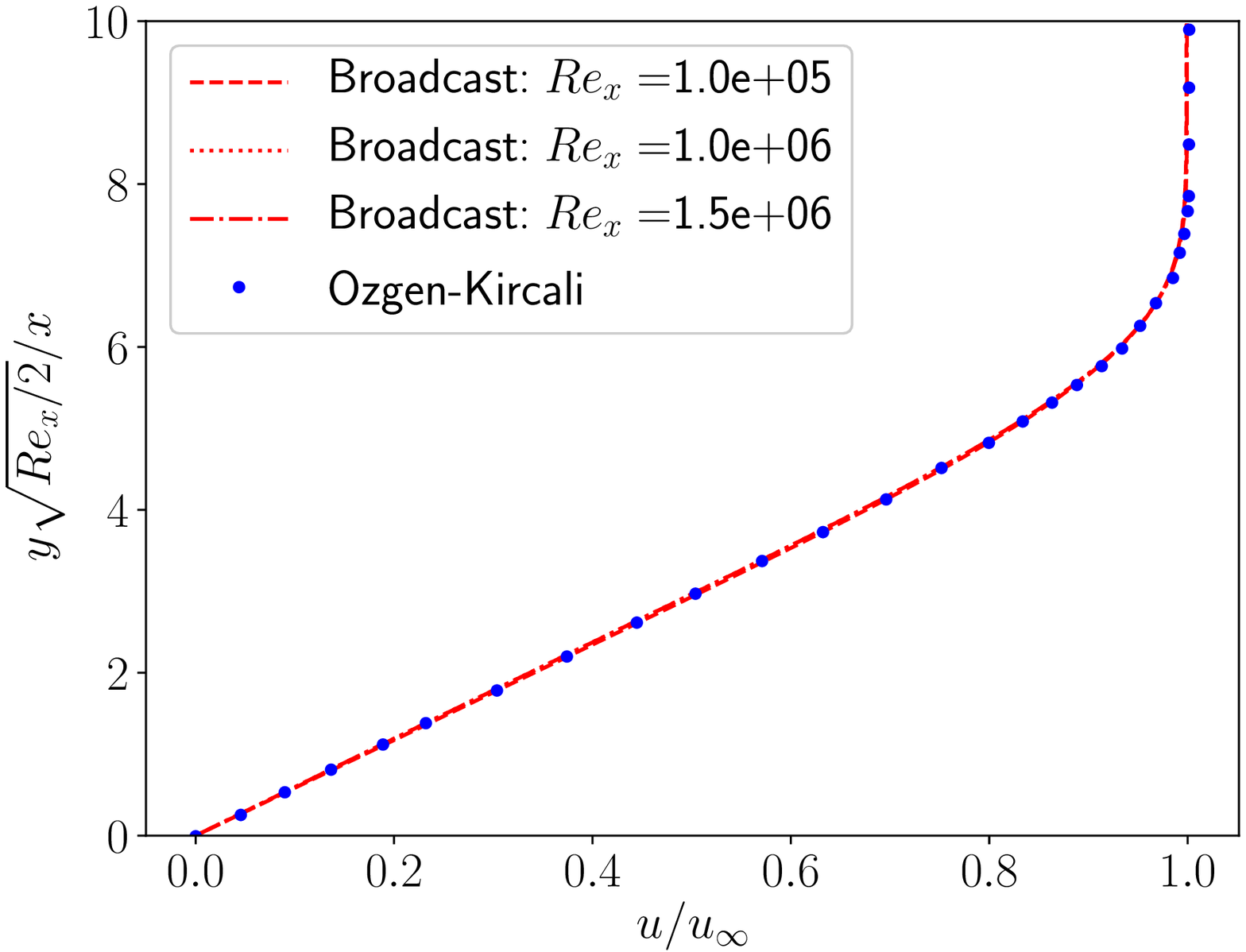}
\caption{Streamwise velocity profile.}
\label{Profu}
\end{subfigure}
\begin{subfigure}{0.49\textwidth}
\includegraphics[width=0.99\textwidth]{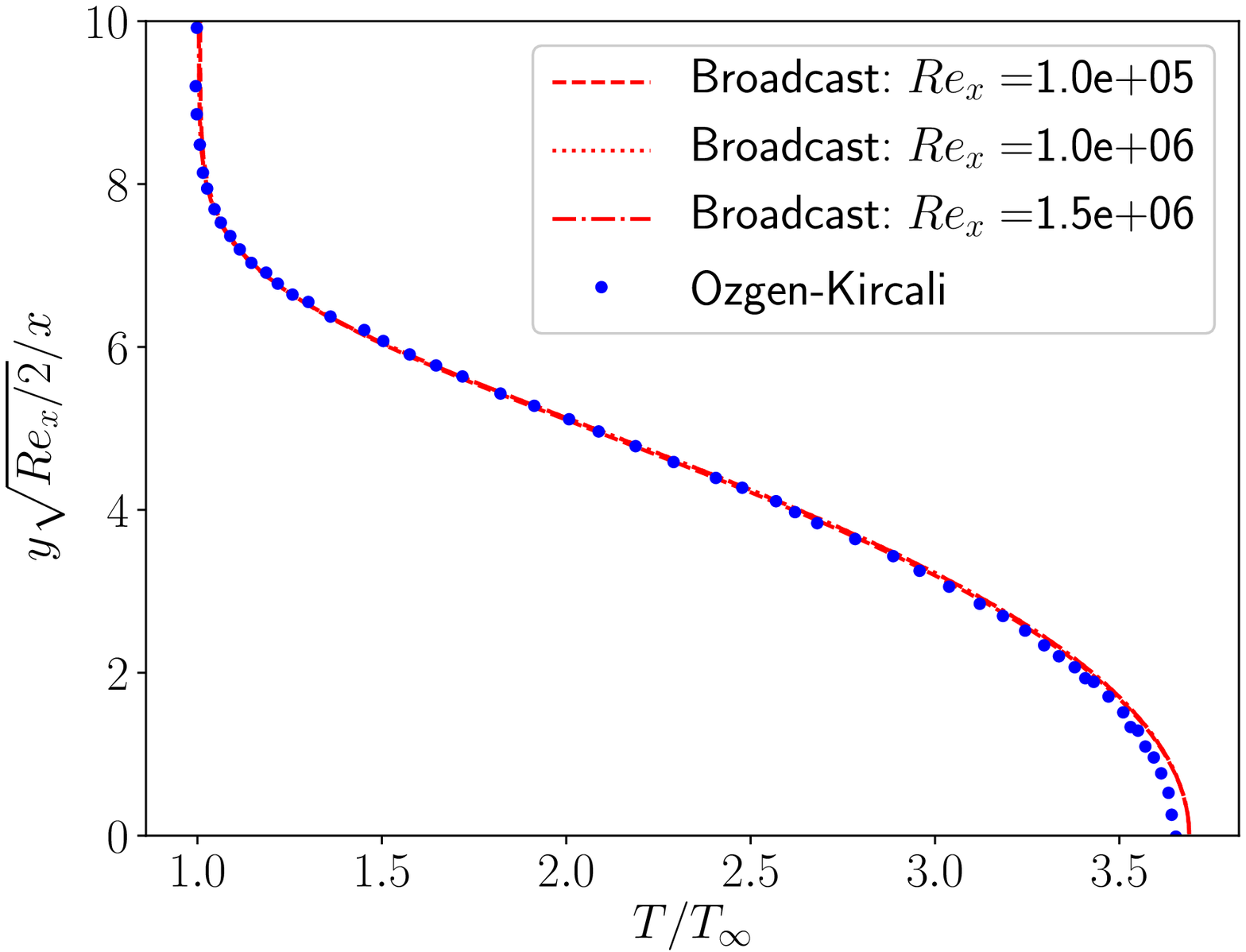}
\caption{Temperature profile.}
\label{ProfT}
\end{subfigure}
\caption{Boundary layer profiles at $M=4$ plotted with respect to the self-similar variable at different streamwise locations. Comparison with \"Ozgen and Kircali \cite{ozgen2008linear}.}
\label{Prof1}
\end{figure}

Boundary layer profiles at Mach number $M=4$ are extracted at different streamwise locations and plotted with respect to the self-similar variable $ \frac{y}{x} \sqrt{\frac{Re_x}{2}}$ on Figures \ref{Prof1}. 
The base-flow computed remains self-similar and the different profiles overlap. 
Furthermore, streamwise velocity and temperature profiles are validated against \"Ozgen and Kircali \cite{ozgen2008linear}. Both show good agreement with a maximum difference of 1\% of the wall temperature prediction.

Then, a base-flow solution is computed at $M=4.5$ to perform the stability analysis. The slope of the ratio of the Reynolds number based on the displacement thickness on the square root of the Reynolds number $Re_{\delta^*} / \sqrt{Re_x}$ is computed to compare the base-flow with Bugeat {\it et al.} \cite{bugeat20193d}. BROADCAST code gives a ratio of 8.29 while Bugeat has 8.32 meaning that both boundary layer base-flows follow a similar evolution.

\subsubsection{Resolvent modes}

For the resolvent modes, we consider Bugeat's case \cite{bugeat20193d} at $M = 4.5$.
The frequency is normalised as $F = \left( \omega L_\text{ref} \right) / U_\infty$ with $L_\text{ref} = Re_{\text{unit}}^{-1}$. The wavenumber $\beta$ is also normalised with the reference length scale $L_\text{ref}$.

\begin{figure}[!ht]
\centering
\includegraphics[width=0.5\textwidth]{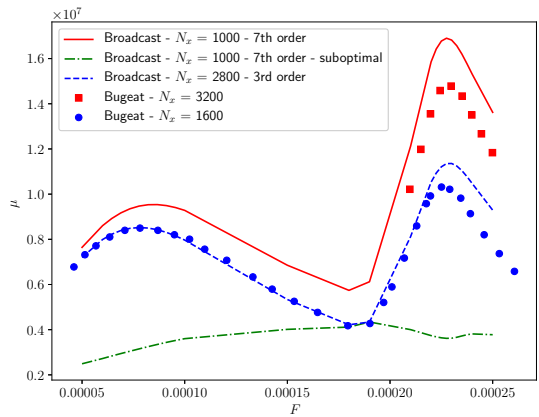}
\caption{Optimal gains $\mu$ for 2D instabilities ($\beta=0$) as a function of normalised frequency $F$. Comparison with Bugeat  {\it et al.}\cite{bugeat20193d}.}
\label{Gain2D}
\end{figure}

The Chu's energy restricted to $Re_{x} \leq 1.75 \times 10^6$ and $ Re_y \leq 59500 $ is chosen for $ Q_E$. For $Q_F$, we consider the discrete norm defined in equation \eqref{Qfbugeat} with momentum forcings and same spatial restriction as for $Q_E$ with the restriction / prolongation matrix $ P $. 
Optimal gains for 2D perturbations ($ \beta=0$) are plotted in Figure \ref{Gain2D} at different frequencies. Above $M=4$, Mack \cite{mack1963inviscid} showed the existence of an infinite number of higher frequency modes. At $M=4.5$, the second Mack mode has the largest amplification rate and corresponds to the frequency with the highest peak in Figure \ref{Gain2D}. Because the second Mack mode appears at high frequency, a fine mesh is necessary to converge well the peak value of the optimal gain. One may notice in Bugeat {\it et al.} \cite{bugeat20193d} that a twice finer mesh (from $N_x = 1600$ to $N_x = 3200$) improved the optimal gain by 50 \%. The suboptimal gain is also plotted in Figure \ref{Gain2D} and appears 4 times smaller than the optimal gain at the frequency of the second Mack mode ($F=2.3 \times 10^{-4}$).

\begin{figure}[!ht]
\centering
\begin{subfigure}{0.49\textwidth}
\includegraphics[width=0.99\textwidth]{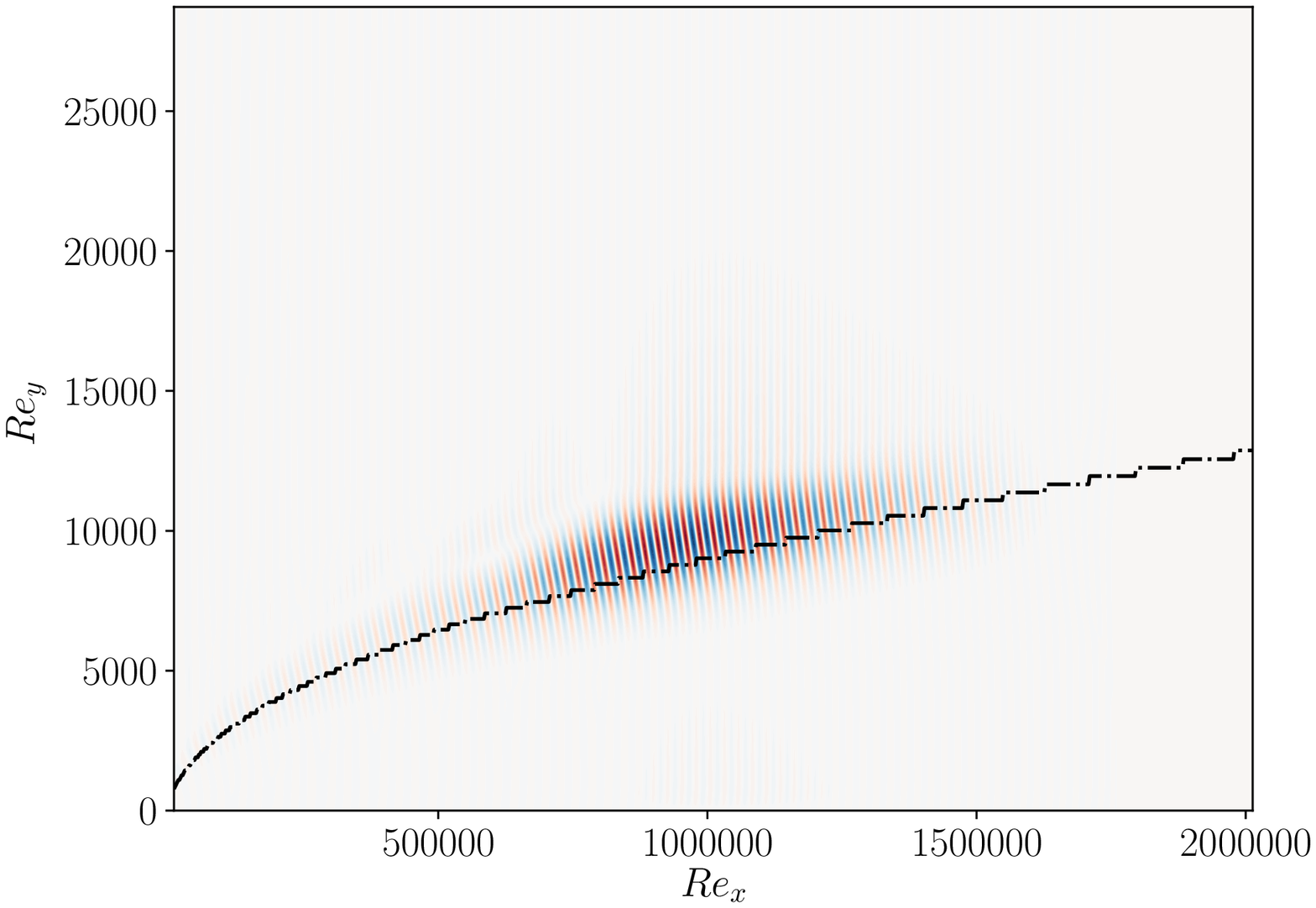}
\caption{Streamwise momentum component of the optimal forcing $\check{f}$.}
\label{forcing}
\end{subfigure}
\begin{subfigure}{0.49\textwidth}
\includegraphics[width=0.99\textwidth]{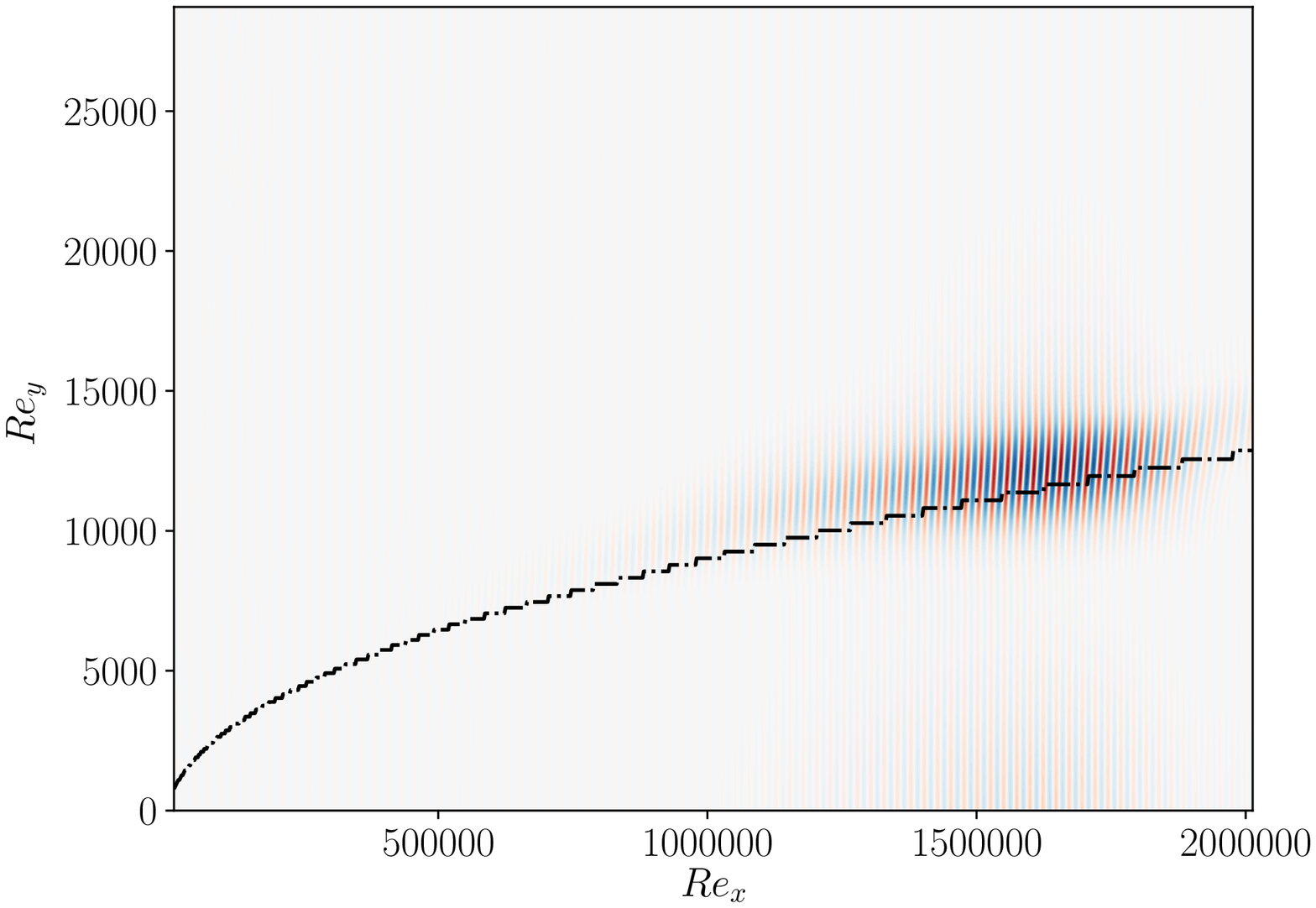}
\caption{Density component of the optimal response $\check{q}$.}
\label{response}
\end{subfigure}
\caption{Optimal forcing and response for the second Mack mode at $F=2.3 \times 10^{-4}$ ($\beta = 0$). Red (resp. blue) color is negative (resp. positive) value. The black dash-dotted line denotes the generalised inflection point.}
\label{forcresp}
\end{figure}

The optimal gain convergence is assessed at the second Mack mode frequency $F$ in Table \ref{table:tabconv2}. The computational performance (RAM and elapsed time) are compared for different mesh refinements and scheme orders. The number of cells per wavelength $ \xi $ in x-direction $N_c/\xi$ is also shown for mesh convergence. According to the optimal forcing and response shapes plotted in Figure \ref{forcresp}, the streamwise direction is the critical one as the wave is mainly oriented in the x-direction. The mesh convergence in y-direction has also been assessed and $N_y=150$ appears to be sufficient. The seventh order scheme with 225000 points ($N_x = 1500$) is enough to reach the peak gain while a coarser mesh with $N_x = 1000$ at the same scheme order gives a gain smaller by only 1\%. Taking the latter as the reference, a comparison of the performance of the different scheme orders is presented in Table \ref{table:tabiso}. All computations are run at the same memory cost (38 GB RAM) by a mesh adjustment in the streamwise direction and the performance criterion is again the optimal gain at the second Mack mode frequency $ F=2.3 \times 10^{-4}$. Results from this iso-computation cost analysis underlines two elements. First, at this frequency, increasing the numerical scheme order from a low order (third order) to a fifth or a seventh order is heavily beneficial as the 3rd order scheme ends up at a smaller gain than all the other schemes (see Table \ref{table:tabconv2} and Figure \ref{Gain2D}). Secondly, increasing to a high order (ninth order) is not always the best solution. Indeed to keep the same memory cost, a high order scheme must be applied on a coarser mesh. However, a minimum number of cells per wavelength is always required, approximately here around $12$ to capture the second Mack mode gain. It means that depending on the considered frequency, it exists an optimal scheme order in terms of memory that the user may have to choose.

\begin{table}[!ht]
\begin{center}
\begin{tabular}{ c|c|c|c|c|c } 
 \hline
Optimal & Scheme  & $N_x$ & $N_c/\xi$ &  RAM  & Elapsed  \\ 
gain $ \mu $ & order &  &   &  (GB) & time (s)  \\
\hline
$1.70 \times 10^7$ & 9th & 2000 & 24 &  113 & 1530  \\
$1.70 \times 10^7$ & 7th & 1500 & 18 &  56 & 480  \\
$1.69 \times 10^7$ & 7th & 1000 & 12 & 38 & 300  \\
$1.23 \times 10^7$ & 3rd & 20000 & 240 & 285 & 10800  \\
$1.22 \times 10^7$ & 3rd & 6000 & 72 & 99 & 1821 
\end{tabular}
\caption{Mesh convergence: optimal gain convergence at the second Mack mode frequency $ F=2.3 \times 10^{-4}$. $N_c/\xi$ is the number of cells per wavelength in the streamwise x-direction. $N_y = 150$. Only one Krylov vector computed to reach $10^{-4}$ convergence in Arnoldi algorithm.}
\label{table:tabconv2}
\end{center}
\end{table}

\begin{table}[!ht]
\begin{center}
\begin{tabular}{ c|c|c|c|c|c|c } 
 \hline
Optimal  & Scheme  & $N_x$ & $N_c/\xi$ & nnz in & RAM & Elapsed \\ 
gain $ \mu $ & order &  &   & Jacobian & (GB) &  time (s)  \\
\hline
$1.66 \times 10^7$  & 9th & 700 & 8.4 & $59 \times 10^6$ & 38 & 290  \\
$1.69 \times 10^7$  & 7th & 1000 & 12 & $72 \times 10^6$ & 38 & 300  \\
$1.68 \times 10^7$ & 5th & 1500 & 18 & $85 \times 10^6$ & 38 & 360  \\
$1.13 \times 10^7$ & 3rd & 2800 & 33.6 & $93 \times 10^6$ & 38 & 440 
\end{tabular}
\caption{Iso-computation cost at $38$ GB RAM: optimal gain at the second Mack mode frequency $ F=2.3 \times 10^{-4}$. nnz in Jacobian represents the total number of non-zero entries in the Jacobian matrix. $N_y = 150$. Only one Krylov vector computed to reach $10^{-4}$ convergence in Arnoldi algorithm.}
\label{table:tabiso}
\end{center}
\end{table}

\begin{figure}[!ht]
\centering
\begin{subfigure}{0.49\textwidth}
\includegraphics[width=0.99\textwidth]{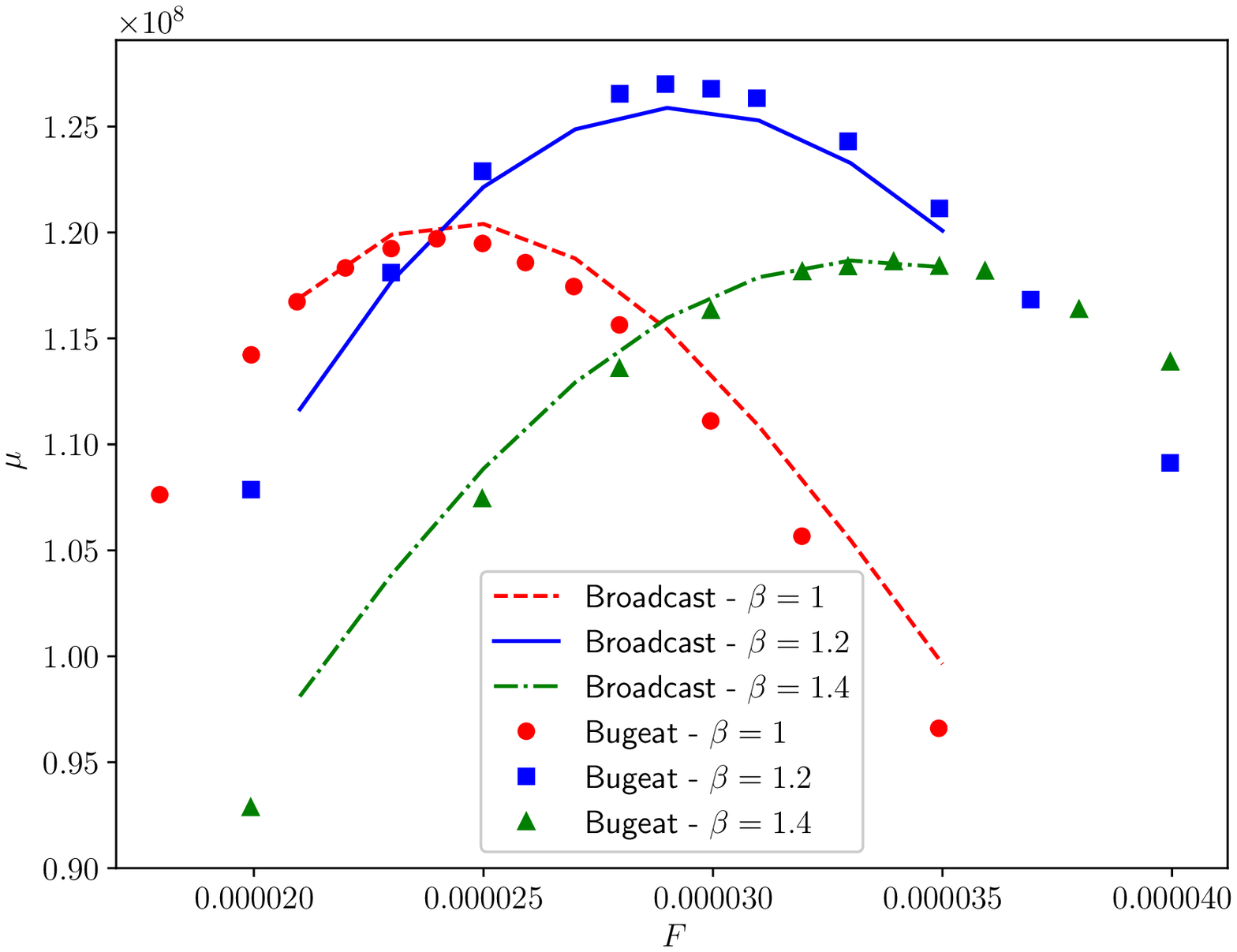}
\caption{Optimal gain with respect to the normalised frequency at the normalised wavenumbers $\beta=1$, $\beta=1.2$ and $\beta=1.4$. Peaks refer to the first Mack mode.}
\label{Gain3D}
\end{subfigure}
\begin{subfigure}{0.49\textwidth}
\includegraphics[width=0.99\textwidth]{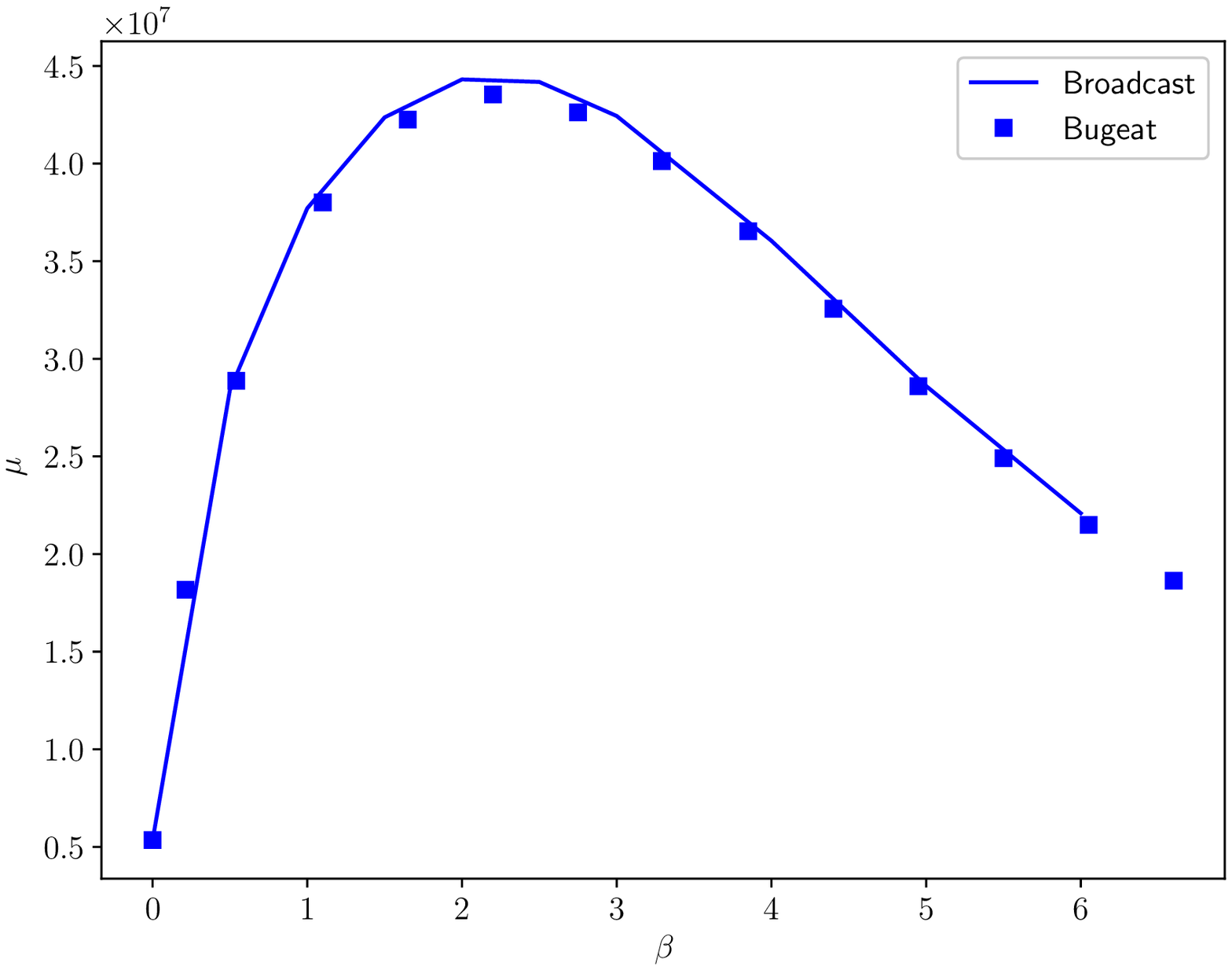}
\caption{Optimal gain with respect to the normalised wavenumber at zero frequency: Streaks. The peak is related to a streak.}
\label{Gain3Dstraks}
\end{subfigure}
\caption{Optimal gain: 3D linear instabilities. Comparison with Bugeat {\it et al.} \cite{bugeat20193d}.}
\label{Gain3Ds}
\end{figure}

Optimal gain is then computed with $\beta \neq 0$. The three-dimensional gains (Figure \ref{Gain3Ds}) compare well with those of Bugeat {\it et al.} \cite{bugeat20193d}, highlighting streaks around $\beta = 2.2$ at zero frequency and the first oblique Mack mode strongest for $\beta = 1.2$ and $F=2.9 \times 10^{-5}$.

\begin{figure}[!ht]
\centering
\begin{subfigure}{0.49\textwidth}
\includegraphics[width=0.99\textwidth]{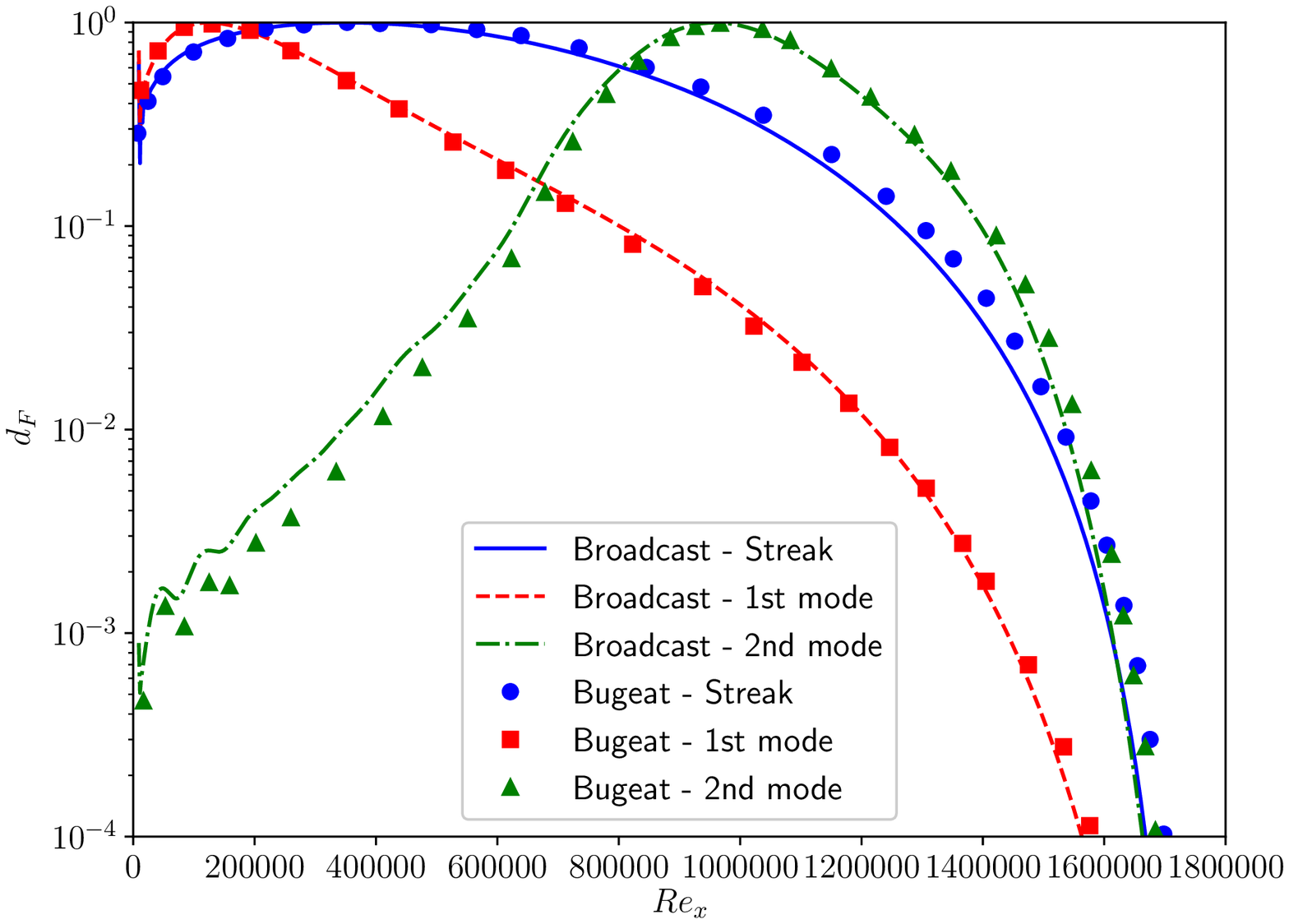}
\caption{Forcing density.}
\label{df}
\end{subfigure}
\begin{subfigure}{0.49\textwidth}
\includegraphics[width=0.99\textwidth]{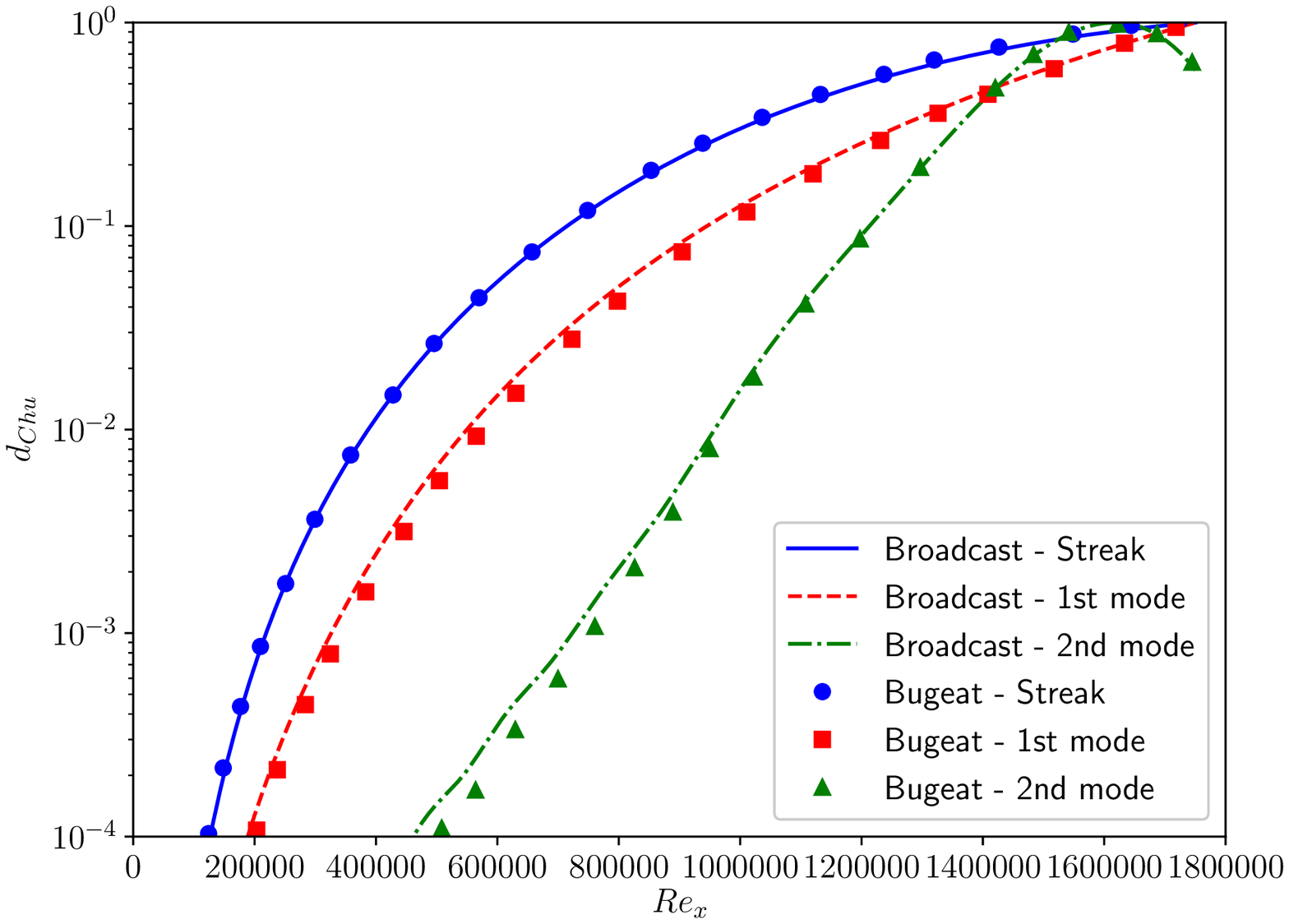}
\caption{Chu's energy density.}
\label{dchu}
\end{subfigure}
\caption{Forcing and Chu's energy density of the streaks, first and second Mack modes normalised by their maximum value with respect to the Reynolds number based on $x$. Comparison with Bugeat {\it et al.} \cite{bugeat20193d}.}
\label{dfdchu}
\end{figure}

Then the eigenvectors for the three predominant modes (streak, first and second Mack modes) are compared through their streamwise energy growth. An energy density is defined as the integral of the local energy in the wall-normal direction. For kinetic energy, it is written $d_\text{K} (x) = \int_0^{y_\text{max}} \overline{\rho} |\check{\mathbf{v}}|^2 \,dy $. From Chu's energy definition (eq. \eqref{gstab7}), a Chu's energy density is defined similarly and written $d_\text{Chu} (x)$. Eventually, a forcing density is defined as $d_\text{F} (x) = \int_0^{y_\text{max}} |\check{f}|^2 \, dy$. Forcing and Chu's energy densities are plotted in Figures \ref{dfdchu}. Forcing and Chu's energy densities have similar trends to the ones predicted by Bugeat {\it et al.} \cite{bugeat20193d}. These results can be linked with local linear stability. The peak of optimal forcing density corresponds with branch I (convectively stable towards unstable bound) while the the peak of optimal response density if it exists inside the domain corresponds with branch II (convectively unstable towards stable bound) \cite{sipp2013characterization}.

\subsubsection{Linear sensitivity of optimal gain}

Linear sensitivity of the optimal gain is computed on an adiabatic flat plate. However, to our knowledge, such sensitivity analysis has not been performed on hypersonic flows yet. For sake of comparison, we consider Brandt {\it et al.} test case \cite{brandt2011effect} at incompressible regime.

The freestream Mach number is $M=0.1$, Reynolds number is $Re=6 \times 10^5$. The computational domain extends from $Re_{x,\text{in}} = 4300$ to $Re_{x,\text{out}} = 7.5 \times 10^5$, the optimisation for the resolvent mode is restrained up to $Re_x \leq 6 \times 10^5$. The height of the domain is about $ 9\,\delta$ with $\delta$ the displacement thickness at $Re_x = 6 \times 10^5$. The following boundary conditions are applied: Blasius solution prescribed at inlet, adiabatic no-slip wall at bottom, non-reflecting condition at top and non-reflecting subsonic outlet with the free-stream pressure taken as reference (zero pressure gradient assumed). A similar mesh to the hypersonic boundary layer is chosen with $(N_x, N_y) = (1000, 150)$. 

\begin{figure}[!ht]
\centering
\begin{subfigure}{0.49\textwidth}
\includegraphics[width=0.99\textwidth]{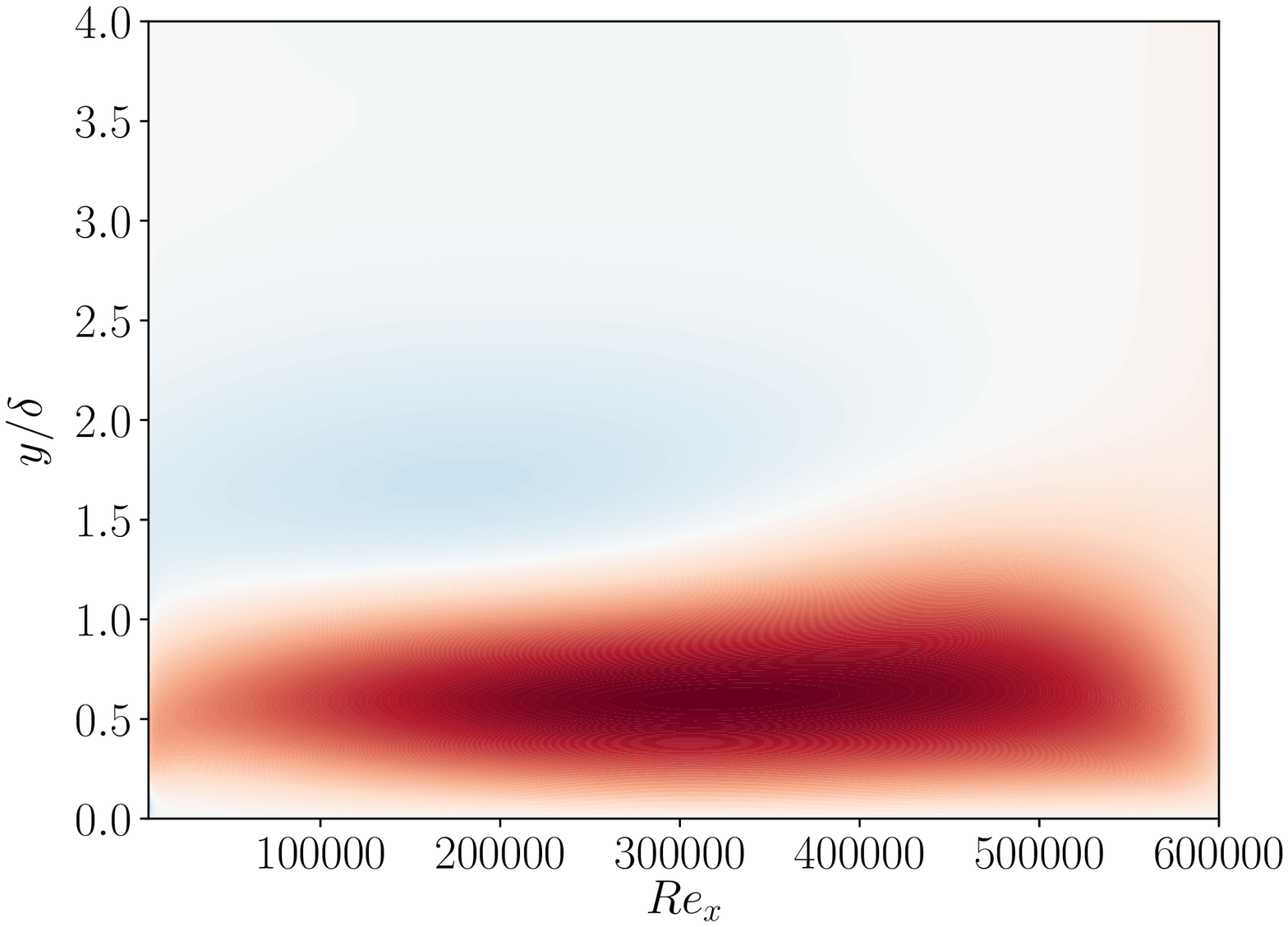}
\caption{Streamwise component of the sensitivity to steady forcing $ \nabla_{f} \left. \mu^2 \right|_Q$. Red (resp. blue) color is negative (resp. positive) gradient. Similar to the Figure 19(a) in Brandt {\it et al.} \cite{brandt2011effect}.}
\label{sensBLforc}
\end{subfigure}
\begin{subfigure}{0.49\textwidth}
\includegraphics[width=0.99\textwidth]{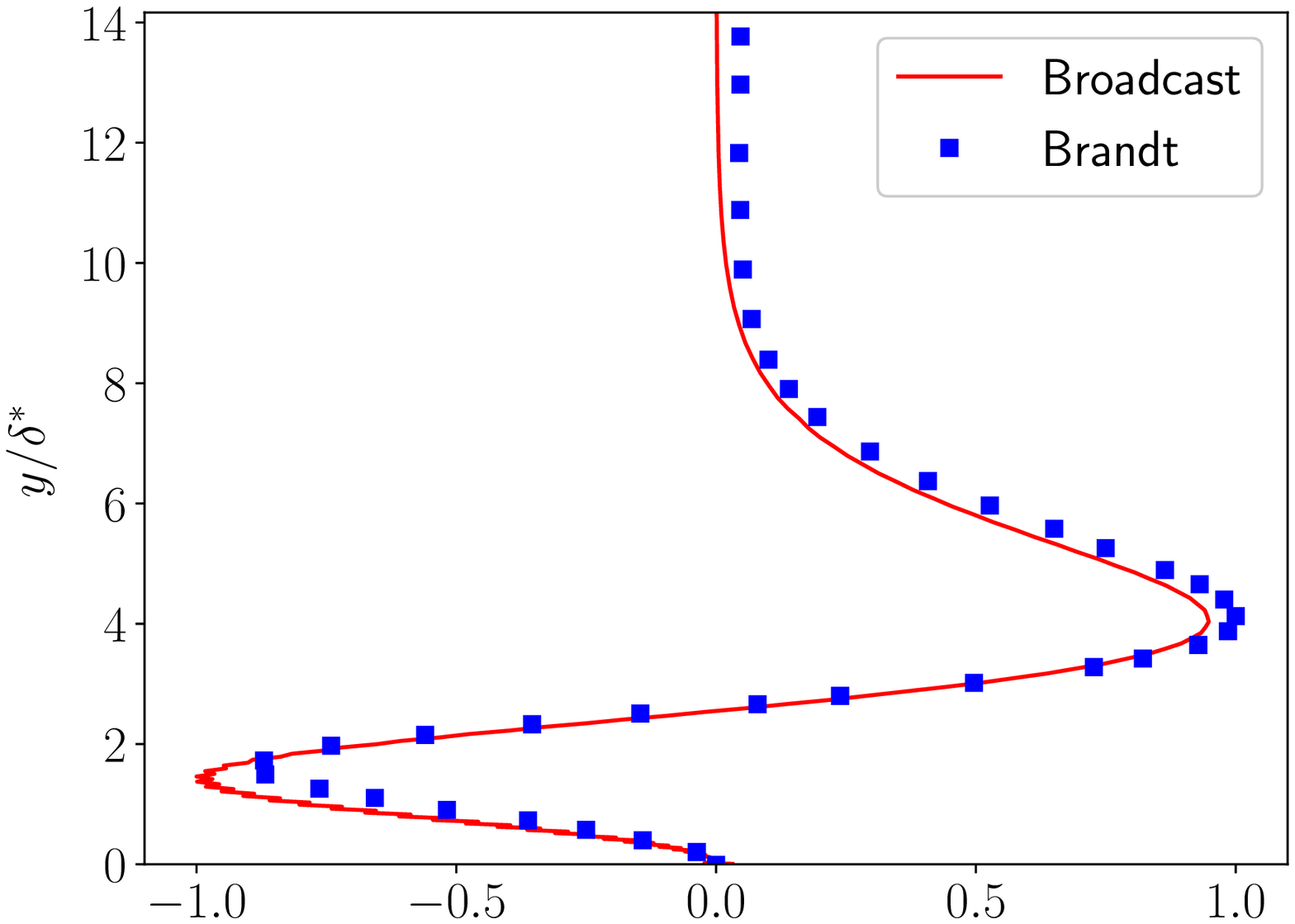}
\caption{Wall-normal profile of the streamwise component of the sensitivity to base-flow modification $ \nabla_{\overline{q}} \left. \mu^2 \right|_Q$ at $Re_{\delta^*} =400$ normalised by its maximum value. Comparison with Brandt {\it et al.} \cite{brandt2011effect}.}
\label{sensBL}
\end{subfigure}
\caption{Sensitivity of the optimal gain for the streaks ($\beta \delta = 0.94$). Comparison with Brandt {\it et al.} \cite{brandt2011effect}.}
\label{sensBLinc}
\end{figure}

In Brandt {\it et al.} \cite{brandt2011effect}, the largest optimal gain at zero frequency is obtained for the streaks at $\beta \delta = 0.94$. We compare the sensitivity of this 3D mode to base-flow modification (Eq. \ref{sensigain}) and to forcing (Eq. \ref{sensigain2}) obtained by BROADCAST with the one from Brandt {\it et al.} \cite{brandt2011effect} (Figure \ref{sensBLinc}). The wall-normal profile of the streamwise component of the sensitivity to base-flow modification is plotted at $Re_{\delta^*} = 400$ with $\delta^*$ the displacement thickness (Figure \ref{sensBL}). They show good agreement.

\section{Conclusion}

\subsection{Conclusion}

The BROADCAST code and the methods implemented have been presented and assessed. The high-order scheme associated with low dissipation has showed accurate results in agreement with literature at a low computational cost for resolvent gains computations at high frequencies. Furthermore, it may be used to Mach numbers as low as $M \approx 0.1$ to explore the incompressible regime both on Cartesian and curvilinear meshes. Finally the Algorithmic Differentiation tool allows accurate and straightforward evaluations of the various residual derivatives that are necessary for stability, sensitivity or weakly nonlinear analyses.

\subsection{Future developments}

The BROADCAST code is still in development. RANS equations with various turbulence models will be implemented as well as multi-block management. The next major stability feature will be the global nonlinear stability tool offered in Rigas {\it et al.} \cite{rigas2021nonlinear}. For a finite amplitude forcing, the response resulting from a limited number of harmonics and their interactions can be studied through an adjoint optimisation algorithm. It will rely on the Time Spectral Method (TSM) framework \cite{canuto2012spectral}.

\section*{Acknowledgements}
This work is funded by the French Agency for Innovation and Defence (AID) as part of the UK-FR PhD program.

\appendix{}

\section{Convective flux: the Flux-Extrapolated MUSCL scheme} \label{sec:convflux}

The FE-MUSCL scheme has been assessed for hypersonic flows by Sciacovelli {\it et al.} \cite{sciacovelli2021assessment}. A description of the scheme and its implementation in BROADCAST follows.
In 1D, assuming a regular grid spacing $h$, from the equation \eqref{FV3}, the discrete scheme for inviscid flow is written for the cell index $i$:

\begin{equation}
\frac{\text{d} q_i}{\text{d} t} + \frac{(\delta F)_i}{h} = 0    
\label{dnc1}
\end{equation}
with $\delta$ the difference operator between two cells: $ \delta(\cdot)_{i+\frac{1}{2}} = (\cdot)_{i+1} - (\cdot)_{i} $ and $F_{i+\frac{1}{2}}$ the inviscid numerical flux at cell interface $i+\frac{1}{2}$. For a ninth-order scheme, this flux is written:

\begin{equation}
F_{i+\frac{1}{2}} = \left( \left(I - \frac{1}{6} \delta^2 + \frac{1}{30} \delta^4 - \frac{1}{140}\delta^6 + \frac{1}{630}\delta^8 \right) \mu f - D \right)_{i+\frac{1}{2}}
\label{dnc2}
\end{equation}
with $f$ the physical flux, $\mu$ the cell average operator: $\mu(\cdot)_{i+\frac{1}{2}} = \frac{1}{2} \left( (\cdot)_{i+1} + (\cdot)_i \right) $ and $D$ the numerical dissipation term.

The third order scheme is for instance constructed keeping only the two first terms in the brackets. As flow discontinuities are inherent in compressible flows, the dissipation coefficient is required and written as a nonlinear artificial viscosity term associated with the spectral radius of the inviscid flux $\lambda_{i+\frac{1}{2}}$. Then, the dissipation coefficient $D_{i+\frac{1}{2}}$ is written:

\begin{equation}
D_{i+\frac{1}{2}} = \lambda_{i+\frac{1}{2}} \left( \epsilon_2 \delta q +  \epsilon_4 \delta^9 q \right)_{i+\frac{1}{2}}   
\label{dnc3}
\end{equation}
with $\epsilon_{2_{i+\frac{1}{2}}} = \kappa_2 \textrm{max} \left( \phi_i, \phi_{i+1} \right) $ and $\epsilon_{4_{i+\frac{1}{2}}} = \textrm{max} \left(0, \kappa_4 - \epsilon_{2_{i+\frac{1}{2}}} \right) $. The dissipation coefficients range respectively $0 \leq \kappa_2 \leq 1$ and $\kappa_4 \approx \frac{1}{1260}$. 

\begin{equation}
\phi_i = \frac{1}{2} \left( 1 - \tanh{ \left( 2.5 + 10\frac{\delta x}{c} \nabla \cdot \mathbf{v} \right) } \right) \times \frac{(\nabla \cdot \mathbf{v})^2}{(\nabla \cdot \mathbf{v})^2 + |\nabla \times \mathbf{v}|^2 + \epsilon} \times \left| \frac{p_{i+1} -2 p_{i} + p_{i-1} }{p_{i+1} +2 p_{i} + p_{i-1}} \right|
\label{dnc4}
\end{equation}
with $c$ the local sound velocity, $\delta x$ the local horizontal cell size, $\epsilon$ a small fixed parameter to avoid a division by zero and $p_i$ the local pressure.
In equation \eqref{dnc4}, the last term is the pressure based shock sensor of Jameson \cite{jameson1981numerical} and the second term is the discrete form of Ducros' vortex sensor which measures the ratio between the local divergence of velocity and the velocity rotational \cite{ducros1999large}. The first term is the sensor modification offered by Bhagatwala {\it et al.} \cite{bhagatwala2009modified} which is switched on only in region of negative dilatation. These three sensors are put in series so that they bypass each other in order to reduce the magnitude of the low-order dissipation. The sensor $\phi$ is then activated in region of strong divergence of velocity typically around shocks so that the flow discontinuities are well-captured. In vortex regions, the sensor tends to zero and the dissipation is driven by the high-order term associated with $\kappa_4$ which damps high-frequency oscillations while minimising the level of numerical dissipation.






The stencil of this scheme needs $(\text{order+1}) / 2 $ cells in each direction to compute the inviscid flux. For instance, in 1D, the ninth order scheme requires 5 cells at the left and 5 cells at the right of the current cell. Close to the computational domain boundaries, the appropriate number of ghost cells is added to the mesh. This treatment is applied for the porous boundary conditions such as inflow, outflow and extrapolation. However, for wall boundaries, another strategy is followed as offered in \cite{cinnella2016high}. The physical flux is imposed at the first mesh interface (first cell layer) and high-order off-centered schemes are used to compute the flux in the next cell layers (from the second to the fifth cell layer for a ninth order scheme). The same order of accuracy is thus preserved everywhere, whether the flux is computed fully inside the domain, using ghost cells or close to a wall boundary condition.

\section{3D contributions for the Jacobian operator} \label{sec:3D}

The 3D contributions gather all the z-derivatives components in the residual. They are listed in the table \ref{table:tabz}. Most of these terms vanish for a two-dimensional base-flow as they involve the transverse velocity $w$.

\begin{table}[!ht]
\begin{center}
\begin{tabular}{ |c|c|c|c|c|c|c|c|c|c| } 
\hline
\multicolumn{10}{|c|}{Continuity equation} \\
\hline
$k$ & $\alpha_k$ & $a_k$ & $l$ & $\lambda_l$ & $b_l$ & $m$ & $\gamma_m$ & $c_m$ & $d_m$ \\
\hline
1 & $1$ & $\rho  w$ &  &  &  &  &  &  &   \\
\hline
\multicolumn{10}{|c|}{Momentum equation in x-direction} \\
\hline
$k$ & $\alpha_k$ & $a_k$ & $l$ & $\lambda_l$ & $b_l$ & $m$ & $\gamma_m$ & $c_m$ & $d_m$ \\
\hline
1 & $1$ & $\rho u w - \mu \frac{\partial w}{\partial x}$ & 1 & $-\mu$ & $u$ & 1 & $-1$ & $\mu$ & $u$  \\
2 & $\frac{2}{3}$ $\frac{\partial \mu}{\partial x}$ & $w$ & & & & & & & \\
3 & $\frac{2}{3}$ $\mu$ & $\frac{\partial w}{\partial x}$ & & & & & & & \\
\hline
\multicolumn{10}{|c|}{Momentum equation in y-direction} \\
\hline
$k$ & $\alpha_k$ & $a_k$ & $l$ & $\lambda_l$ & $b_l$ & $m$ & $\gamma_m$ & $c_m$ & $d_m$ \\
\hline
1 & $1$ & $\rho v w - \mu \frac{\partial w}{\partial y}$ & 1 & $-\mu$ & $v$ & 1 & $-1$ & $\mu$ & $v$ \\
2 & $\frac{2}{3} \frac{\partial \mu}{\partial y}$ & $w$ & & & & & & & \\
3 & $\frac{2}{3} \mu$ & $\frac{\partial w}{\partial y}$ & & & & & & & \\
\hline
\multicolumn{10}{|c|}{Momentum equation in z-direction} \\
\hline
$k$ & $\alpha_k$ & $a_k$ & $l$ & $\lambda_l$ & $b_l$ & $m$ & $\gamma_m$ & $c_m$ & $d_m$ \\
\hline
1 & $1$ & $\rho w^2 + p + \frac{2}{3}\mu \left( \frac{\partial u}{\partial x} + \frac{\partial v}{\partial y} \right) $ & 1 & $-\frac{4}{3} \mu$ & $w$ & 1 & $-\frac{4}{3}$ & $\mu$ & $w$ \\
2 & $-\frac{\partial \mu}{\partial x}$ & $u$ & & & & & & & \\
3 & $-\frac{\partial \mu}{\partial y}$ & $v$ & & & & & & & \\
4 & $-\mu$ & $\frac{\partial u}{\partial x} + \frac{\partial v}{\partial y}$ & & & & & & & \\
\hline
\multicolumn{10}{|c|}{Energy equation} \\
\hline
$k$ & $\alpha_k$ & $a_k$ & $l$ & $\lambda_l$ & $b_l$ & $m$ & $\gamma_m$ & $c_m$ & $d_m$ \\
\hline
1 & $1$ & $(\rho E + p) w$ & 1 & $-\lambda$ & $T$ & 1 & $-1$ & $\lambda$ & $T$  \\
2 & $\frac{2}{3} \frac{\partial (\mu u)}{\partial x}$ & $w$ & 2 & $- \mu u$ & $u$ & 2 & $-\mu$ & $u$ & $u$ \\
3 & $\frac{2}{3} \mu u$ & $\frac{\partial w}{\partial x}$ & 3 & $-\mu v$ & $v$ & 3 & $-u$ & $\mu$ & $u$ \\
4 & $-\frac{\partial (\mu w)}{\partial x}$ & $u$ & 4 & $-\frac{4}{3} \mu w$ & $w$ & 4 & $-\mu$ & $v$ & $v$ \\
5 & $-\mu w$ & $\frac{\partial u}{\partial x}$ & & & & 5 & $-v$ & $\mu$ & $v$ \\
6 & $\frac{2}{3} \frac{\partial (\mu v)}{\partial y}$ & $w$ & & & & 6 & $-\frac{4}{3} \mu$ & $w$ & $w$ \\
7 & $\frac{2}{3} \mu v$ & $\frac{\partial w}{\partial y}$ & & & & 7 & $-\frac{4}{3} w$ & $\mu$ & $w$ \\
8 & $-\frac{\partial (\mu w)}{\partial y}$ & $v$  & & & & & & & \\
9 & $-\mu w$ & $\frac{\partial v}{\partial y}$   & & & & & & & \\
10 & $-\mu \frac{\partial w}{\partial x}$ & $u$ & & & & & & & \\
11 & $-u$ & $\mu \frac{\partial w}{\partial x}$  & & & & & & & \\
12 & $-\mu \frac{\partial w}{\partial y}$ & $v$ & & & & & & & \\
13 & $-v$ & $\mu  \frac{\partial w}{\partial y}$ & & & & & & & \\
14 & $\frac{2}{3} \mu \left( \frac{\partial u}{\partial x} + \frac{\partial v}{\partial y} \right)$ & $w$  & & & & & & & \\
15 & $\frac{2}{3} w \left( \frac{\partial u}{\partial x} + \frac{\partial v}{\partial y} \right)$ & $\mu$ & & & & & & & \\
16 & $\frac{2}{3} \mu w$ & $\frac{\partial u}{\partial x} + \frac{\partial v}{\partial y}$ & & & & & & & \\
\hline
\end{tabular}
\caption{3D contributions in Navier-Stokes equations.}
\label{table:tabz}
\end{center}
\end{table}

\section{Weakly nonlinear analysis equations} \label{sec:wnl}

The method described in Sipp and Lebedev \cite{sipp2007global} for incompressible flow, developing the Taylor expansion of the residual up to the third order, is applied here on the general form given by the discretised equation \eqref{Newton0}. First, one assumes that the analysis is performed in the vicinity of the critical Reynolds number so that a small parameter $\epsilon$ can be defined as $\epsilon^2 = Re_c^{-1} - Re^{-1}$ with $0 < \epsilon \ll 1$. As it is dependent on the Reynolds number, the state vector $q$ can be written as a Taylor expansion of the parameter $\epsilon$:

\begin{equation}
q(t,\epsilon) = q_0 + \epsilon q_1(t) + \epsilon^2 q_2(t) + \epsilon^3 q_3(t) = q_0 + \epsilon q
\label{wna1}
\end{equation}
with $q = q_1(t) + \epsilon q_2(t) + \epsilon^2  q_3(t)$. 

The discretised residual $R=R(q(t),\epsilon)$ is a function of the state vector $q(t)$ and the parameter $\epsilon$. It is expanded up to the third order for these two variables in equation \eqref{wna2}. Some notations and definitions are introduced for clarity. The Jacobian operator around the base-flow, written $A$, is equal to $\partial_q R(\overline{q},0)$ where $\partial_q(\cdot) = \partial (\cdot) / \partial q$. Then, the Hessian operator appears too as $H(q,q) = \left[\partial_{qq} R(\overline{q},0)q\right]q$. In discrete form, it can be expressed as $H_i(q,q) = \sum_k \sum_j (\partial_{q_jq_k} R_i(\overline{q},0)) q_j q_k$. As this is a symmetric operator, one gets $H(q_1,q_2) = H(q_2,q_1)$. Similarly the third-order derivative operator, written $T(q,q,q) = \left[ \left[\partial_{qqq} R(\overline{q},0)q\right]q\right]q$, has the same symmetrical properties and is written in discrete form as $T_i(q,q,q) = \sum_l \sum_k \sum_j (\partial_{q_jq_kq_l} R_i(\overline{q},0)) q_j q_k q_l$.


\begin{equation}
R(q(t),\epsilon) = R(q_0 + \epsilon q, \epsilon) = R(q_0) + \epsilon A q + \frac{\epsilon^2}{2} H(q,q) + \\ \frac{\epsilon^3 }{6} T(q,q,q) + \epsilon^2 \partial_{\epsilon} R(q_0,0) + \epsilon^3 \partial_{q\epsilon} R(q_0,0) q
\label{wna2}
\end{equation}

By introducing the equations \eqref{wna1} and \eqref{wna2} into the equation \eqref{Newton0}, written as $\text{d} q(t) / \text{d}t = R(q(t), \epsilon)$, one gets the equation \eqref{wna3} below which will be studied at the different orders of $\epsilon$. 

\begin{multline}
\frac{\text{d} q_0}{\text{d} t} + \epsilon \frac{\text{d} q_1}{\text{d} t} + \epsilon^2 \frac{\text{d} q_2}{\text{d} t} + \epsilon^3 \frac{\text{d} q_3}{\text{d} t} = R(q_0) +  \epsilon A q_1 +  \epsilon^2 \left[ A q_2 + \frac{1}{2} H(q_1, q_1) + \partial_{\epsilon} R(q_0,0) \right] + \\ \epsilon^3 \left[ A q_3 + H(q_1,q_2) + \right. \left. \frac{1}{6} T(q_1, q_1, q_1) + \partial_{q\epsilon} R(q_0,0) q_1 \right]
\label{wna3}
\end{multline}
First, at order $\epsilon^0$, the base-flow equation $\text{d} q_0 / \text{d} t = 0 = R(q_0,0)$ is recovered and then $q_0$ is the base-flow solution.
Secondly, at order $\epsilon^1$, one gets the eigenvalue problem \eqref{wna4}:

\begin{equation}
\frac{\text{d} q_1}{\text{d} t} - A q_1 = 0
\label{wna4}
\end{equation}
The general solution is of the following form:

\begin{equation}
q_1(t) = \mathcal{A} e^{i\omega t} \hat{q}_1 + \overline{\mathcal{A}} e^{-i\omega t} \overline{\hat{q}_1}
\label{wna5}
\end{equation}
with $\mathcal{A}$ a complex scalar which represents the amplitude and depends on the slow time scale $t_1 = \epsilon^2 t$ according to the Stuart-Landau amplitude equation \eqref{landau}, $\omega$ the critical frequency and $\hat{q}$ the first eigenmode solution of the equation \eqref{wna55}.

\begin{equation}
\left( i\omega I - A\right) \hat{q}_1 = 0
\label{wna55}
\end{equation}
Thirdly, at order $\epsilon^2$, the equation becomes:

\begin{multline}
\frac{\text{d} q_2}{\text{d} t} - A q_2 = \frac{1}{2} H(q_1,q_1) + \partial_{\epsilon} R(q_0,0) = \frac{1}{2} \mathcal{A}^2 e^{2i\omega t} H(\hat{q}_1,\hat{q}_1) +  \frac{1}{2} \overline{\mathcal{A}}^2 e^{-2i\omega t} H(\overline{\hat{q}_1}, \overline{\hat{q}_1}) + |\mathcal{A}|^2  H(\hat{q}_1, \overline{\hat{q}_1}) + \partial_{\epsilon} R(q_0,0)
\label{wna6}
\end{multline}
From the form of the equation \eqref{wna6}, the solution is sought as:

\begin{equation}
q_2(t) = \mathcal{A}^2 e^{2i\omega t} \hat{q}_{22} + \overline{\mathcal{A}}^2 e^{-2i\omega t} \overline{\hat{q}_{22}} + |\mathcal{A}|^2  \hat{q}_{20} + \hat{q}_{21}
\label{wna7}
\end{equation}
with $\hat{q}_{22}$ the second harmonic, $\hat{q}_{20}$ the zeroth (mean flow) harmonic and $\hat{q}_{21}$ the base flow modification due to a $\epsilon$ Reynolds increase, respectively solutions of the equations \eqref{wna8}, \eqref{wna9} and \eqref{wna10}.

\begin{equation}
2i\omega \hat{q}_{22} - A \hat{q}_{22} = \frac{1}{2} H(\hat{q}_{1}, \hat{q}_{1})
\label{wna8}
\end{equation}

\begin{equation}
-A \hat{q}_{20} = H (\hat{q}_{1},\overline{\hat{q}_{1}})
\label{wna9}
\end{equation}

\begin{equation}
- A \hat{q}_{21} = \partial_{\epsilon} R(q_0,0)
\label{wna10}
\end{equation}
Fourthly, at order $\epsilon^3$, the remaining terms in the equation \eqref{wna3} give the equation \eqref{wna11}.

\begin{equation}
\frac{\text{d} q_3}{\text{d} t} - A q_3 = H(q_1,q_2) + \frac{1}{6} T(q_1, q_1, q_1) + \partial_{q\epsilon} R(q_0,0) q_1
\label{wna11}
\end{equation}
In the equation \eqref{wna11}, by computing only the terms proportional to $e^{i\omega t}$, it reduces to the equation \eqref{wna12}.

\begin{multline}
i\omega q_3 - A q_3 = - \frac{\text{d} \mathcal{A}}{\text{d} t_1} \:\hat{q}_{1} + \mathcal{A} H(\hat{q}_{1}, \hat{q}_{21}) + \mathcal{A} \partial_{q\epsilon} R(q_0,0) \hat{q}_{1} +  \mathcal{A} |\mathcal{A}|^2 \left[ H(\hat{q}_{1},\hat{q}_{20}) + H(\overline{\hat{q}_{1}},\hat{q}_{22}) + \frac{3}{6} T(\hat{q}_{1}, \hat{q}_{1}, \overline{\hat{q}_{1}}) \right]
\label{wna12}
\end{multline}
Let's introduce the adjoint mode $\tilde{q}_{1}$ solution of the adjoint eigenvalue problem \eqref{wna13}.

\begin{equation}
\left( -i\omega I - A^*\right) \tilde{q}_{1} = 0
\label{wna13}
\end{equation}

Then, the goal is to recover the Stuart-Landau amplitude equation from the equation \eqref{wna12}. As the latter is a "vector" equation and the amplitude equation a "scalar" equation, we compute the dot product of the equation \eqref{wna12} with the adjoint mode. The left hand-side of the new equation becomes zero as described in the equation \eqref{wna14}.

\begin{multline}
i\omega \langle \tilde{q}_{1}, q_3 \rangle - \langle \tilde{q}_{1}, A q_3 \rangle = i\omega \langle \tilde{q}_{1}, q_3 \rangle - \langle A^*\tilde{q}_{1}, q_3 \rangle =  i\omega \langle \tilde{q}_{1}, q_3 \rangle - \langle -i\omega\tilde{q}_{1}, q_3 \rangle = i\omega \langle \tilde{q}_{1}, q_3 \rangle - i\omega \langle \tilde{q}_{1}, q_3 \rangle = 0
\label{wna14}
\end{multline}
And the right hand-side gives the amplitude equation.

\begin{multline}
0 =  - \frac{\text{d} \mathcal{A}}{\text{d} t_1} \: \langle \tilde{q}_{1}, \hat{q}_{1} \rangle + \mathcal{A} \left( \langle \tilde{q}_{1}, H(\hat{q}_{1}, \hat{q}_{21}) \rangle + \langle \tilde{q}_{1}, \partial_{q\epsilon} R(q_0,0) \hat{q}_{1} \rangle \right) + \\ \mathcal{A} |\mathcal{A}|^2 \left[ \langle \tilde{q}_{1},H(\hat{q}_{1},\hat{q}_{20}) \rangle + \langle \tilde{q}_{1}, H(\overline{\hat{q}_{1}},\hat{q}_{22}) \rangle + \right. \left. \frac{1}{2} \langle \tilde{q}_{1},T(\hat{q}_{1}, \hat{q}_{1}, \overline{\hat{q}_{1}}) \rangle \right] 
\label{wna15prev}
\end{multline}

\begin{multline}
\Rightarrow \frac{\text{d} \mathcal{A}}{\text{d} t_1} = \mathcal{A} \left( \frac{\langle \tilde{q}_{1}, H(\hat{q}_{1}, \hat{q}_{21}) \rangle}{\langle \tilde{q}_{1}, \hat{q}_{1} \rangle} + \frac{\langle \tilde{q}_{1}, \partial_{q\epsilon} R(q_0,0) \hat{q}_{1} \rangle}{\langle \tilde{q}_{1}, \hat{q}_{1} \rangle} \right) + \\ \mathcal{A} |\mathcal{A}|^2 \left[ \frac{\langle \tilde{q}_{1},H(\hat{q}_{1},\hat{q}_{20}) \rangle}{\langle \tilde{q}_{1}, \hat{q}_{1} \rangle} + \frac{\langle \tilde{q}_{1}, H(\overline{\hat{q}_{1}},\hat{q}_{22}) \rangle}{\langle \tilde{q}_{1}, \hat{q}_{1} \rangle} + \right.  \left. \frac{1}{2} \frac{\langle \tilde{q}_{1},T(\hat{q}_{1}, \hat{q}_{1}, \overline{\hat{q}_{1}}) \rangle}{\langle \tilde{q}_{1}, \hat{q}_{1} \rangle} \right]
\label{wna15}
\end{multline}

By comparison with the Stuart-Landau amplitude equation \eqref{landau}, one gets the expression of the coefficients $\kappa$, $\mu$, $\nu$ and $\xi$ given in the equations \eqref{kappa}-\eqref{xi}.






\bibliographystyle{plainnat}
\bibliography{biblioAP}

\begin{thebibliography}{64}
\providecommand{\natexlab}[1]{#1}
\providecommand{\url}[1]{\texttt{#1}}
\expandafter\ifx\csname urlstyle\endcsname\relax
  \providecommand{\doi}[1]{doi: #1}\else
  \providecommand{\doi}{doi: \begingroup \urlstyle{rm}\Url}\fi

\bibitem[Amestoy et~al.(2001)Amestoy, Duff, L'Excellent, and
  Koster]{amestoy2001fully}
Patrick~R Amestoy, Iain~S Duff, Jean-Yves L'Excellent, and Jacko Koster.
\newblock A fully asynchronous multifrontal solver using distributed dynamic
  scheduling.
\newblock \emph{SIAM Journal on Matrix Analysis and Applications}, 23\penalty0
  (1):\penalty0 15--41, 2001.

\bibitem[Balay et~al.(2019)Balay, Abhyankar, Adams, Brown, Brune, Buschelman,
  Dalcin, Dener, Eijkhout, Gropp, et~al.]{balay2019petsc}
Satish Balay, Shrirang Abhyankar, Mark Adams, Jed Brown, Peter Brune, Kris
  Buschelman, Lisandro Dalcin, Alp Dener, Victor Eijkhout, W~Gropp, et~al.
\newblock Petsc users manual.
\newblock \emph{ANL-95/11}, 2019.

\bibitem[B{\'e}gou(2018)]{begou2018prevision}
Guillaume B{\'e}gou.
\newblock \emph{Pr{\'e}vision de la transition laminaire-turbulent dans le code
  elsA par la m{\'e}thode des paraboles}.
\newblock PhD thesis, Toulouse, ISAE, 2018.

\bibitem[Bhagatwala and Lele(2009)]{bhagatwala2009modified}
Ankit Bhagatwala and Sanjiva~K Lele.
\newblock A modified artificial viscosity approach for compressible turbulence
  simulations.
\newblock \emph{Journal of Computational Physics}, 14\penalty0 (228):\penalty0
  4965--4969, 2009.

\bibitem[Bogey and Bailly(2004)]{bogey2004family}
Christophe Bogey and Christophe Bailly.
\newblock A family of low dispersive and low dissipative explicit schemes for
  flow and noise computations.
\newblock \emph{Journal of Computational physics}, 194\penalty0 (1):\penalty0
  194--214, 2004.

\bibitem[Bottaro et~al.(2003)Bottaro, Corbett, and Luchini]{bottaro2003effect}
Alessandro Bottaro, Peter Corbett, and Paolo Luchini.
\newblock The effect of base flow variation on flow stability.
\newblock \emph{Journal of Fluid Mechanics}, 476:\penalty0 293--302, 2003.

\bibitem[Brandt et~al.(2011)Brandt, Sipp, Pralits, and
  Marquet]{brandt2011effect}
Luca Brandt, Denis Sipp, Jan~O Pralits, and Olivier Marquet.
\newblock Effect of base-flow variation in noise amplifiers: the flat-plate
  boundary layer.
\newblock \emph{Journal of Fluid Mechanics}, 687:\penalty0 503--528, 2011.

\bibitem[Bugeat et~al.(2019)Bugeat, Chassaing, Robinet, and
  Sagaut]{bugeat20193d}
Benjamin Bugeat, J-C Chassaing, J-C Robinet, and Pierre Sagaut.
\newblock 3d global optimal forcing and response of the supersonic boundary
  layer.
\newblock \emph{Journal of Computational Physics}, 398:\penalty0 108888, 2019.

\bibitem[Canuto et~al.(2012)Canuto, Hussaini, Quarteroni, Thomas~Jr,
  et~al.]{canuto2012spectral}
Claudio Canuto, M~Yousuff Hussaini, Alfio Quarteroni, A~Thomas~Jr, et~al.
\newblock \emph{Spectral methods in fluid dynamics}.
\newblock Springer Science \& Business Media, 2012.

\bibitem[Chapman and Rubesin(1949)]{chapman1949temperature}
Dean~R Chapman and Morris~W Rubesin.
\newblock Temperature and velocity profiles in the compressible laminar
  boundary layer with arbitrary distribution of surface temperature.
\newblock \emph{Journal of the Aeronautical Sciences}, 16\penalty0
  (9):\penalty0 547--565, 1949.

\bibitem[Cho and Aessopos(2004)]{cho2004similarity}
Yeunwoo Cho and Angelica Aessopos.
\newblock Similarity transformation methods in the analysis of the two
  dimensional steady compressible laminar boundary layer.
\newblock \emph{Term paper}, 2, 2004.

\bibitem[Chomaz(2005)]{chomaz2005global}
Jean-Marc Chomaz.
\newblock Global instabilities in spatially developing flows: non-normality and
  nonlinearity.
\newblock \emph{Annu. Rev. Fluid Mech.}, 37:\penalty0 357--392, 2005.

\bibitem[Cinnella and Content(2016)]{cinnella2016high}
Paola Cinnella and C{\'e}dric Content.
\newblock High-order implicit residual smoothing time scheme for direct and
  large eddy simulations of compressible flows.
\newblock \emph{Journal of Computational Physics}, 326:\penalty0 1--29, 2016.

\bibitem[Crivellini and Bassi(2011)]{crivellini2011implicit}
Andrea Crivellini and Francesco Bassi.
\newblock An implicit matrix-free discontinuous galerkin solver for viscous and
  turbulent aerodynamic simulations.
\newblock \emph{Computers \& fluids}, 50\penalty0 (1):\penalty0 81--93, 2011.

\bibitem[Crouch et~al.(2007)Crouch, Garbaruk, and
  Magidov]{crouch2007predicting}
JD~Crouch, A~Garbaruk, and D~Magidov.
\newblock Predicting the onset of flow unsteadiness based on global
  instability.
\newblock \emph{Journal of Computational Physics}, 224\penalty0 (2):\penalty0
  924--940, 2007.

\bibitem[Dalcin et~al.(2011)Dalcin, Paz, Kler, and Cosimo]{dalcin2011parallel}
Lisandro~D Dalcin, Rodrigo~R Paz, Pablo~A Kler, and Alejandro Cosimo.
\newblock Parallel distributed computing using \textsc{P}ython.
\newblock \emph{Advances in Water Resources}, 34\penalty0 (9):\penalty0
  1124--1139, 2011.

\bibitem[De~Pando et~al.(2012)De~Pando, Sipp, and Schmid]{de2012efficient}
Miguel~Fosas De~Pando, Denis Sipp, and Peter~J Schmid.
\newblock Efficient evaluation of the direct and adjoint linearized dynamics
  from compressible flow solvers.
\newblock \emph{Journal of Computational Physics}, 231\penalty0 (23):\penalty0
  7739--7755, 2012.

\bibitem[Ducros et~al.(1999)Ducros, Ferrand, Nicoud, Weber, Darracq, Gacherieu,
  and Poinsot]{ducros1999large}
F~Ducros, V~Ferrand, Franck Nicoud, C~Weber, D~Darracq, C~Gacherieu, and
  Thierry Poinsot.
\newblock Large-eddy simulation of the shock/turbulence interaction.
\newblock \emph{Journal of Computational Physics}, 152\penalty0 (2):\penalty0
  517--549, 1999.

\bibitem[Fabre et~al.(2018)Fabre, Citro, Ferreira~Sabino, Bonnefis, Sierra,
  Giannetti, and Pigou]{fabre2018practical}
David Fabre, Vincenzo Citro, D~Ferreira~Sabino, Paul Bonnefis, Javier Sierra,
  Flavio Giannetti, and Maxime Pigou.
\newblock A practical review on linear and nonlinear global approaches to flow
  instabilities.
\newblock \emph{Applied Mechanics Reviews}, 70\penalty0 (6), 2018.

\bibitem[Fosso et~al.(2010)Fosso, Deniau, Sicot, and
  Sagaut]{fosso2010curvilinear}
Arnaud Fosso, Hugues Deniau, Fr{\'e}d{\'e}ric Sicot, and Pierre Sagaut.
\newblock Curvilinear finite-volume schemes using high-order compact
  interpolation.
\newblock \emph{Journal of Computational Physics}, 229\penalty0 (13):\penalty0
  5090--5122, 2010.

\bibitem[George and Sujith(2011)]{george2011chu}
K~Joseph George and RI~Sujith.
\newblock On chu's disturbance energy.
\newblock \emph{Journal of Sound and Vibration}, 330\penalty0 (22):\penalty0
  5280--5291, 2011.

\bibitem[Giannetti and Luchini(2007)]{giannetti2007structural}
Flavio Giannetti and Paolo Luchini.
\newblock Structural sensitivity of the first instability of the cylinder wake.
\newblock \emph{Journal of Fluid Mechanics}, 581:\penalty0 167--197, 2007.

\bibitem[Giles and Pierce(2001)]{giles2001analytic}
Michael~B Giles and Niles~A Pierce.
\newblock Analytic adjoint solutions for the quasi-one-dimensional
  \textsc{E}uler equations.
\newblock \emph{Journal of Fluid Mechanics}, 426:\penalty0 327--345, 2001.

\bibitem[Gopinath and Jameson(2005)]{gopinath2005time}
Arathi Gopinath and Antony Jameson.
\newblock Time spectral method for periodic unsteady computations over two-and
  three-dimensional bodies.
\newblock In \emph{43rd AIAA aerospace sciences meeting and exhibit}, page
  1220, 2005.

\bibitem[Griewank and Walther(2008)]{griewank2008evaluating}
Andreas Griewank and Andrea Walther.
\newblock \emph{Evaluating derivatives: principles and techniques of
  algorithmic differentiation}.
\newblock SIAM, 2008.

\bibitem[Gunzburger et~al.(1991)Gunzburger, Hou, and
  Svobodny]{gunzburger1991analysis}
Max~D Gunzburger, L~Hou, and Th~P Svobodny.
\newblock Analysis and finite element approximation of optimal control problems
  for the stationary navier-stokes equations with distributed and neumann
  controls.
\newblock \emph{Mathematics of Computation}, 57\penalty0 (195):\penalty0
  123--151, 1991.

\bibitem[Hanifi et~al.(1996)Hanifi, Schmid, and
  Henningson]{hanifi1996transient}
Ardeshir Hanifi, Peter~J Schmid, and Dan~S Henningson.
\newblock Transient growth in compressible boundary layer flow.
\newblock \emph{Physics of Fluids}, 8\penalty0 (3):\penalty0 826--837, 1996.

\bibitem[Hascoet and Pascual(2013)]{hascoet2013tapenade}
Laurent Hascoet and Val{\'e}rie Pascual.
\newblock The tapenade automatic differentiation tool: principles, model, and
  specification.
\newblock \emph{ACM Transactions on Mathematical Software (TOMS)}, 39\penalty0
  (3):\penalty0 1--43, 2013.

\bibitem[Hecht(2012)]{hecht2012new}
Fr{\'e}d{\'e}ric Hecht.
\newblock New development in freefem++.
\newblock \emph{Journal of numerical mathematics}, 20\penalty0 (3-4):\penalty0
  251--266, 2012.

\bibitem[Hern{\'a}ndez et~al.(2007)Hern{\'a}ndez, Rom{\'a}n, Tom{\'a}s, and
  Vidal]{hernandez2007krylov}
Vicente Hern{\'a}ndez, Jose~E Rom{\'a}n, Andr{\'e}s Tom{\'a}s, and Vicente
  Vidal.
\newblock Krylov-\textsc{S}chur methods in slepc.
\newblock \emph{Universitat Politecnica de Valencia, Tech. Rep. STR-7}, 2007.

\bibitem[Huerre and Monkewitz(1990)]{huerre1990local}
Patrick Huerre and Peter~A Monkewitz.
\newblock Local and global instabilities in spatially developing flows.
\newblock \emph{Annual review of fluid mechanics}, 22\penalty0 (1):\penalty0
  473--537, 1990.

\bibitem[Jackson(1987)]{jackson1987finite}
CP~Jackson.
\newblock A finite-element study of the onset of vortex shedding in flow past
  variously shaped bodies.
\newblock \emph{Journal of fluid Mechanics}, 182:\penalty0 23--45, 1987.

\bibitem[Jameson and Turkel(1981)]{jameson1981implicit}
Antony Jameson and Eli Turkel.
\newblock Implicit schemes and \textsc{LU} decompositions.
\newblock \emph{Mathematics of Computation}, 37\penalty0 (156):\penalty0
  385--397, 1981.

\bibitem[Jameson et~al.(1981)Jameson, Schmidt, and
  Turkel]{jameson1981numerical}
Antony Jameson, Wolfgang Schmidt, and Eli Turkel.
\newblock Numerical solution of the \textsc{E}uler equations by finite volume
  methods using runge kutta time stepping schemes.
\newblock In \emph{14th fluid and plasma dynamics conference}, page 1259, 1981.

\bibitem[Knoll and Keyes(2004)]{knoll2004jacobian}
Dana~A Knoll and David~E Keyes.
\newblock Jacobian-free \textsc{N}ewton-\textsc{K}rylov methods: a survey of
  approaches and applications.
\newblock \emph{Journal of Computational Physics}, 193\penalty0 (2):\penalty0
  357--397, 2004.

\bibitem[Lerat and Corre(2003)]{lerat2003approximations}
A~Lerat and C~Corre.
\newblock Approximations d’ordre {\'e}lev{\'e} pour les {\'e}coulements
  compressibles.
\newblock \emph{Ecole de Printemps de M{\'e}canique des Fluides Num{\'e}rique,
  Fr{\'e}jus}, 2003.

\bibitem[Lerat and Corre(2006)]{lerat2006}
A.~Lerat and C.~Corre.
\newblock Higher order residual-based compact schemes on structured grids.
\newblock \emph{VKI LS 2006-1, von Karman Institute for Fluid Dynamics}, 2006.

\bibitem[Logg and Wells(2010)]{logg2010dolfin}
Anders Logg and Garth~N Wells.
\newblock Dolfin: Automated finite element computing.
\newblock \emph{ACM Transactions on Mathematical Software (TOMS)}, 37\penalty0
  (2):\penalty0 1--28, 2010.

\bibitem[Mack(1963)]{mack1963inviscid}
LM~Mack.
\newblock The inviscid stability of the compressible laminar boundary layer.
\newblock \emph{Space Programs Summary}, 37:\penalty0 23, 1963.

\bibitem[Marquet et~al.(2008{\natexlab{a}})Marquet, Sipp, Chomaz, and
  Jacquin]{marquet2008amplifier}
Olivier Marquet, Denis Sipp, Jean-Marc Chomaz, and Laurent Jacquin.
\newblock Amplifier and resonator dynamics of a low-reynolds-number
  recirculation bubble in a global framework.
\newblock \emph{Journal of Fluid Mechanics}, 605:\penalty0 429--443,
  2008{\natexlab{a}}.

\bibitem[Marquet et~al.(2008{\natexlab{b}})Marquet, Sipp, and
  Jacquin]{marquet2008sensitivity}
Olivier Marquet, Denis Sipp, and Laurent Jacquin.
\newblock Sensitivity analysis and passive control of cylinder flow.
\newblock \emph{Journal of Fluid Mechanics}, 615:\penalty0 221--252,
  2008{\natexlab{b}}.

\bibitem[Martini et~al.(2021)Martini, Rodr{\'\i}guez, Towne, and
  Cavalieri]{martini2021efficient}
Eduardo Martini, Daniel Rodr{\'\i}guez, Aaron Towne, and Andr{\'e}~VG
  Cavalieri.
\newblock Efficient computation of global resolvent modes.
\newblock \emph{Journal of Fluid Mechanics}, 919, 2021.

\bibitem[Maugars(2016)]{maugars2016methodes}
Bruno Maugars.
\newblock \emph{M{\'e}thodes de volumes finis d'ordre {\'e}lev{\'e} en
  maillages non co{\"\i}ncidents pour les {\'e}coulements dans les
  turbomachines}.
\newblock PhD thesis, Paris, ENSAM, 2016.

\bibitem[Meliga et~al.(2014)Meliga, Boujo, Pujals, and
  Gallaire]{meliga2014sensitivity}
Philippe Meliga, Edouard Boujo, Gregory Pujals, and Fran{\c{c}}ois Gallaire.
\newblock Sensitivity of aerodynamic forces in laminar and turbulent flow past
  a square cylinder.
\newblock \emph{Physics of Fluids}, 26\penalty0 (10):\penalty0 104101, 2014.

\bibitem[Mettot(2013)]{mettot2013linear}
Cl{\'e}ment Mettot.
\newblock \emph{Linear stability, sensitivity, and passive control of turbulent
  flows using finite differences}.
\newblock PhD thesis, Palaiseau, Ecole polytechnique, 2013.

\bibitem[Mettot et~al.(2014)Mettot, Renac, and Sipp]{mettot2014computation}
Cl{\'e}ment Mettot, Florent Renac, and Denis Sipp.
\newblock Computation of eigenvalue sensitivity to base flow modifications in a
  discrete framework: Application to open-loop control.
\newblock \emph{Journal of Computational Physics}, 269:\penalty0 234--258,
  2014.

\bibitem[Moulin(2020)]{moulin2020flutter}
Johann Moulin.
\newblock \emph{On the flutter bifurcation in laminar flows: linear and
  nonlinear modal methods}.
\newblock PhD thesis, Institut polytechnique de Paris, 2020.

\bibitem[{\"O}zgen and K{\i}rcal{\i}(2008)]{ozgen2008linear}
Serkan {\"O}zgen and Senem~Atalayer K{\i}rcal{\i}.
\newblock Linear stability analysis in compressible, flat-plate
  boundary-layers.
\newblock \emph{Theoretical and Computational Fluid Dynamics}, 22\penalty0
  (1):\penalty0 1--20, 2008.

\bibitem[Palacios et~al.(2013)Palacios, Alonso, Duraisamy, Colonno, Hicken,
  Aranake, Campos, Copeland, Economon, Lonkar, Lukaczyk, and
  Taylor]{palacios_stanford_2013}
Francisco Palacios, Juan Alonso, Karthikeyan Duraisamy, Michael Colonno, Jason
  Hicken, Aniket Aranake, Alejandro Campos, Sean Copeland, Thomas Economon,
  Amrita Lonkar, Trent Lukaczyk, and Thomas Taylor.
\newblock Stanford {University} {Unstructured} ({SU}{\textasciicircum}2): {An}
  open-source integrated computational environment for multi-physics simulation
  and design.
\newblock In \emph{51st {AIAA} {Aerospace} {Sciences} {Meeting} including the
  {New} {Horizons} {Forum} and {Aerospace} {Exposition}}. American Institute of
  Aeronautics and Astronautics, 2013.
\newblock \doi{10.2514/6.2013-287}.

\bibitem[Poinsot and Lele(1992)]{poinsot1992boundary}
T.J. Poinsot and S.K. Lele.
\newblock Boundary conditions for direct simulations of compressible viscous
  flows.
\newblock \emph{Journal of computational physics}, 101\penalty0 (1):\penalty0
  104--129, 1992.

\bibitem[Rezgui et~al.(2001)Rezgui, Cinnella, and Lerat]{rezgui2001third}
Ali Rezgui, Paola Cinnella, and Alain Lerat.
\newblock Third-order accurate finite volume schemes for \textsc{E}uler
  computations on curvilinear meshes.
\newblock \emph{Computers \& fluids}, 30\penalty0 (7-8):\penalty0 875--901,
  2001.

\bibitem[Rigas et~al.(2021)Rigas, Sipp, and Colonius]{rigas2021nonlinear}
Georgios Rigas, Denis Sipp, and Tim Colonius.
\newblock Nonlinear input/output analysis: application to boundary layer
  transition.
\newblock \emph{Journal of Fluid Mechanics}, 911, 2021.

\bibitem[Roman et~al.(2015)Roman, Campos, Romero, and
  Tom{\'a}s]{roman2015slepc}
Jose~E Roman, Carmen Campos, Eloy Romero, and Andr{\'e}s Tom{\'a}s.
\newblock Slepc users manual.
\newblock \emph{D. Sistemes Inform{\`a}tics i Computaci{\'o} Universitat
  Polit{\`e}cnica de Val{\`e}ncia, Valencia, Spain, Report No. DSIC-II/24/02},
  2015.

\bibitem[Sartor et~al.(2015)Sartor, Mettot, Bur, and
  Sipp]{sartor2015unsteadiness}
Fulvio Sartor, Cl{\'e}ment Mettot, Reynald Bur, and Denis Sipp.
\newblock Unsteadiness in transonic shock-wave/boundary-layer interactions:
  experimental investigation and global stability analysis.
\newblock \emph{Journal of Fluid Mechanics}, 781:\penalty0 550--577, 2015.

\bibitem[Sciacovelli et~al.(2021)Sciacovelli, Passiatore, Cinnella, and
  Pascazio]{sciacovelli2021assessment}
Luca Sciacovelli, Donatella Passiatore, Paola Cinnella, and Giuseppe Pascazio.
\newblock Assessment of a high-order shock-capturing central-difference scheme
  for hypersonic turbulent flow simulations.
\newblock \emph{Computers \& Fluids}, 230:\penalty0 105134, 2021.

\bibitem[Shen et~al.(2009)Shen, Zha, and Chen]{shen2009high}
Yiqing Shen, Gecheng Zha, and Xiangying Chen.
\newblock High order conservative differencing for viscous terms and the
  application to vortex-induced vibration flows.
\newblock \emph{Journal of Computational Physics}, 228\penalty0 (22):\penalty0
  8283--8300, 2009.

\bibitem[Sipp and Lebedev(2007)]{sipp2007global}
Denis Sipp and Anton Lebedev.
\newblock Global stability of base and mean flows: a general approach and its
  applications to cylinder and open cavity flows.
\newblock \emph{Journal of Fluid Mechanics}, 593:\penalty0 333--358, 2007.

\bibitem[Sipp and Marquet(2013)]{sipp2013characterization}
Denis Sipp and Olivier Marquet.
\newblock Characterization of noise amplifiers with global singular modes: the
  case of the leading-edge flat-plate boundary layer.
\newblock \emph{Theoretical and Computational Fluid Dynamics}, 27\penalty0
  (5):\penalty0 617--635, 2013.

\bibitem[Sipp et~al.(2010)Sipp, Marquet, Meliga, and
  Barbagallo]{sipp2010dynamics}
Denis Sipp, Olivier Marquet, Philippe Meliga, and Alexandre Barbagallo.
\newblock Dynamics and control of global instabilities in open-flows: a
  linearized approach.
\newblock \emph{Applied Mechanics Reviews}, 63\penalty0 (3), 2010.

\bibitem[Sutherland(1893)]{sutherland1893lii}
William Sutherland.
\newblock Lii. the viscosity of gases and molecular force.
\newblock \emph{The London, Edinburgh, and Dublin Philosophical Magazine and
  Journal of Science}, 36\penalty0 (223):\penalty0 507--531, 1893.

\bibitem[Thomas et~al.(2010)Thomas, Lazarus, and Touz{\'e}]{thomas2010harmonic}
Olivier Thomas, Arnaud Lazarus, and Cyril Touz{\'e}.
\newblock A harmonic-based method for computing the stability of periodic
  oscillations of non-linear structural systems.
\newblock In \emph{International Design Engineering Technical Conferences and
  Computers and Information in Engineering Conference}, volume 44137, pages
  883--892, 2010.

\bibitem[Trefethen et~al.(1993)Trefethen, Trefethen, Reddy, and
  Driscoll]{trefethen1993hydrodynamic}
Lloyd~N Trefethen, Anne~E Trefethen, Satish~C Reddy, and Tobin~A Driscoll.
\newblock Hydrodynamic stability without eigenvalues.
\newblock \emph{Science}, 261\penalty0 (5121):\penalty0 578--584, 1993.

\bibitem[Yoon and Kwak(1991)]{yoon1991three}
Seokkwan Yoon and Dochan Kwak.
\newblock Three-dimensional incompressible navier-stokes solver using
  lower-upper symmetric-gauss-seidel algorithm.
\newblock \emph{AIAA journal}, 29\penalty0 (6):\penalty0 874--875, 1991.

\bibitem[Zingg et~al.(2000)Zingg, De~Rango, Nemec, and
  Pulliam]{zingg2000comparison}
DW~Zingg, S~De~Rango, M~Nemec, and TH~Pulliam.
\newblock Comparison of several spatial discretizations for the navier--stokes
  equations.
\newblock \emph{Journal of computational Physics}, 160\penalty0 (2):\penalty0
  683--704, 2000.

\end{thebibliography}







\end{document}